\documentstyle[11pt,epsf]{article}

\newcommand\ee{\end{equation}}
\newcommand\be{\begin{equation}}
\newcommand\eea{\end{eqnarray}}
\newcommand\bea{\begin{eqnarray}}

\newcommand\GeV{\,\mbox{GeV}}
\newcommand\MeV{\,\mbox{MeV}}

\newcommand\mpl{M_{\rm P}}

\newcommand\lsim{\mathrel{\rlap{\lower4pt\hbox{\hskip1pt$\sim$}}
    \raise1pt\hbox{$<$}}}
\newcommand\gsim{\mathrel{\rlap{\lower4pt\hbox{\hskip1pt$\sim$}}
    \raise1pt\hbox{$>$}}}

\def\mstop{m_{\,\widetilde{t}}}

\def\dslash{\not{\hbox{\kern-2pt $\partial$}}}
\def\Dslash{\not{\hbox{\kern-4pt $D$}}}
\def\Oslash{\not{\hbox{\kern-4pt $O$}}}
\def\Qslash{\not{\hbox{\kern-4pt $Q$}}}
\def\pslash{\not{\hbox{\kern-2.3pt $p$}}}
\def\kslash{\not{\hbox{\kern-2.3pt $k$}}}
\def\qslash{\not{\hbox{\kern-2.3pt $q$}}}
 \newtoks\slashfraction
 \slashfraction={.13}
 \def\slash#1{\setbox0\hbox{$ #1 $}
 \setbox0\hbox to \the\slashfraction\wd0{\hss \box0}/\box0 }
 
\def\eeq{\end{equation}}
\def\beq{\begin{equation}}

\newcommand\sub[1]{_{\rm #1}}

\begin{document}

\begin{titlepage}
\begin{flushright}
CERN-TH/98-204\\
hep-ph/9807454\\
 \today
\end{flushright}
\begin{center}
{\Large \bf Theories of Baryogenesis}

\vspace{2cm}
{\large\bf   Antonio 
Riotto $^{*,}$\footnote{On leave of absence {}from Theoretical 
Physics Department, University of Oxford,U.K.}
\\}
\vspace{.4 cm}
{\em $^*$  CERN, Theory Division,\\
CH-1211, Geneva 23, Switzerland.\\
{\tt E-mail: riotto@nxth04.cern.ch}}
\end{center}

\vspace{.6cm}
\begin{abstract}
\noindent
These   lectures  provide a pedagogical review of the present status of   theories explaining the observed baryon asymmetry of the Universe. Particular 
 emphasis is given  on GUT baryogenesis and  electroweak baryogenesis. The    key issues, the  unresolved problems and the very recent developments, such as GUT baryogenesis during preheating, are explained.  Some exercises (and their solution) are also provided.

\end{abstract}

\vspace{2cm}

\centerline{ Lectures delivered at the {\it Summer School in High Energy Physics and Cosmology,}}
\vspace{0.2cm}
\centerline{Miramare-Trieste, Italy, 29 June -17 July 1998.}

\end{titlepage}
                                                                         
\tableofcontents
\newpage
\setcounter{page}{2}
\section{Introduction}

We do not know the history of the observable Universe before the epoch
of nucleosynthesis, but it is widely believed that there was an
early era of cosmological inflation. The attraction of this paradigm is that it can 
set the initial conditions for the subsequent hot big-bang, which 
otherwise have to be imposed by hand. One of these is 
that there be no unwanted relics (particles or topological defects
which survive to the present and contradict observation).
Another is that the initial density parameter should have the value 
$\Omega=1$ to very high accuracy, to ensure that its present value
has at least roughly this value. There is also the requirement
that the Universe be homogeneous and isotropic to high accuracy.
The flatness and the horizon problems of the standard big bang
cosmology are -- indeed --  elegantly solved if during the evolution of the early
Universe the energy density happens to be dominated by the vacuum energy of a scalar field -- the inflaton -- and comoving scales grow quasi-exponentially. 

At the end of inflation the energy density of the Universe is locked up in a combination of kinetic energy and potential energy of the inflaton field, with the bulk of the inflaton energy density in the
zero-momentum mode of the field.  Thus, the Universe at the end of inflation is in a cold, low-entropy state with few degrees of freedom, very much unlike the present hot, high-entropy universe.  The process by which the inflaton energy density is converted into radiation is known as reheating.  
What is crucial about these considerations is that,  at the end of inflation, the Universe does not contain any matter and -- even more important -- the  Universe looks perfectly baryon symmetric -- there is no dominance of matter over antimatter. 

The observed Universe -- however -- is drastically different. We  do not observe any bodies of antimatter around us  within the solar system 
and if domains of  antimatter exist in the Universe, they are separated from us on scales certainly larger than  the Virgo cluster $(\sim 10$ Mpc). The Universe looks baryon asymmetric to us. Considerations about how the light element abundances were formed when the Universe was about 1 MeV hot lead us to conclude that the difference between the number density of baryons and that of antibaryons is about $10^{-10}$ if normalized to the entropy density of the Universe. 

Theories that explain how to produce such a tiny number go generically  under the name of {\it Theories of  baryogenesis} and they represent
 perhaps the best example of the perfect interplay between particle physics and cosmology.  Until now, many mechanisms for the generation of the baryon asymmetry have been proposed and we have no idea which is the correct one. Grand Unified Theories (GUTs) unify the strong and the electroweak interactions and predict baryon number violation at the tree level. They are -- therefore -- perfect candidates for a theory of baryogenesis. There, the out-of-equilibrium decay of 
superheavy particles can explain the observed baryon asymmetry, even though there remain problems strictly related to the dynamics of reheating after inflation. In the theory of electroweak baryogenesis, baryon number violation takes place at the quantum level due to the chiral anomaly. Baryogenesis scenarios at the electroweak scale have been the subject of intense activity in the last few years. They are certainly attractive because they can be tested at the current and future accelerator experiments. 

The bottom line of all this intense  research   is that, within  the standard model of weak interactions, it is difficult, if not impossible, to explain how the generation of the baryon asymmetry took place. Therefore, the observation of a baryon asymmetry in the Universe is an indication that the description of Nature cannot be limited to the Weinberg-Salam theory,  something else is called for. 

The goal  of these  lectures is to provide a pedagogical review of the present state of  baryogenesis, with particular emphasis  on GUT baryogenesis and  electroweak baryogenesis. The technical details of the numerous models considered in the literature are not elaborated, but  the   key points, the  unresolved problems and the very recent developments -- such as GUT baryogenesis during preheating -- are presented. We hope that this approach will help the reader to get interested in this fascinating subject. A different focus may be found in other accounts of the subject \cite{review1,review2,review3,review4}. Some exercises (and their solution) are also provided.

The review is laid out as follows. Section 2 describes some necessary tools of equilibrium thermodynamics. Section 3 contains some considerations about the baryon symmetric Universe and explains the three basic conditions necessary to generate the baryon asymmetry. The standard out-of-equilibrium scenario is addressed in Section 4, while Section 5 contains informations about GUT baryogenesis and the thermal history of the Universe, with particular
attention paid to the 
recent developments related to the theory of preheating. Section 6 is dedicated to the issue of baryon number violation in the standard model and its possible implications for GUT baryogenesis and leptogenesis. Finally, section 7 addresses the rapidly moving subject of electroweak baryogenesis.

A note about conventions. We employ units such that $\hbar=c=k=1$ and references are listed in alphabetic order.

\section{Some necessary notions of equilibrium thermodynamics}

\subsection{Expansion rate, number density,  and entropy}

Before  launching ourselves into the issue of baryon asymmetry production in the early Universe, let us just remind the reader a few  notions about thermodynamics in an expanding Universe that will turn out to be useful in the following. According to general relativity,  the space-time evolution is determined via the Einstein equation by the matter content of the Universe, which differs from epoch to epoch depending on what kind of energy dominates the energy density of the Universe at that time. There are three important epochs characterized by different relation between the energy density $\rho$ and the pressure $p$: {\it 1)} vacuum energy dominance with $p=-\rho$, {\it 2)} massless (relativistic) particle dominance with $p=\rho/3$ and {\it 3)} nonrelativistic particle dominance with $p\ll \rho$. 
The Einstein equation reads
\be
\label{ein}
R_{\mu\nu}-\frac{1}{2}g_{\mu\nu} {\cal R}=8\pi G_N T_{\mu\nu},
\ee
where $R_{\mu\nu}$ is the Ricci tensor, ${\cal R}$ is the Ricci scalar, $g_{\mu\nu}$ is the metric, $G_N=\mpl^{-2}=(1.2\times 10^{19})^{-2}\:{\rm GeV}^{-2}$ is the Newton constant and $T_{\mu\nu}$ is the stress-energy tensor.

With the homogeneity and isotropy of the three-space, the Einstein equation is much simplified with the Robertson-Walker metric 
\be ds^2= dt^2-a^2(t)\vec{x}^2,
\ee 
where $a(t)$ is the cosmic scale factor and the stress-energy tensor is reduced to $T_{\mu\nu}=-p g_{\mu\nu}+(p+\rho)u_\mu u_\nu$. Here  $u^\mu$ is the velocity vector which in the rest frame of the plasma reads $u^\mu=(1,{\bf 0})$ and  has the property $u^\mu u_\mu=1$. The  
 $0-0$ component  of eq. (\ref{ein}) becomes  
 the so-called Friedmann equation
\be
H^2+\frac{k}{a^2}=\frac{8\pi G_N}{3}\rho,
\ee
where $k$ can be chosen to be $+1$, $-1$ or $0$ for spaces of constant positive, negative or zero spatial curvature, respectively, and we have defined the Hubble parameter 
\be
H\equiv \frac{\dot{a}}{a},
\ee
which meaures how fast the Universe is expanding during the different stages of its evolution.

The $\mu=0$ component of the conservation of the stress-energy tensor ($T^{\mu\nu}_{;\nu}=0$) gives the first law of thermodynamics in the familiar form
\be
d(\rho a^3)=-p\:d(a^3),
\ee
that is, the change in energy in a comiving volume element, $d(\rho a^3)$ is equal to minus the pressure times the change in volume, $p\:d(a^3)$.
For a simple equation of state $p=w\rho$, where $w$ is independent of time, the energy density evolves like $\rho\propto a^{-3(1+w)}$. Examples of interest include radiation ($\rho\propto a^{-4}$), matter  ($\rho\propto a^{-3}$), vacuum energy ($\rho\propto$ constant).
The time-behaviour of the scale factor $a(t)$ then is
\begin{eqnarray}
\label{time}
{\it 1)} \:\:\:a&\propto& e^{Ht},\:\:H=\sqrt{\frac{8\pi V}{3\mpl^2}}, \nonumber\\
{\it 2)} \:\:\:a&\propto& t^{1/2}, \nonumber\\
{\it 3)} \:\:\:a&\propto& t^{2/3}. \nonumber\\
\end{eqnarray}
The  first stage is the inflationary epoch where the constant vacuum energy $V$ gives the exponential growth of the scale factor, which is believed to solve the horizon and the flatness problems of the standard big-bang theory of cosmology \cite{guth} (for a review, see \cite{abook}). Of great importance is the transient stage from inflation to radiation dominance. This epoch is called reheating after inflation and we shall come back to it   later in these lectures.

What is relevant for us is that the early Universe was to a good approximation in thermal equilibrium at  temperature $T$ \cite{book} and we can define the equilibrium number density $n^{{\rm EQ}}_X$ of a generic interacting species $X$ as
\be
\label{number}
n_X^{{\rm EQ}}=\frac{g_X}{(2\pi)^3}\int\:f_{{\rm EQ}}({\bf p},\mu_X)\:d^3p,
\ee
where $g_X$ denotes the number of degrees of freedom of the species $X$ and  the phase space occupancy $f_{{\rm EQ}}$ is given by the familiar Fermi-Dirac or Bose-Einstein distributions
\be
f_{{\rm EQ}}({\bf p},\mu_X)=\left[{\rm exp}((E_X-\mu_X)/T)\pm 1\right],
\ee
where $E_X=\left({\bf p}^2+m_X^2\right)^{1/2}$ is the energy, 
$\mu_X$ is the chemical potential of the species and $+1$ pertains to the Fermi-Dirac species and $-1$ to the Bose-Einstein species. 

In the relativistic regime $T\gg m_X,\mu_X$ formula (\ref{number}) reduces to 
\be
n_X^{{\rm EQ}}=\left\{
\begin{array}{cc}
(\zeta(3)/\pi^2)g_X T^3 & ({\rm Bose}),\\
(3/4)(\zeta(3)/\pi^2)g_X T^3  & ({\rm Fermi}),
\end{array}\right.
\ee
where $\zeta(3)\simeq 1.2$ is the Riemann function of 3. In the non-relativitic limit, $T\ll m_X$, the number density is the same for Bose and Fermi species and reads
\be
n_X^{{\rm EQ}}=g_X\left(\frac{m_X T}{2\pi}\right)^{3/2}\:e^{-\frac{m_X}{T}+\frac{\mu_X}{T}}.
\ee
It is also important to define the  number density of particles minus the number density of antiparticles 
\begin{eqnarray}
\label{chem}
n_X^{{\rm EQ}}-n_{\overline{X}}^{{\rm EQ}}&=&\frac{g_X}{(2\pi)^3}\int\:f_{{\rm EQ}}({\bf p},\mu_X)\:d^3p - (\mu_X\leftrightarrow -\mu_X)\nonumber\\
&=&\left\{
\begin{array}{cc}
\frac{g_X T^3}{6\pi^2}\left[\pi^2\left(\frac{\mu_X}{T}\right)+\left(\frac{\mu_X}{T}\right)^3\right] &(T\gg m_X)\\
2 g_X(m_X T/2\pi)^{3/2}\:{\rm sinh}(\mu_X/T)\:{\rm exp}(-m_X/T)& (T\ll m_X).
\end{array}\right.
\end{eqnarray}
Notice that, in the relativistic limit $T\gg m_X$, this difference scales linearly for $T\gsim \mu_X$. This means that detailed balances among particle number asymmetries may be expressed in terms of linear equations in the chemical potentials.

We can similarly define the equilibrium energy density $\rho_X^{{\rm EQ}}$ of a  species $X$ as
\be
\rho_X^{{\rm EQ}}=\frac{g_X}{(2\pi)^3}\int\:E\:f_{{\rm EQ}}({\bf p})\:d^3p,
\end{equation}
which reads in the relativistic limit
\be
\rho_X^{{\rm EQ}}=\left\{
\begin{array}{cc}
(\pi^2/30)g_X T^4 & ({\rm Bose}),\\
(7/8)(\pi^2/30)g_X T^4  & ({\rm Fermi}).
\end{array}\right.
\ee
Since the energy density of a non-relativistic particle species is exponentially smaller than that of a relativisitic species, it is a very convenient approximation to  include only relativistic species with energy density $\rho_R$  in the total energy density $\rho$ of the Universe at temperature $T$
\be
\rho\simeq \rho_R=\frac{\pi^2}{30}g_* T^4,
\ee
where $g_*$ counts the total number of effectively massless degrees of freedom of the plasma 
\be
g_*= \sum_{i={\rm bos}}g_i\left(\frac{T_i}{T}\right)^4+
\frac{7}{8}\sum_{i={\rm fer}}g_i\left(\frac{T_i}{T}\right)^4.
\ee
Here $T_i$ denotes the effective temperature of any species $i$ (which might be decoupled from the thermal bath at temperature $T$). In the rest of these lectures we will be always concern with temperatures higher than about 100 GeV. At these temperatures,  all the degrees of freedom of the standard model are in equilibrium and $g_*$ is at least equal to 106.75.

From this expression we derive that, when the energy density of the Universe was dominated by a gas of relativistic particles,  $\rho\propto a^{-4}\propto T^4$ and, therefore \cite{book}
\be
T\propto a^{-1}.
\ee 
Assuming that during the early radiation-dominated epoch $(t\lsim 4\times 10^{10}$ sec), 
the scale factor  scales like  $t^\alpha$, where $\alpha$ is a constant, the Hubble parameter   scales like $t^{-1}\propto T^2\propto a^{-2}$. This means that 
the scale factor $a(t)$ scales like $t^{1/2}$ and we recover {\it 2)} of Eq. (\ref{time}). More precisely, the expansion rate  of the Universe $H$ is  \cite{book}
\be
\label{h}
H=\left(\frac{8\pi}{3\mpl^2}\rho\right)^{1/2}
\simeq 1.66\:g_*^{1/2}\frac{T^2}{\mpl}.
\ee
Using the fact that $H=(1/2 t)$ and Eq. (\ref{h}), we can easily relate time and temperature as
\be
t\simeq 0.301\frac{\mpl}{g_*^{1/2}T^2}\simeq\left(\frac{T}{{\rm MeV}}\right)^{-2}\:{\rm sec}.
\ee
Another quantity that will turn out to be useful in the following is the entropy density. Throughout most of the history of the Universe, local thermal equilibrium is attained and the entropy in a comoving volume element $s$ remains constant. Since it is dominated by the contribution of relativisitic particles, to a very good approximation
\be
s=\frac{2\pi^2}{45}g_{*S}T^3,
\ee
where
\be
g_{*S}= \sum_{i={\rm bos}}g_i\left(\frac{T_i}{T}\right)^3+
\frac{7}{8}\sum_{i={\rm fer}}g_i\left(\frac{T_i}{T}\right)^3.
\ee
For most of the history of the Universe, however, all the particles have the same temperature and we can safely replace $g_{*S}$ with $g_*$. Notice that the conservation of entropy implies that $s\propto a^{-3}$ and therefore $g_{*S}T^3a^3$ remains a constant as the Universe expands. This means that the number of some species $X$ in a comoving volume $N_X\equiv a^3 n_X$ is proportional  to the number density of that species divided by $s$, $N_X\propto n_X/s$.

\subsection{Local thermal equilibrium and chemical equilibrium}

So far we have been using the fact that, throughout most of the history of the Universe, thermal equilibrium was attained. The characteristic time $\tau_X$ for particles of a species $X$ with respect to the process $X+A\cdots \rightarrow C+D+\cdots$ is defined by the rate of change of the number of particles per unit volume $n_X$ due to this process
\be
\frac{1}{\tau_X}=-\frac{1}{n_X}\left(\frac{dn_X}{dt}\right)_{X+A\cdots \rightarrow C+D+\cdots}.
\ee
In the early Universe, if $\tau_X$ is smaller than the characteristic time of the expansion $H^{-1}$, then there is enough time for the process to occur and the particles $X$'s are said to be {\it thermally} coupled to the cosmic fluid. By contrast, if $\tau_X\gg H^{-1}$, for {\it every} process in which the particles
$X$'s are involved, then they are not in thermal equilibrium and they are said to be decoupled. 

In order to analyze the evolution of the particle populations which constitute the cosmic fluid, it is necessary to compare $H^{-1}$ with $\tau_X$ at different temperatures. This is done through the Boltzmann equation \cite{kin}, which, in an expanding Universe,  reads
\be
\label{bol}
\frac{1}{a^3}\frac{d}{dt}(a^3 n_X)=\int\:\pi_X\:C[f_X],
\ee
where $C$ is the collision operator. Eq. (\ref{bol}) may be rewritten as
\begin{eqnarray}
\label{aa}
\frac{dn_X}{dt}&+&3H n_X=\sum_{j,\ell,m,\cdots}\:
\int\:\pi\: f_\ell f_m\cdots(1\pm f_X)(1\pm f_j)\cdots
W\left(\ell +m+\cdots\rightarrow X+j+\cdots\right)\nonumber\\
&-&f_X f_j\cdots (1\pm f_\ell)(1\pm f_m)\cdots W\left(X+j+\cdots\rightarrow \ell +m+\cdots\right),
\end{eqnarray}
where $\pi\equiv \pi_X\pi_j\cdots \pi_\ell\pi_m\cdots$, $\pi_i=(2\pi)^{-3} g_i (d^3 p/2 E_i)$ is the volume element in the phase space, $W$ is the matrix element of the given process and $(+)$ applies to bosons and $(-)$ to fermions..
The second term in the left-hand side of  Eq. (\ref{aa}) accounts for the $n_X$ diluition due to the cosmic expansion and the right-hand side accounts for the $n_X$ variations due to any elemenatry process $X+j+\cdots\rightarrow \ell +m+\cdots$ in which the $X$ particles are involved. As it stands, Eq. (\ref{aa}) is rather formidable and complicated, but some approximations can be made to transform it in  a simpler form. 

Let us consider, for example, a process like $X+f\rightarrow X^\prime+f$, where the number of $X$ particles does  change in the scatterings and let us also suppose that the $f$ particles are light ($T\gg m_f)$ and that the corresponding population  is in thermal equilibrium. In the case in which the $X$ distribution function is   described by a Maxwell-Boltzmann distribution, {\it i.e.} the $X$ particles are in equilibrium at temperatures smaller than $m_X$, it is easy to see that the right-hand side of  Eq. (\ref{aa}) may be expressed in the form
\be
{\rm r.h.s.}\:\:\:{\rm of}\:\:\:{\rm Eq.}\:\: (\ref{aa})=
-\left[n_X-n_X^{{\rm EQ}}\right] \:S
\ee
where
\begin{eqnarray}
S&=&\int\:\pi_f\:E_f\:f_f^{{\rm EQ}} \: \sigma(X+f\rightarrow X^\prime
+f)\:\nonumber\\
&\simeq& n_f^{{\rm EQ}}\:\langle \sigma(X+f\rightarrow X^\prime
+f)\: v\rangle.
\end{eqnarray}
The notation $\langle\sigma v\rangle$ stands for the thermal average cross section times the relative velocity $v$. The inverse time scale $\tau_X^{-1}=\Gamma_X$  associated to the elastic process is therefore 
\be
 \Gamma_X=\tau_X^{-1}\simeq n_f^{{\rm EQ}}\langle\sigma(X+f\rightarrow X^\prime
+f) v\rangle.
\ee
From these very simple considerations, we may conclude that the $X$ degrees of freedom are in thermal equilibrium if 
\be
\Gamma_X\simeq n_f^{{\rm EQ}}\langle\sigma(X+f\rightarrow X^\prime
+f) v\rangle \gsim H \:\:\: \:\:\:({\rm thermal}\:{\rm  equilibrium}\:{\rm  is}\: {\rm attained}).
\ee
Departure from thermal equilibrium is expected whenever a rate crucial for mantaining thermal equilibrium becomes smaller than the expansion rate, $\Gamma_X \lsim H$.

Another useful concept is that of {\it chemical} equilibrium. In general, a species $X$ is in chemical equilibrium if the inelastic scatterings which change the number of $X$ particles in the plasma, $X+j\rightarrow \ell+m$,  have a rate $\Gamma_{{\rm inel}}$ larger than the expansion rate of the Universe. In such a case, one is allowed to  write down a relation between the different chemical potentials $\mu$'s 
\be
\mu_X+\mu_j=\mu_\ell+\mu_m
\ee
of the particles involved in the process. With these simple notions in mind we may start our voyage towards the country of baryogenesis.

\section{The graveyard for a baryon symmetric Universe}

The $CPT$ theorem assures that any particle species $X$ there exists the antiparticle $\overline{X}$ with exactly the same mass, $m_X=m_{\overline{X}}$, and decay width, $\Gamma_X=\Gamma_{\overline{X}}$, and eventually opposite charges associated to these particles, $Q_X=-Q_{\overline{X}}$. This striking symmetry would naturally lead us to conclude that the Universe contains particles and antiparticles in equal number densities, $n_X=n_{\overline{X}}$. The observed Universe, however, is drastically different. We do not observe any bodies of antimatter within the solar system and only antiprotons $\overline{p}$ in the cosmic rays, which are believed to be of extra solar origin.  
Antiprotons are likely to be produced as secondaries in collisions
$pp\rightarrow 3p+ \overline{p}$ at a rate similar to the observed one
\be
\frac{n_{{\bar p}}}{n_p}\sim 3\times 10^{-4}.
\ee
The experimental   limit on $\overline{n}_{^4{\rm He}}/n_{^4{\rm He}}$ is similarly of the order of $10^{-5}$. 
We cannot exclude, of course, that the dominance of matter over antimatter is only local and is only realized  up to a certain length scale $\ell_B$, beyond which the picture is reversed and islands of antimatter are found. However, the size of our matter domain must be quite large, roughly speaking 
$\ell_B\gsim$ 10 Mpc \cite{stec,steg}  (for more restrictive bounds see \cite{alv}). Indeed, for smaller scales one would expect a significant amount of energetic $\gamma$-rays  coming from the reaction of annihilation of $p\overline{p}$ into $\pi$-mesons followed by the subsequent decay $\pi^0\rightarrow 2\gamma$, which would take in the boundary area separating the matter and antimatter islands. Another signature for the presence of domains of antimatter would be the distortion of the spectrum of the cosmic microwave background radiation. In such a case, the permitted value of $\ell_B$ might be smaller if voids separate matter and antimatter domains. These voids might be created
 because of an excessive pressure produced by the annihilations at earlies stages of the evolution of the Universe or because of low density matter and 
antimatter in the boundary regions, provided that the baryon asymmetry changes sign locally so that in the boundaries it is zero or very small. 

All these considerations lead us to conclude that, if domains of matter and antimatter exist in the Universe, they are separated on scales certainly larger than the radius of our own galaxy $(\sim 3$ Kpc) and most probably on scales larger than the Virgo cluster $(\sim 10$ Mpc). A much more  severe bound on $\ell_B$ ($\sim 300$ Mpc) is  potentially  reachable by the  Alpha Magnetic Spectrometer (AMS)  \cite{ams}, a   detector for extraterrestrial study of antimatter, matter and missing matter which,  after a precursor flight on STS91 in May 1998,  will be installed on the International Space Station where it should operate for three years.

\subsection{Some considerations on nucleosynthesis and the baryon number}

The baryon number density does not keep constant during the evolution of the Universe because it scales like $a^{-3}$, where $a$ is the cosmological scale factor \cite{book}. It is therefore convinient to define the baryon asymmetry of the Universe in terms of the quantity
\be
\label{eta}
\eta\equiv \frac{n_B}{n_\gamma},
\ee
where $n_B=n_b-n_{{\bar b}}$ is the difference between the number of baryons and antibaryons per unit volume and $
n_\gamma=2\frac{\zeta(3)}{\pi^2}T^3$
is the photon number density at a certain temperature $T$. The parameter $\eta$ is essential for determining the present light element abundances produced at the nucleosynthesis epoch. The parameter $\eta$ may have not changed since nucleosynthesis. At these energy scales ($\sim 1$ MeV) the baryon number is conserved if there are no processes which would have produced entropy to change the photon number. 

Let us now estimate $\eta$. The baryon number density is
\be
 n_B=\frac{\rho_B}{m_B}=\frac{\Omega_B}{m_B}\:\rho_c,
\ee 
where $\rho_B$ is the baryonic energy density and $\Omega_B\equiv \rho_B/\rho_c$. Using the critical density
\be
\rho_c= 1.88\times 10^{-29}\;h^2 \:{\rm gr}\:{\rm cm}^{-3},
\ee
where $0.5\lsim h\lsim 0.9$ parametrizes the present value of the Hubble parameter $H_0$, $h\equiv H/100$ Km Mpc$^{-1}$ sec $^{-1}$, we obtain
\be
\label{nB}
n_B=1.1\times 10^{-5} \:h^2 \:\Omega_B\:{\rm cm}^{-3}.
\ee
On the other hand, the present  temperature of the background radiation  is $T_0=2.735\:^0$K giving rise to
\be
\label{ngamma}
n_\gamma\simeq 415\:\left(\frac{T_0}{2.735\:^0K}\right)^3\:{\rm cm}^{-3}.
\ee
Putting (\ref{nB}) and (\ref{ngamma}) together, we obtain
\be
\eta=2.65\times 10^{-8}\:\Omega_B\:h^2\:\left(\frac{T_0}{2.735\:^0K}\right)^{-3}.
\ee
The range of $\eta$ consistent with the deuterium and $^3$He primordial abundances is \cite{book}
\be
\label{asy}
4(3)\times 10^{-10}\lsim\eta\lsim 7 (10)\times 10^{-10},
\ee
where the most conservative bounds are in parenthesis. Conversely we may write the range for $\Omega_B \:h^2$ to be
\be
0.015(0.011)\lsim \Omega_B\:h^2\lsim 0.026 (0.038).
\ee
Sometimes it is useful to describe the baryon asymmetry in terms of $B\equiv n_B/s$, where $s$ is the entropy density of the Universe at a certain temperature $T$.  The range (\ref{asy}) translates into 
\be 
5.7 (4.3)\times 10^{-11}\lsim B\lsim 9.9 (14)\times 10^{-11}.
\ee
Now, the fundamental question is: are we able to explain the tiny value of $\eta$ within the standard cosmological model?

Suppose  that initially we start with $\eta=0$. We can compute the final number density of nucleons $b$  that are left over after annihilations have frozen out. At temperatures $T\lsim$ 1 GeV the equilibrium abundance of nucleons and antinucleons is \cite{book}
\be
\label{a}
\frac{n_b}{n_\gamma}\simeq\frac{n_{{\bar b}}}{n_\gamma}\simeq
\left(\frac{m_p}{T}\right)^{3/2}\:e^{-\frac{m_p}{T}}.
\ee
When the Universe cools off, the number of nucleons and antinucleons decreases as long as the annihilation rate $\Gamma_{{\rm ann}}\simeq n_b\langle \sigma_A v\rangle$ is larger than the expansion rate of the Universe $
H\simeq 1.66\:g_*^{1/2}\frac{T^2}{\mpl}$. The thermally averaged annihilation cross section $\langle \sigma_A v\rangle$ is of the order of $m_\pi^2$. 
At $T\simeq$ 20 MeV, $\Gamma_{{\rm ann}}\simeq H$ and annihilations freeze out, nucleons and antinucleons being so rare that they cannot annihilate any longer. Therefore, from (\ref{a}) we obtain
\be
\frac{n_b}{n_\gamma}\simeq \frac{n_{{\bar b}}}{n_\gamma}\simeq 10^{-18},
\ee
which is much smaller than the value required by nucleosynthesis. In order to avoid the annihilation catastrophe, we may suppose that hypothetical new interactions separated matter from antimatter before $T\simeq$ 38 MeV, when $\eta\simeq 10^{-10}$. At that time, $t\simeq 10^{-3}$ sec,  however, the causal region (horizon) was small and contained only $\sim 10^{-7} M_{\odot}$. Hence we cannot explain the asymmetry over the galaxy scales. This argument is not valid, however, in cosmological models invoking inflation. Indeed, in these models the region of the Universe which is causally connected today was connected even at times $\sim 10^{-3}$ sec. These scenarios pose other serious
cosmological drawbacks, though. If the processes responsible for the separation of matter from antimatter took place before inflation, then the baryon number was diluted by an enormous factor $\sim$ exp(200),  because of the entropy production  due to inflation. On the other side, if the separation took place after inflation, then it is not clear how to eliminate the boundaries separating matter from antimatter islands.

Another possibility may be represented by explaining the tiny value of $\eta$ via statistical fluctuations in the baryon and antibaryon distributions. Our own galaxy containes at the present epoch approximately $10^{79}$ photons. The comoving volume $V$  that encompasses our galaxy today contains about $10^{69}$ baryons, but when the temperature was $T\gsim 1$ GeV, it contained about $10^{79}$ baryons and antibaryons. Frome pure statistical fluctuations one may expect an asymmetry $(n_b-n_{\overline{b}})/n_b\simeq (n_b V)^{-1/2}\simeq 10^{-39.5}$,   which is again far too small to explain the observed baryon asymmetry.

In conclusion, in the standard cosmological model there is no explanation for the smallness of the ratio (\ref{asy}), if we start from $\eta=0$. An initial asymmetry may be imposed by hand as an initial condition, but this would violate any naturalness principle and would be extremely boring!

\subsection{The three basic conditions for baryogenesis}

As we have already learned, the Universe was initially baryon symmetric ($n_b=n_{\overline{b}}$) although the matter-antimatter asymmetry appears to be large today ($n_b\gg n_{{\bar b}}$). In the standard cosmological model there is no explanation for such a small value of the baryon asymmetry consistent  with nucleosynthesis and  it has to be imposed by hand as an initial condition. This option is far from being appealing. However, it has been suggested by Sakharov long ago \cite{sak} that a tiny baryon asymmetry $B$ may have been produced in the early Universe. Three are the necessary conditions for this to happen.

\vspace{0.5cm}
\centerline{{\it Exercise 1}}
\vspace{0.5cm}
Show that the baryon asymmetry is zero  if there is no baryon number violation. 

\subsubsection{Baryon number violation}

This condition is somehow obvious since we want to start from a baryon symmetric Universe ($B=0$) and to evolve it to a Universe where $B\neq 0$. Baryon number violation interactions are therefore mandatory. They might also mediate proton decay; in such a case phenomenological constraints are provided by the lower bound on the proton lifetime $\tau_p\gsim 5\times 10^{32}$ years.. 

\subsubsection{ $C$ and $CP$ violation}

$C$ (charge conjugation symmetry) and $CP$ (the product of charge conjugation and parity) are not exact symmetries. Indeed, were $C$ an exact symmetry, the probability of the process $i\rightarrow f$ would be equal to the one of the process $\overline{i}\rightarrow \overline{f}$. Since the baryon number of $f$ is equal in absolute value and opposite in sign to that of $\overline{f}$, the net baryon number $B$ would vanish. $C$ is maximally violated by the weak interactions.

Furthermore, because of the $CPT$ theorem, $CP$ invariance is equivalent to time-invariance (time reversal). The latter assures that the rate of the process
\begin{equation}
i({\bf r}_i, {\bf p}_i,{\bf s}_i)\rightarrow  f({\bf r}_j, {\bf p}_j,{\bf s}_j)
\end{equation}
and that of its time-reversed process
\begin{equation}
f({\bf r}_j, -{\bf p}_j,-{\bf s}_j)\rightarrow  i({\bf r}_i, -{\bf p}_i,-{\bf s}_i)
\end{equation}
are equal. 
Thus, even though it is possible to create a baryon asymmetry in a certain region of the phase space, integrating over all momenta ${\bf p}$ and summing over all spins ${\bf s}$ would produce a vanishing baryon asymmetry. $CP$ violation has been observed  in the kaon system. However, a fundamental understanding of $CP$ violation is still lacking. Hopefully, studies of baryogenesis may  shed some light on it.

\subsubsection{Departure from thermal equilibrium}

If all the particles in the Universe remained in thermal equilibrium, then no preferred direction for time may be defined and the $CPT$ invariance would prevent the appearance of any baryon excess, making the presence of $CP$ violating interactions irrelevant. 

Let us suppose that a certain species $X$ with mass $m_X$ is in thermal equilibrium at temperatures $T\ll m_X$. Its number density will be given by
\begin{equation}
n_X\simeq g_X (m_X T)^{3/2}\: e^{-\frac{m_X}{T}+\frac{\mu_X}{T}},
\end{equation}
where $\mu_X$ is the associated chemical potential. 

As we have mentioned in the previous Section, a species $X$ is in {\it chemical} equilibrium if the inelastic scatterings which change the number of $X$ particles in the plasma, $X+A\rightarrow B+C$,  have a rate $\Gamma_{{\rm inel}}$ larger than the expansion rate of the Universe. In such a case, one can write down a relation among the different chemical potentials of the particles involved in the process
\be
\mu_X+\mu_A=\mu_B+\mu_C.
\ee
In this way the number density in thermal equilibrium of the antiparticle $\overline{X}$ ($m_X=m_{{\bar X}}$) is
\begin{equation}
n_{{\bar X}}\simeq g_X (m_X T)^{3/2}\: e^{-\frac{m_X}{T}-\frac{\mu_X}{T}},
\end{equation}
where we have made use of the fact that $\mu_{{\bar X}}=-\mu_X$ because of the process
\be 
\overline{X} X\rightarrow \gamma \gamma,
\ee
and $\mu_\gamma=0$. If the $X$ particle carries baryon number, then $B$ will get a contribution from
\be
B\propto n_X-n_{{\bar X}}= 2 g_X (m_XT)^{3/2}e^{-\frac{m_X}{T}}{\rm sinh}\left(\frac{\mu_X}{T}\right).
\ee
The crucial point is now that, if $X$ and $\overline{X}$ undergo $B$-violating reactions, as required by the first Sakharov condition, 
\be
XX\rightarrow \overline{X}\overline{X},
\ee
then $\mu_X=0$ and the relative contribution of the $X$ particles to the net baryon number vanishes. Only a departure from thermal equilibrium can allow 
for a finite baryon excess. 

\section{The standard out-of-equilibrium decay scenario}

Out of the three Sakharov conditions that we  discussed in the previous section, the baryon number violation and $C$ and $CP$ violation may be investigated thoroughlly only within a given particle physics model, while the third condition -- the departure from thermal equilibrium --may be discussed in a more general way.  Very roughly speaking, the various models of baryogenesis that have been proposed so far fall into  two categories: 

--  models where the out-of-equilibrium condition is attained thanks  to the expansion of the Universe and the presence of heavy decaying particles;

--  models where the departure from thermal equilibrium is attained during the    phase transitions which lead   to  the  breaking of some global and/or gauge symmetry.

In this lecture we will analyse the first category --the standard out-of-equilibrium decay scenario \cite{book,decay}.

\subsection{The conditions for the out-of-equilibrium decay scenario}

 It is obvious that in a static Universe any particle, even very weakly interacting, will attain sooner or later thermodynamical equilibrium with the surroinding plasma. The expansion of the Universe, however, introduces a finite time-scale, $\tau_U\sim H^{-1}$.
Let suppose that $X$ is a {\it baryon number violating} superheavy boson field (vector or scalar) which is coupled to lighter fermionic degrees of freedom with a strength $\alpha_X^{1/2}$ (either a gauge coupling $\alpha_{{\rm gauge}}$ or a Yukawa coupling $\alpha_Y$).

In the case in which the couplings are renormalizable,  the decay rate $\Gamma_X$  of the superheavy boson may be easily estimated to be
\be
\Gamma_X\sim \alpha_X\: M_X,
\ee
where $M_X$ is the mass of the particle $X$. In the opposite case in which the boson is a gauge {\it singlet} scalar field and it only couples to light matter through gravitational interactions -- this is the case of singlets in the hidden sector of supergravity models \cite{nilles} -- the decay rate is from dimensional arguments
\be
\Gamma_X\sim \frac{M_X^3}{\mpl^2}.
\ee
At very large temperatures $T\gg M_X$, it is assumed that all the particles species are in thermal equilibrium, {\it i.e.} $n_{X}\simeq  n_{\overline{X}}\simeq n_\gamma$ (up to statistical factors) and that $B=0$. At $T\lsim M_X$ the equilibrium abundance of $X$ and $\overline{X}$ relative to photons is given by
\be
\frac{n^{{\rm EQ}}_X}{n_\gamma}\simeq \frac{n^{{\rm EQ}}_{\overline X}}{n_\gamma}\simeq
\left(\frac{M_X}{T}\right)^{3/2}\: e^{-\frac{M_X}{T}},
\ee
where we have neglected the chemical potential $\mu_X$. 

For the $X$ and $\overline{X}$ particles to mantain their equilibrium abundances, they must be able to diminish their number rapidly with respect to the Hubble rate $H(T)$. The conditions necessary for doing so are easily quantified. The superheavy $X$ and $\overline{X}$ particles may attain equilibrium through decays with rate $\Gamma_X$, inverse decays with rate $\Gamma_X^{{\rm ID}}$ 
\be
\Gamma_X^{{\rm ID}}\simeq \Gamma_X\left\{
\begin{array}{cc}
1 & T\gsim M_X,\\
(M_X/T)^{3/2}\:{\rm exp}(-M_X/T) & T\lsim M_X,
\end{array}
\right.
\ee
and annihilation processes
with rate $\Gamma_X^{{\rm ann}}\propto n_X$. The latter, however are ``self-quenching'' and therefore less important than the decay and inverse decay processes. They will be ignored from now on. Of crucial interest are the $B$-nonconserving scattering processes $2\leftrightarrow 2$   mediated by the $X$ and $\overline{X}$ particles with rate $\Gamma_X^{{\rm S}}$
\be  
\Gamma_X^{{\rm S}}\simeq n\sigma\simeq \alpha^2 T^3\frac{T^2}{(M_X^2+T^2)^2},
\ee
where $\alpha\simeq g^2/4\pi$ denotes the coupling strength of the $X$ boson. At high temperatures, the $2\leftrightarrow 2$ scatterings      cross section is $\sigma\simeq \alpha^2/T^2$, while at low temperatures $\sigma\simeq \alpha^2 T^2/M_X^4$. 

For baryogenesis, the most important rate is the decay rate, as decays (and inverse decays) are the mechanism that regulates the number of $X$ and $\overline{X}$ particles in the plasma. It is therefore useful to define the following  quantity 
\be 
K\equiv \left.\frac{\Gamma_X}{H}\right|_{T=M_X}
\ee
which measures the effectiveness of decays at the crucial epoch ($T\sim M_X$) when the $X$ and $\overline{X}$ particles must decrease in number if they are to stay in equilibrium. Note also that for $T\lsim M_X$, $K$ determines the effectiveness of inverse decays and $2\leftrightarrow 2$ scatterings as well: $\Gamma_X^{{\rm ID}}/H\simeq (M_X/T)^{3/2}\:{\rm exp}(-M_X/T)\:K$ and 
$\Gamma_X^{{\rm S}}/H\simeq\alpha (T/M_X)^5\:K$. 

Now, if  $K\gg 1$, and therefore
\be
\label{no}
\Gamma_X\gg \left. H\right|_{T=M_X},
\ee
then the $X$ and $\overline{X}$ particles will adjust their abundances by decaying to their equilibrium abundances and no baryogenesis can be induced by their decays --this is simply because out-of-equilibrium conditions are not attained. Given the expression (\ref{h}) for the expansion rate of the Universe, the condition (\ref{no}) is equivalent to 
\be
M_X\ll g_{*}^{-1/2} \:\alpha_X\:\mpl
\ee
for strongly coupled scalar bosons, and to
\be
M_X\gg g_*^{1/2}\: \mpl,
\ee
for gravitationally coupled $X$ particles. Obviously, this last condition is never satisfied for $M_X\lsim \mpl$.

However, if the decay rate is such that $K\ll 1 $, and therefore
\be
\label{yes}
\Gamma_X\lsim  \left. H\right|_{T=M_X},
\ee
then the $X$ and $\overline{X}$ particles cannot decay on the expansion time-scale $\tau_U$ and so {\it they remain as abundant as photons} for $T\lsim M_X$. In other words, at some  temperature $T > M_X$, the superheavy bosons 
$X$ and $\overline{X}$ are so weakly interacting that they cannot catch up with the expansion of the Universe and they decouple from the thermal bath when still {\it relativistic}, $n_{X}\simeq  n_{\overline{X}}\simeq n_\gamma$ at the time of decoupling. Therefore, at temperature $T\simeq M_X$, they will populate the Universe with an abundance which is much larger than the equilibrium one. 
This {\it overbundance} with respect to the equilibrium abundance is precisely the departure from thermal equilibrium needed to produce a final nonvanishing baryon asymmetry. Condition (\ref{yes}) is equivalent to 
\be
\label{yes1}
M_X\gsim  g_{*}^{-1/2}\: \alpha_X\:\mpl
\ee
for strongly coupled scalar bosons, and to
\be
\label{sup}
M_X\lsim g_*^{1/2}\: \mpl,
\ee
for gravitationally coupled $X$ particles. It is clear that this last condition       is always satisfied,  whereas the condition (\ref{yes1}) is based on the smallness of the quantity $g_{*}^{-1/2} \alpha_X$. In particular, if the $X$ particle is a gauge boson, $\alpha_X\sim \alpha_{{\rm gauge}}$ can span the range $(2.5\times 10^{-2}-10^{-1})$, while $g_*$ is about $10^2$. In this way we obtain from (\ref{yes1}) that the condition of out-of-equilibrium can be satisfied for
\be
\label{gauge}
M_X\gsim (10^{-4}-10^{-3})\:\mpl\simeq (10^{15}-10^{16})\:{\rm GeV}.
\ee
If $X$ is a scalar boson, its coupling $\alpha_Y$ to fermions $f$ with mass $m_f$  is proportional to the squared mass of the fermions
\be
\alpha_Y\sim \left(\frac{m_f}{m_W}\right)^2\;\alpha_{{\rm gauge}},
\ee
where $m_W$ is the $W$-boson mass and $\alpha_Y$  is typically in the range $(10^{-2}-10^{-7})$, from where
\be
\label{scalar}
M_X\gsim (10^{-8}-10^{-3})\:\mpl\simeq (10^{10}-10^{16})\:{\rm GeV}.
\ee
Obviously, condition (\ref{scalar}) is more easily satisfied than condition (\ref{gauge}) and we conclude that baryogenesis is more easily produced through
the decay of superheavy scalar bosons. On the other hand, as we have seen above, the condition (\ref{sup}) tells us that the out-of-equilibrium condition is automatically satisfied for  gravitationally interacting
particles.  

\subsection{The production of the baryon asymmetry}

Let us now follow the subsequent evolution of the $X$ and $\overline{X}$ particles. When the Universe becomes as old as the lifetime of these particles, $t\sim H^{-1}\sim \Gamma_X^{-1}$, they start decaying. This takes place at a temperature $T_D$ defined by the condition 
\be
\Gamma_X\simeq \left. H\right|_{T=T_D},
\ee
{\it i.e.} at
\be
T_D\simeq g_{*}^{-1/4}\:\alpha_X^{1/2}\:(M_X\mpl)^{1/2}<M_X,
\ee
where the last inequality comes from (\ref{yes1}) and is valid for   particles with unsuppressed couplings. For particles with only gravitational interactions  
\be
T_D\sim g_*^{-1/4}\:M_X\:\left(\frac{M_X}{\mpl}\right)^{1/2}<M_X,
\ee
the last inequality coming from (\ref{sup}). At $T\sim T_D$, $X$ and $\overline{X}$ particles  start to decay and  their number decrease. If their decay violate the baryon number, they will generate a net baryon number per decay. 

Suppose now that  the $X$ particle may decay into two channels, let us denote them by $a$ and $b$, with different baryon numbers $B_a$ and $B_b$, respectively. Correspondingly, the decay channels of $\overline{X}$, $\overline{a}$ and $\overline{b}$, have baryon numbers $-B_a$ and $-B_b$, respectively. Let $r (\overline{r})$ be the branching ratio of the $X(\overline{X})$ in channel $a(\overline{a})$ and $1-r(\overline{r})$ the branching ratio of $X(\overline{X})$ in channel $b(\overline{b})$,
\begin{eqnarray}
r &=&\frac{\Gamma(X\rightarrow a)}{\Gamma_X},\nonumber\\
\overline{r} &=&\frac{\Gamma(\overline{X}\rightarrow \overline{a})}{\Gamma_X},\nonumber\\
1-r &=&\frac{\Gamma(X\rightarrow b)}{\Gamma_X},\nonumber\\
1-\overline{r} &=&\frac{\Gamma(\overline{X}\rightarrow \overline{b})}{\Gamma_X},
\end{eqnarray}
where we have been using the fact that the {\it total} decay rates of $X$ and $\overline{X}$  are equal because of the $CPT$ theorem plus unitarity.

The average net baryon number produced in the $X$ decays is
\be
r B_a+(1-r) B_b,
\ee
and that produced by $\overline{X}$ decays is 
\be
-\overline{r} B_a-(1-\overline{r})B_b.
\ee
Finally, the mean net baryon number produced in $X$ and $\overline{X}$ decays is
\be
\label{gamma}
\Delta B= (r-\overline{r}) B_a +\left[(1-r)-(1-\overline{r})\right]B_b=
(r-\overline{r})(B_a-B_b).
\ee
Equation (\ref{gamma}) may be easily generalized to the case in which $X(\overline{X})$ may decay into a set of final states $f_n(\overline{f}_n)$ with baryon number $B_n(-B_n)$
\be
\label{bn}
\Delta B=\frac{1}{\Gamma_X}\sum_n B_n\:\left[\Gamma(X\rightarrow f_n)-
\Gamma(\overline{X}\rightarrow \overline{f}_n)\right].
\ee
At the decay temperature, $T_D\lsim M_X$, because $K\ll 1$ both inverse decays and $2\leftrightarrow 2$ baryon violating scatterings are impotent and can be safely ignored and thus the net baryon number produced per decay $\Delta B$ is not destroyed by the net baryon number $-\Delta B$ produced by the inverse decays and by the baryon number violating scatterings.

 At $T\simeq T_D$, $n_{X}\simeq  n_{\overline{X}}\simeq n_\gamma$ and therefore the net baryon number density produced by the out-of-equilibrium decay is
\be
n_B=\Delta B\: n_X,
\ee
from where we can see that $\Delta B$ coincides with the parameter $\eta$ defined in (\ref{eta}) if $n_X\simeq n_\gamma$.

The three Sakharov ingredients for producing a net baryon asymmetry can be easily traced back here:

-- If $B$ is not violated, then $B_n=0$ and $\Delta B=0$.

--If $C$ and $CP$ are not violated, then $\Gamma(X\rightarrow f_n)=\Gamma(\overline{X}\rightarrow \overline{f}_n)$, and also $\Delta B=0$.

-- In thermal equilibrium, the inverse processes are not suppressed and the net baryon number produced by decays will be erased by the inverse decays.

Since each decay produces a mean net baryon number density $n_B=\Delta B n_X\simeq \Delta B n_\gamma$ and since the entropy density is $s\simeq g_* n_\gamma$, the net baryon number produced is
\be
\label{finalresult}
B\equiv \frac{n_B}{s}\simeq\frac{\Delta B n_\gamma}{g_* n_\gamma}\simeq
\frac{\Delta B}{g_*}.
\ee
Taking $g_*\sim 10^2$, we see that only  tiny $C$ and $CP$ violations are required to generate $\Delta B\sim 10^{-8}$, and thus $B\sim 10^{-10}$.

To obtain (\ref{finalresult}) we have assumed that the entropy realese in $X$ decays is negligible. However, sometimes, this is not a good approximation (especially if the $X$ particles decay very late, at $T_D\ll M_X$,  which is the case of gravitationally interacting particles). In that case, assuming that the energy density of the Universe at $T_D$ is dominated by $X$ particles
\be
\rho_X\simeq M_X \: n_X,
\ee
and that it is converted entirely into radiation at the reheating temperature $T_{RH}$
\be
\rho=\frac{\pi^2}{30}\:g_*\:T_{RH}^4,
\ee
we obtain
\be
n_X\simeq \frac{\pi^2}{30}\:g_*\:\frac{T_{RH}^4}{M_X}.
\ee
We can therefore write the baryon number as
\be
B\simeq \frac{3}{4}\frac{T_{RH}}{M_X}\Delta B.
\ee
We can relate $T_{RH}$ with the decay rate $\Gamma_X$ using the decay condition
\be
\Gamma_X^2\simeq H^2(T_D)\simeq\frac{8\pi\rho_X}{3\mpl^2}
\ee
and so we can write
\be
B\simeq\left(\frac{g_*^{-1/2}\Gamma_X\mpl}{M_X^2}\right)^{1/2}\:\Delta B.
\ee
For the case of strongly decaying particles (through renormalizable interactions) we obtain
\be 
B\simeq\left(\frac{g_*^{-1/2}\alpha\mpl}{M_X}\right)^{1/2}\:\Delta B.
\ee
while for the case of weakly decaying particles (through gravitational interactions) we obtain
\be
B\simeq \left(\frac{g_*^{-1/2} M_X}{\mpl}\right)^{1/2}\:\Delta B.
\ee
In the other extreme regime $K\gg 1$, one expects the abundance of $X$ and $\overline{X}$ bosons to track the equilibrium values as $\Gamma_X\gg H$ for $T\sim M_X$. If the equilibrium is tracked precisely enough, there will be no departure from thermal equilibrium and no baryon number may evolve. The intermediate regime, $K\sim 1$, is more interesting and to address it one has to invoke numerical analysis involving Boltzmann equations for the evolution of $B$. This has been done in refs. \cite{wolfram,fry,harvey,decay}. The numerical analysis essentially confirms the qualitative picture we have described so far and its discussion is beyond the scope of these lectures.

\subsubsection{An explicit example}

Let us consider first two massive boson fields $X$ and $Y$ coupled to four fermions $f_1$, $f_2$, $f_3$ and $f_4$ through the vertices of Fig. 1 and describing the decays $X\rightarrow \overline{f}_1 f_2, \overline{f}_3 f_4$  and $Y\rightarrow \overline{f}_3 f_1, \overline{f}_4 f_2$. We will refer to these vertices as $\langle f_2|X|f_1\rangle$, $\langle f_4|X|f_3\rangle$, $\langle f_1|Y|f_3\rangle$ and $\langle f_2|Y|f_4\rangle$, and their $CP$ conjugate $\overline{X}\rightarrow \overline{f}_2 f_1, \overline{f}_4 f_3$ and 
$\overline{Y}\rightarrow \overline{f}_1 f_3, \overline{f}_2 f_4$ by their complex conjugate. In the Born approximation $\Delta B=0$ because from (\ref{bn}) one finds
\be
\Gamma(X\rightarrow \overline{f}_1 f_2)_{{\rm Born}}=I_X^{12}\left|\langle f_2|X|f_1\rangle\right|^2=\Gamma(\overline{X}\rightarrow \overline{f}_2 f_1)_{{\rm Born}},
\ee
where $I_X^{12}$ accounts for the kinematic structures of the processes 
$X\rightarrow \overline{f}_1 f_2$ and $\overline{X}\rightarrow \overline{f}_2 f_1$ and the same may be found for the other processes contributing to $\Delta B$. This shows that, to obtain a non-zero result for $\Delta B$, one must include (at least) corrections arising from the interference of Born amplitudes
of Fig. 1 with the one-loop amplitude of Fig. 2. For example, the interference of the diagrams in Fig. 1(a) and Fig. 2(a) (in the square amplitude) is shown in Fig. 3(a), where the thick dashed line is the unitarity cut (equivalent to say that each cut line represents on-shell mass particles). The amplitude of the diagram in Fig. 3(a) is given by $I_{XY}^{1234}\Omega_{1234}$, where the kinematic factor  
$I_{XY}^{1234}$ accounts for the integration over the final state phase space of $f_2$ and $\overline{f}_1$ and over momenta of the internal states $f_4$ and $\overline{f}_3$, and
\be
\Omega_{1234}=\langle f_1|Y|f_3\rangle^* \langle f_4|X|f_3\rangle \langle f_2|Y|f_4\rangle \langle f_2|X|f_1\rangle^*.
\ee
\begin{figure}
\centering
\leavevmode\epsfysize=3.6in \epsfbox{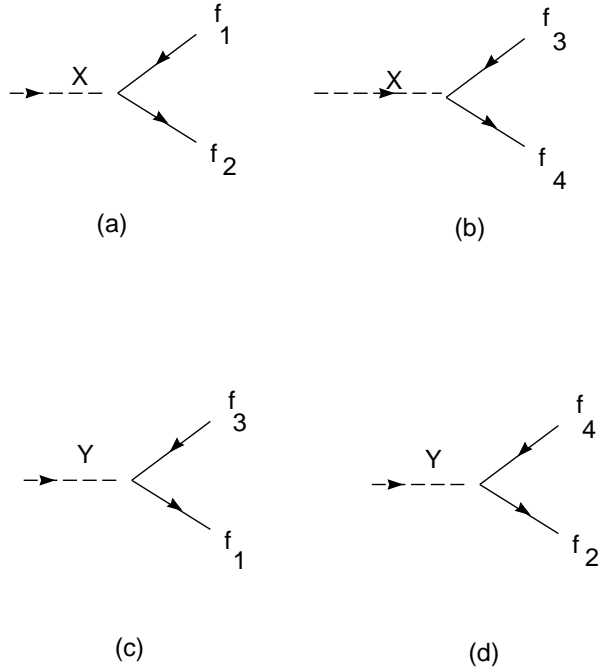}
\caption{Couplings of $X$ and $Y$ to fermions $f_i$. }

\end{figure}

\begin{figure}
\centering
\leavevmode\epsfysize=3.6in \epsfbox{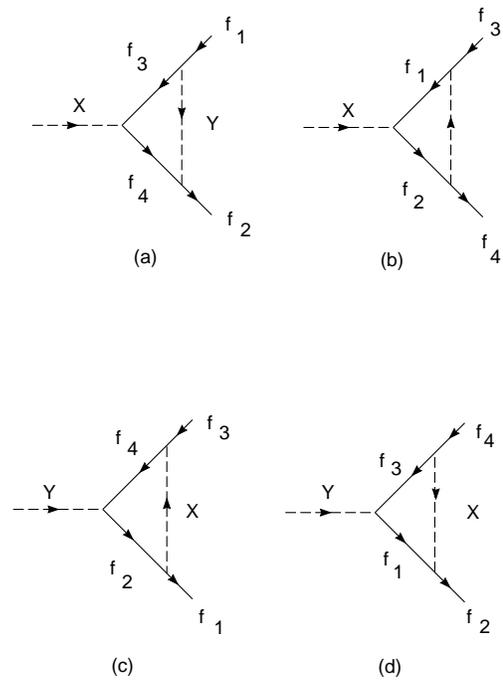}
\caption{One-loop corrections to the Born amplitude of Fig. 1. }

\end{figure}

\begin{figure}
\centering
\leavevmode\epsfysize=3.6in \epsfbox{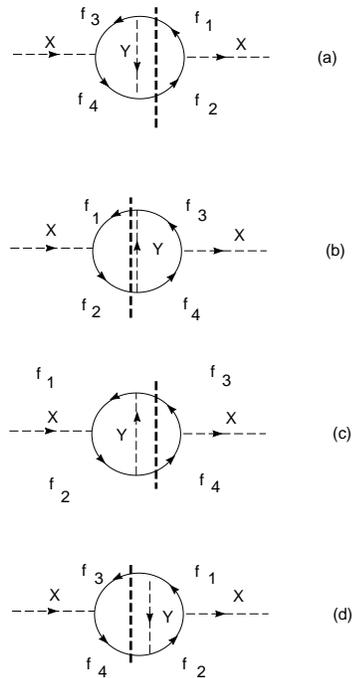}
\caption{Intereference between the diagrams of Fig. 1 and Fig. 2 for the square amplitudes of $X$ decay. }

\end{figure}

The complex conjugate diagram of Fig. 3(b) has the complex conjugate amplitude. Therefore, the contribution from the diagrams in Figs. 3(a) and 3(b) to the decay $X\rightarrow \overline{f}_1 f_2$ is
\be
\Gamma(X\rightarrow \overline{f}_1 f_2)_{{\rm interference}}=I_{XY}^{1234}\Omega_{1234}+{\rm h.c.}
\ee
To obtain the $CP$ conjugate amplitude $\overline{X}\rightarrow \overline{f}_2 f_1$ all couplings must be complex conjugated, although the kinematic factors $I_{XY}$ are unaffected by $CP$ conjugation. Therefore the interference contribution to the $\overline{X}\rightarrow \overline{f}_2 f_1$ decay rate is given by
\be
\Gamma(\overline{X}\rightarrow \overline{f}_2 f_1)_{{\rm interference}}=I_{XY}^{1234}\Omega_{1234}^*+{\rm h.c.}
\ee
and the relevant quantity for baryogenesis is given by
\be
\Gamma(X\rightarrow \overline{f}_1 f_2)-\Gamma(\overline{X}\rightarrow \overline{f}_2 f_1)=-4\: {\rm Im}\:\left[I_{XY}^{1234}\right]\:
{\rm Im}\:\left[\Omega_{1234}\right].
\ee
The diagrams of the decays $X\rightarrow \overline{f}_3 f_4$ and $\overline{X}\rightarrow \overline{f}_3 f_4$  differ from the one in Figs. 3(a) and 3(b) only in that the unitarity cut is taken through $f_3$ and $f_4$ instead of $f_1$ and $f_2$. One easily obtains
\be
\Gamma(X\rightarrow \overline{f}_3 f_4)-\Gamma(\overline{X}\rightarrow \overline{f}_4 f_3)=-4\:{\rm Im}\:\left[I_{XY}^{3412}\right]\:
{\rm Im}\:\left[\Omega_{1234}^*\right].
\ee
The kinematic factors $I_{XY}$ for loop diagrams may have an imaginary part
whenever any internal lines may propagate on their mass shells in the intermediate states, picking the pole of the propagator
\be
\frac{1}{p^2-m^2+i\epsilon}=\frac{{\rm PP}}{p^2-m^2}+i\pi\delta(p^2-m^2),
\ee
where PP stands for the principal part. This happens if $M_X>m_1+m_2$ and $M_X>m_3+m_4$. This means that with light fermions, the imaginary part of $I_{XY}$ will be always nonzero. The kinematic factors ${\rm Im}\:\left[I_{XY}^{1234}\right]$ and ${\rm Im}\:\left[I_{XY}^{3412}\right]$ are therefore obtained from diagrams involving two unitarity cuts: one through the lines $f_1$ and $f_2$ and the other through the lines $f_3$ and $f_4$. The resulting quantities are invariant under the interchanges $f_1\leftrightarrow f_3$ and $f_2\leftrightarrow f_4$ and consequently
\be
{\rm Im}\:\left[I_{XY}^{1234}\right]={\rm Im}\:\left[I_{XY}^{3412}\right]={\rm Im}\:\left[I_{XY}\right].
\ee
Defining $B_i$ the baryon number of the fermion $f_i$, the net baryon number produced in the $X$ decays is therefore
\be
(\Delta B)_X=\frac{4}{\Gamma_X}\:{\rm Im}\:\left[I_{XY}\right]
\:{\rm Im}\:\left[\Omega_{1234}\right]\left[B_4-B_3-(B_2-B_1)\right].
\ee
To compute the baryon asymmetry $(\Delta B)_Y$ one may observe that the set of vertices in Fig. 1 is invariant under the transformations $X\leftrightarrow Y$ and $f_1\leftrightarrow f_4$. These rules yield
\be
(\Delta B)_Y=\frac{4}{\Gamma_Y}\:{\rm Im}\:\left[I_{YX}\right]
\:{\rm Im}\:\left[\Omega^*_{1234}\right]\left[B_4-B_3-(B_2-B_1)\right]
\ee
and the total baryon number is therefore given by
\be
\label{sum}
(\Delta B)=(\Delta B)_X+(\Delta B)_Y=4\left\{\frac{{\rm Im}\:\left[I_{XY}\right]}{\Gamma_X}-\frac{{\rm Im}\:\left[I_{YX}\right]}{\Gamma_Y}\right\}\:{\rm Im}\:
\left[\Omega_{1234}\right]\left[B_4-B_3-(B_2-B_1)\right].
\ee
We can notice a few things:

-- If the $X$ and $Y$ couplings were $B$ conserving, the two possible final states in $X$ and $Y$ decays would have the same baryon number, {\it i.e.} $B_4-B_3=B_2-B_1$ and therefore $\Delta B=0$. Therefore the baryon number must be violated not only in $X$ decays but also in the decays of the particle exchanged in the loop. 

-- Some coupling constants in the Lagrangian must be complex to have $
{\rm Im}\:\left[\Omega_{1234}\right]$.

-- Even if $(\Delta B)_X$ and $(\Delta B)_Y$ are both nonvanishing, the sum can be vanish if the first bracket in (\ref{sum}) cancels out. This happens if 
the $X$ and $Y$ particles have the same mass and $\Gamma_X=\Gamma_Y$.

\subsection{Baryon number violation in Grand Unified Theories}

The Grand Unified Theories (for a review, see \cite{lan}) try to describe the fundamental interactions by means of a unique gauge group $G$ which contains the Standard Model (SM) gauge group $SU(3)_C\otimes SU(2)_L\otimes U(1)_Y$. The fundamental idea of GUTs is that at energies higher than a certain energy threshold $M_{{\rm GUT}}$ the group symmetry is $G$ and that, at lower energies, the symmetry is broken down to the SM gauge symmetry, possibly through a chain of symmetry breakings
\be
G\stackrel{M_{{\rm GUT}}}{\rightarrow} G_{(1)}\stackrel{M_1}{\rightarrow} G_{(2)}
\stackrel{M_2}{\rightarrow} \cdots \stackrel{M_n}{\rightarrow} SU(3)_C\otimes SU(2)_L\otimes U(1)_Y,
\ee
corresponding to
\be
M_{{\rm GUT}}> M_1> M_2>\cdots M_W, G\supset G_{(1)}\supset G_{(2)}\supset\cdots SU(3)_C\otimes SU(2)_L\otimes U(1)_Y.
\ee
What is the main motivation for invoking GUTs? Gauge couplings (couplings to gauge fields)
are charactherized by a dimensionless constant $g$,
or equivalently by $\alpha=g^2/4\pi$. (For electromagnetism,
$g$ is the electron charge and $\alpha$ evaluated at low energy
is the fine structure constant $\alpha\sub{em}=1/137$.)
Gauge couplings are not supposed to be extremely small,
and one should take $g\sim 1$ for crude order of magnitude estimates
(making $\alpha$ one or two orders of magnitude below 1).
Assuming small couplings, the
perturbative effects usually dominate, and we focus on them for the 
moment. With perturbative quantum effects included,
the effective  masses and couplings depend 
on the relevant energy scale $Q$. The dependence on $Q$ (called `running')
can be calculated 
through the renormalization group equations (RGE's), and is 
logarithmic.
In the context of collider physics, $Q$ can be taken to be 
the collision energy, if there are no bigger relevant scales
(particle masses).
For the Standard Model
there are three gauge couplings, $\alpha_i$ where $i=3,2,1$, corresponding 
for respectively to the strong interaction (colour $SU(3)_C$)
the non-abelian electroweak interaction ($SU(2)_L$) and 
electroweak hypercharge ($U(1)_Y$). (The electromagnetic gauge coupling 
is given by $\alpha^{-1}= \alpha_1^{-1}+ \alpha_2^{-1}$.) 
In the one-loop approximation, ignoring the Higgs field,
their running is given (at one loop) by
\be
\frac{d\alpha_i}{d\ln(Q^2)}  = \frac{b_i}{4\pi}\alpha_i^2 \,.
\label{70}
\ee
The coefficients $b_i$ depend on the number of particles with
mass $\ll Q$. Including all particles in the minimal supersymmetric standard model gives $b_1= 11$, $b_2=1$ and $b_3=-3$.

Using the values of $\alpha_i$ measured by collider experiments at 
a scale $Q\simeq 100\MeV$, one finds that all three couplings
become equal at a scale \cite{couplingunity,n2,n3}\footnote
{To be precise, $\frac53\alpha_1=\alpha_2=\alpha_3=
\alpha\sub{GUT}$, the factor
$5/3$ arising because the historical definition of 
$\alpha_1$ is not very sensible. In passing we note that the unification 
fails by many standard deviations in the absence of supersymmetry,
which may be construed as evidence for supersymmetry and 
anyhow highlights the remarkable accuracy of the experiments leading to 
this result.}
$Q=M\sub{GUT}$, where
\be
M\sub{GUT} \simeq 2\times 10^{16}\GeV.
\ee
The unified value is 
\be
\alpha\sub{GUT}\simeq 1/25.
\ee
One explanation of this remarkable
experimental result may be that there is a GUT, 
involving a higher symmetry with a single gauge coupling,
which is unbroken above the scale $M\sub{GUT}$.
Another might be that field theory becomes invalid above the 
unification scale, to be replaced by something like
weakly coupled string theory 
or M-theory \cite{mtheory} which 
is the source of unification. At the time of writing there is no
consensus about which explanation is correct, but in this section we will focus on Grand Unified Theories and their relevance for baryogenesis.

It is a general property of GUTs that the same representation may contain both quarks and leptons and therefore there exist gauge bosons which mediate gauge interactions among fermions having different baryon number. This is not enough --though-- to conclude that in GUTs the baryon number is violated, because it might be possible to assign a baryonic charge to the gauge bosons in such a way that each vertex boson-fermion-fermion the baryon number is conserved. Let us discuss this crucial point in more detail.

The fundamental fermions of the SM are 
\begin{eqnarray}
\label{list}
\ell_L &=& (1, 2, -1/2),\nonumber\\
Q_L & =& (3,2,1/6),\nonumber\\
e^c_L &=& (1,1,1),\nonumber\\
u^c_L &=& (\overline{3},1,-2/3),\nonumber\\
d^c_L &=& (\overline{3}, 1,1/3),
\end{eqnarray}
where in parenthesis we have written then  $SU(3)_C\otimes SU(2)_L\otimes U(1)_Y$ quantum numbers and all the spinors are left-handed. Given two spinors $\psi_L$ and $\chi_L$,  it is possible to define a renormalizable  coupling to a gauge boson $V_\mu$ by
\be
\label{cou}
i\: \psi_L^\dagger \sigma^\mu\: \chi_L\: V_\mu +{\rm h.c.},
\ee
where $\sigma^\mu=({\bf 1},\vec{\sigma})$ and $\vec{\sigma}$ are the Pauli matrices. At this point one may try to write down  all the couplings of the form (\ref{cou}) starting from the spinors of the SM and identify all the possible gauge bosons which may be present in a GUT having the same spinors of the SM. Of course, the same gauge boson may be coupled to more than one pair of spinors. If all the spinor pairs have the same baryon number $B$, then it suffices to assign a baryon number $-B$ to the gauge boson and obtain a baryon number conserving theory. If there exist gauge bosons which couple to spinor pairs having different baryon number, one may write down baryon number violating interactions. These bosons are given in Table 1, where we have indicated, for every gauge boson, all the possible interactions and the corresponding baryon numbers $B$ and baryon minus lepton numbers $B-L$

\vspace{0.3cm}
\centerline{Table 1}
\vspace{0.1cm}\centerline{
\begin{tabular}{|| c|c|c|c||}\hline
{\rm Gauge boson} & {\rm spinors} & $B$ & $B-L$ \\ \hline
$V^1=(3,2,-5/6)$ & $u_L^{c\dagger} Q_L$ & 2/3 & 2/3\\
& $Q_L^\dagger e_L^c$ & -1/3 & 2/3\\
& $\ell_L^\dagger d_L^c$ & -1/3 & 2/3\\ \hline
$V^2=(3,2,1/6)$  & $\ell_L^\dagger u_L^c$ & -1/3 & 2/3\\ 
& $d_L^{c\dagger} Q_L$ & 2/3 & 2/3 \\ \hline
\end{tabular}
}
\vspace{0.3cm}

Of course, every gauge boson listed in Table 1 has the corresponding antiboson.
One can repeat the same procedure to identify the scalar bosons $S$ which may mediate baryon number violation interactions via fermions. The generic coupling
reads
\be
i\:\chi_L^T\:\sigma^2\:\psi_L\:S+{\rm h.c.}
\ee
If we consider all the spinor pairs $\chi_L^T\psi_L$, even belonging to different families, we get the following possibilities

\vspace{0.3cm}
\centerline{Table 2}
\vspace{0.1cm}
\centerline{
\begin{tabular}{|| c|c|c|c||}\hline
{\rm Scalar boson} & {\rm spinors} & $B$ & $B-L$ \\ \hline
$S^1=(3,1,-1/3)$ & $ \ell^\dagger Q_L^\dagger$ & -1/3 & 2/3\\
& $Q_L Q_L$ & 2/3 & 2/3\\
& $e^c_L u_L^c$ & -1/3 & 2/3\\ 
& $d_L^{c\dagger}u_L^{c\dagger}$ & 2/3 & 2/3\\ \hline
$S^2=(3,1,-4/3)$  & $e_L^c d_L^c$ & -1/3 & 2/3\\ 
& $u_L^{c\dagger} u_L^{c\dagger}$ & 2/3 & 2/3 \\ \hline
$S^3=(3,3,-1/3)$ & $\ell_L^\dagger Q_L^\dagger$ & -1/3 & 2/3 \\
& $Q_L Q_L$ & 2/3 & 2/3 \\ \hline
\end{tabular}
}
\vspace{0.3cm}

Out of all possible scalar and gauge bosons which may couple to the fermions of the SM, only the five that we have listed may give rise to interactions which violate the baryon number. A crucial point for what we will be discussing in the following is that each of these bosons have the same combination $B-L$, which means that this combination may be not violated in any vertex boson-fermion-fermion. This is quite a striking result and originates  only from having required the invariance under the SM gauge group and that the only fermions of the theory are those of the SM. 

The extension of the fermionic content of the theory may allow the presence of more heavy bosons which will possibly violate $B$ and even $B-L$. In the Grand Unified Theories based on $SO(10)$ -- for instance -- there is another fermion which is a singlet under the SM gauge group and is identified with the antineutrino $N_L^c=(1,1,0)$. It carries lepton number equal to $L=-1$. It is possible to introduce a new scalar field $S^4$ which may couple to $N_L^c u_L^c$ and $d_L^{c\dagger}d_L^{c\dagger}$, thus violating the baryon number. It is remarkable that the choice for the lepton number of $N_L^c$ leads to no new gauge boson which violates $B-L$. These considerations do not apply to supersymmetric models  though (for a review see \cite{haber}). Indeed, for every fermionic degree of freedom there exist a superpartner (squark or slepton) which does have the same quantum number. Furthermore, in the Minimal Supersymmetric Standard Model (MSSM) one has to introduce two Higgs doublets $H_1=(1,2,-1/2)$ and $H_2=(1,2,1/2)$ and the corresponding fermionic superpartners, the so-called higgsinos $\widetilde{H}_{1,2}$. Finally, every gauge boson has its own superparner, the gaugino. In this large zoo of new particles, one can easily find couplings that violate $B$ and $B-L$. For instance, the higgsino $\widetilde{H}_1$ may couple to the quark doublet $Q_L$ and to the scalars
$S^1$ and $S^3$ of the Table 2. The pair $\widetilde{H}_1^\dagger Q_L^\dagger$ has baryon number $B=-1/3$ and $B-L=-1/3$ and both quantum numbers are not conserved. Nevertheless, in the supersymmetric models which are phenomenologically acceptable, even without considering the presence of superheavy particles, it is necessary to suppress some supersymmetric couplings which would lead at the weak scale to a  proton decay at a rate which is too fast for being in agreement with the tight experimental constraints . One commonly accepted solution is to introduce a discrete symmetry $Z_2$, called $R$-parity, under which all the fields of the SM are even and all the superpartners are odd. The scalar component of any chiral supermultiplet has the following $R$-parity number
\be
R=(-1)^{3(B-L)},
\ee
while the corresponding fermion has the same number multiplied by $-1$. If we impose that $R$-parity is exact, then it is easy to check that, besides
suppressing the fast proton decay at the weak scale, one avoids the presence $B$ and $L$ violating couplings of heavy fields with the light fermionic fields
of the MSSM. Indeed, all the heavy bosons of Tables 1 and 2 have $R$-parity $R=1$, while -- for instance -- the pair $\widetilde{H}_1^\dagger Q_L^\dagger$ has parity $R=-1$. Similar considerations apply to other fermionic pairs. 

We conclude that in the GUTs, both supersymmetric and non-supersymmetric, no $B-L$ asymmetry may be generated through the out-of-equilibrium decay of gauge boson fields. We will mention in the following -- though -- that the generation of such an asymmetry is possible in the framework of (supersymmetric)  $SO(10)$ via the out-of-equilibrium decay of the right-handed (s)neutrino, {\it i.e.} via the decay of a superheavy fermion (scalar). 

After having learned that GUTs are the perfect arena for baryon number violating interactions, we will illustrate now some features of the out-of-equilibrium decay scenario within some specific GUTs,  like $SU(5)$ and $SO(10)$. 

\subsubsection{The case of $SU(5)$}

The gauge group $SU(5)$ is the smallest group containing the SM gauge group and as such it represents the most appealing candidate to build up a Grand Unified Theory. The non-supersymmetric version of $SU(5)$ is -- however -- already ruled out by its prediction of the proton lifetime $\tau_p\sim 10^{30}$ years, which is in disagreement with the experimental lower bound $\tau_p\gsim 10^{32}$ years \cite{be}. Recent precise measurements of coupling constants at LEP suggest that the supersymmetric extension of $SU(5)$ gives a consistent picture of coupling unification \cite{couplingunity,n2,n3} and is a viable possibility.

The fermionic content of $SU(5)$ is the same as the one in the SM. Therefore, as we explained in Section 4.3,  it is not possible to create any asymmetry in $B-L$. Fermions are assigned to the reducible representation $\overline{5}_f\oplus 10_f$ as
\be
\overline{5}_f=\left\{ d_L^c, \ell_L\right\}
\ee
and
\be
10_f=\left\{Q_L, u_L^c, e_L^c\right\}.
\ee
There are 24 gauge bosons which belong to the adjoint representation $24_V$ and may couple to the fermions through the couplings
\be
\frac{g}{\sqrt{2}} 24_V\:\left[(\overline{5}_f)^\dagger\: (\overline{5}_f)+
(10_f)^\dagger \:10_f\right].
\ee
Among the 24 gauge bosons there are the bosons $XY=V^1=(3,2,-5/6)$ (and their $CP$-conjugate) which may decay violating the baryon number: $XY\rightarrow
QL, \overline{Q}\overline{Q}$, where $Q$ and $L$ denotes an arbitrary quark and lepton, respectively. They
 have electric charges $Q_X=-1/3$ and $Q_Y=-4/3$. The mass and the couplings of these
bosons are determined by the gauge coupling unification
\begin{eqnarray}
M_{XY}&\simeq& 5\times 10^{14}\:{\rm GeV}, \:\:\:\alpha_{{\rm GUT}}\simeq 1/45,
\:\:\:{\rm non-supersymmetric}\:\:\:SU(5),\nonumber\\
M_{XY}&\simeq&  10^{16}\:{\rm GeV}, \:\:\:\alpha_{{\rm GUT}}\simeq 1/24,
\:\:\:{\rm supersymmetric}\:\:\:SU(5).
\end{eqnarray}
While in the gauge sector the structure is uniquely determined by the gauge group, in the Higgs sector the results depend upon the choice of the representation. The Higgs fields which couple to the fermions may be  in the 
representation $5_H$ or in the representations $10_H$, $15_H$, $45_H$ and $50_H$. If we consider the minimal choice $5_H$,
we obtain
\be
h_U\:(10_f)^T\:(10_f)\:5_H+ h_D\:(\overline{5}_f)^T\:(10_f)\:\overline{5}_H,
\ee
where $h_{U,D}$ are matrices in the flavor space. The representation $5_H$ contains the Higgs doublet of the SM, (1,2,1/2) and the triplet $S^1=(3,1,-1/3)$ which is $B$-violating. Unfortunately, this minimal choice of the Higgs sector does not suffice to explain the baryon number of the Universe. The $CP$ violation is due to the complex phases which cannot be reabsorbed by field redifinition (they are physical) in the Yukawa sector. At the tree-level these phases do not give  any contribution to the baryon asymmetry and at the one-loop level the asymmetry is proportional to
\be
{\rm Im}\:{\rm Tr}\:\left(h_U^\dagger h_U h_D^\dagger h_D\right)=0,
\ee
where the trace is over generation indices. This is because the Higgs on the external and internal legs of the one-loop interference diagrams is the same.
A net baryon number only appears at three-loop, resulting in a baryon asymmetry $\sim 10^{-16}$ which is far too small to explain the observed one.  The same problem in present in the supersymmetric version of $SU(5)$ where one has to introduce two Higgs supeerfields $5_H$ and $\overline{5}_{\overline{H}}$ \cite{habergut}. 

The problem of too tiny  $CP$ violation in $SU(5)$ may be solved by complicating further the Higgs sector. One may introduce an extra scalar $5_H^\prime$ with the same quantum numbers of $5_H$, but with a different mass and/or lifetime \cite{wayout}.   In that case one-loop diagrams with exchange of $5_H^\prime$ instead of $5_H$ can give rise to a net baryon number proportional 
\be
{\rm Im}\:{\rm Tr}\:\left(h_U^{\prime\dagger} h_U h_D^{\prime\dagger} h_D\right),
\ee
where $h^\prime_{U,D}$ are the couplings of $5_H^\prime$ to $QL$ and $\overline{Q}\overline{Q}$, respectively. A second alternative  is to introduce a different second Higgs representation. For example, adding a Higgs in the $45$ representation of $SU(5)$ an adequate baryon asymmetry may be producedb for a wide range of the parameters \cite{harvey}.

\subsubsection{The case of $SO(10)$}

In the GUT based on $SO(10)$ the spontaneous breaking down to the SM gauge group is generally obtained through different steps (for a general review, see \cite{sl}). The main two channels are
\begin{eqnarray}
&SO(10)&\stackrel{M_{{\rm GUT}}}{\rightarrow}G_{224}\stackrel{M_{R}}{\rightarrow}G_{214}
\stackrel{M_C}{\rightarrow}G_{2113}\stackrel{M_{B-L}}{\rightarrow}
SU(3)_C\otimes SU(2)_L\otimes U(1)_Y,\nonumber\\
&SO(10)&\stackrel{M_{{\rm GUT}}}{\rightarrow}G_{224}\stackrel{M_C}{\rightarrow}G_{2213}
\stackrel{M_{R}}{\rightarrow}G_{2113}\stackrel{M_{B-L}}{\rightarrow}
SU(3)_C\otimes SU(2)_L\otimes U(1)_Y,
\end{eqnarray}
where
\begin{eqnarray}
G_{224} &=& SU(2)_L\otimes SU(2)_R\otimes SU(4),\nonumber\\
G_{214} &=& SU(2)_L\otimes U(1)_{I_{3R}}\otimes SU(4),\nonumber\\
G_{2113} &=& SU(2)_L\otimes U(1)_{I_{3R}}\otimes U(1)_{B-L}\otimes  SU(3)_C,\nonumber\\
G_{2213} &=& SU(2)_L\otimes SU(2)_R\otimes U(1)_{B-L}\otimes  SU(3)_C,
\end{eqnarray}
where the four intermediate scales have not to be necessarily different from each other. We notice that, 

-- if we  are interested in the  generation of an asymmetry in $B-L$, the relevant scale is the scale at which the abelian group $U(1)_{B-L}$ breaks down, {\it i.e.} $M_{B-L}$,  and not the Grand Unification scale $M_{{\rm GUT}}$; 

-- it is usually  not possible to generate any baryon asymmetry at the scale $M_{{\rm GUT}}$. Indeed, the fermionic content of $SO(10)$ is the one of the SM plus a right-handed neutrino $N_L^c=(1,1,0)$. All the fermions belonging to the same generation are contained in the spinorial representation $16_f$. Differently from what happens for the case of $SU(5)$, now all the fermions posses the corresponding antifermions and it is possible to define a conjugation operator of the charge $C$  starting from the operators of $SO(10)$, in such a way that, if $SO(10)$ is not broken, then $C$ is conserved \cite{masiero}. In the simplest mechanism for the breaking of $SO(10)$, the one into $G_{224}$, a crucial role is played by the $54_H$. In such a case there is a symmetry $
SU(2)_L\otimes SU(2)_R$ \cite{pati,pati2,goran} with equal coupling constants $g_L$ and $g_R$ and consequently, $C$ is still a symmetry of the theory. It is possible to see that, with this choice of the Higgs representation, $C$ is not broken until $U(1)_{B-L}$ is broken, {\it i.e.} at the scale $M_{B-L}$ \footnote{In fact, if one uses the $210_H$ to break $SO(10)$, one can maintain the gauge part 
of the left-right symmetry  but not $C$. Th eimplications for baryogenesis are discussed in \cite{chang}.}. At this scale, the right-handed neutrino acquires a Majorana mass $M_N={\cal O}(M_{B-L})$ and its out-of-equilibrium decays may generate a nonvanishing $B-L$ asymmetry \cite{fy}. We will return to this point later. With a more complicated choice of the Higgs representation it is possible to break $C$ at the scale $M_{R}$ where $SU(2)_R$ is broken and in such a case baryogenesis may take place at that scale. 

\section{The out-of-equilibrium decay scenario and the thermal history of the Universe}

The out-of-equilibrium scenario that we have depicted in the previous section is operative only if a nonequilibrium number density of $X$ heavy bosons was present in the early Universe. Usually massive particles are in equilibrium at at high temperatures, $T\gg M_X$ and their number density exceeds the equilibrium one when $T$ becomes of the same order of the mass $M_X$. We have seen that, if the decay rate is small enough around $T\sim M_X$, see Eq. (\ref{yes}),  then departure from equilibrium is attained and the subsequent decays of $X$ and $\overline{X}$ particles may produce the observed baryon number asymmetry. The basic assumption -- however -- of this picture is that the superheavy bosons were as abundant as photons at very high temperatures $T\gsim M_X$.
  
If the $X$ particles are gauge or Higgs bosons of Grand Unification, the situation is somewhat more complicated because they might have never been in thermal equilibrium at the very early stages of the evolution of the Universe. Even if the temperature of the primeval plasma was higher than the Grand Unified scale $M_{{\rm GUT}}\sim 10^{16}$ GeV, the rate of production of superheavy particles would be smaller than the expansion rate of the Universe and the number density of superheavy bosons could always be smaller than the equilibrium one. Secondly, the temperature of the Universe might be always smaller than  $M_{{\rm GUT}}$ and correspondingly the thermally produced $X$ bosons might be never as abundant as photons, making their role in baryogensis negligible. All these considerations depend crucially upon the thermal history of the Universe  and deserve a closer look. 

\subsection{Inflation and reheating: the old days}

The flatness and the horizon problems of the standard big bang
cosmology are elegantly solved if during the evolution of the early
Universe the energy density happened to be dominated by some form of
vacuum energy and comoving scales grow quasi-exponentially \cite{guth}.
An inflationary stage is also required to dilute any undesirable
topological defects left as remnants after some phase transition taking
place at early epochs. 

The vacuum energy driving inflation is generally
assumed to be associated to the potential $V(\phi)$ of  some scalar field $\phi$, the {\it inflaton},
which is initially displaced from the minimum of its potential. As a
by-product, quantum fluctuations of the inflaton field may be the seeds
for the generation of structure and the 
 fluctuations observed in the cosmic microwave background radiation,
$\delta T/T\sim 10^{-5}$ \cite{ll,lr,liddle}.

Inflation ended when the potential energy associated with the inflaton
field became smaller than the kinetic energy of the field.  By that
time, any pre-inflation entropy in the Universe had been inflated
away, and the energy of the universe was entirely in the form of
coherent oscillations of the inflaton condensate around the minimum of
its potential.  The Universe may be said to be frozen after the end of
inflation. We know that somehow the low-entropy cold Universe
dominated by the energy of coherent motion of the $\phi$ field must be
transformed into a high-entropy hot Universe dominated by
radiation. The process by which the energy of the inflaton field is
transferred from the inflaton field to radiation has been dubbed
{\it reheating}. In the old theory of reheating \cite{dolgov,abbot}, 
the simplest way to envision this process is if the comoving energy
density in the zero mode of the inflaton decays into normal particles,
which then scatter and thermalize to form a thermal background.  It is
usually assumed that the decay width of this process is the same as
the decay width of a free inflaton field.

Of particular interest is a quantity known as the reheat temperature,
denoted as $T_{RH}$. The reheat temperature is calculated by assuming 
an instantaneous conversion of the energy density in the inflaton 
field into radiation when the decay width of the inflaton energy,
$\Gamma_\phi$, is equal to $H$, the expansion rate of the universe. 

The reheat temperature is calculated quite easily.   After inflation
the inflaton field executes coherent oscillations about the minimum
of the potential.  Averaged over several oscillations, the coherent
oscillation energy density redshifts as matter: $\rho_\phi \propto
a^{-3}$, where $a$ is the Robertson--Walker scale factor.  If we
denote as $\rho_I$ and $a_I$ the total inflaton energy density 
and the scale factor at the initiation of coherent oscillations,
then the Hubble expansion rate as a function of $a$ is 
\begin{equation}
H^2(a) = \frac{8\pi}{3}\frac{\rho_I}{\mpl^2}
	\left( \frac{a_I}{a} \right)^3.
\end{equation}
Equating $H(a)$ and $\Gamma_\phi$ leads to an expression for $a_I/a$.
Now if we assume that all available coherent energy density is
instantaneously converted into radiation at this value of $a_I/a$, we
can find the reheat temperature by setting the coherent energy
density, $\rho_\phi=\rho_I(a_I/a)^3$, equal to the radiation energy
density, $\rho_R=(\pi^2/30)g_*T_{RH}^4$, where $g_*$ is the effective
number of relativistic degrees of freedom at temperature $T_{RH}$.
The result is
\begin{equation}
\label{eq:TRH}
T_{RH} = \left( \frac{90}{8\pi^3g_*} \right)^{1/4}
	\sqrt{ \Gamma_\phi \mpl } \
       = 0.2 \left(\frac{200}{g_*}\right)^{1/4}
      \sqrt{ \Gamma_\phi \mpl } \ .
\end{equation}

In the simplest
chaotic inflation model,  the inflaton potential is given by 
\be
V(\phi) =
\frac{1}{2}M_\phi^2\phi^2,
\ee
with 
\be
M_\phi\sim 10^{13}\:{\rm GeV}
\ee
in order to reproduce the
observed temperature anisotropies in the microwave background \cite{lr}.
Writing $\Gamma_\phi=\alpha_\phi M_\phi$, one finds 
\be
T_{RH}\simeq
10^{15}\sqrt{\alpha_\phi}\:{\rm  GeV}.
\ee

\subsection{GUT baryogenesis and the old theory of reheating: a Herculean task}

There are very good reasons to suspect that GUT baryogenesis is not in a good shape in the old theory of reheating.

\subsubsection{Kinematical suppression of superheavy particles}

The density and temperature fluctuations observed in the present
universe, $\delta T/T\sim 10^{-5}$, require the inflaton
potential to be extremely flat  -- that is $\alpha_\phi\ll 1$. This means that the couplings of the
inflaton field to the other degrees of freedom  cannot be too large, since large couplings would induce
large loop corrections to the inflaton potential, spoiling its
flatness.  As a result, $T_{RH}$ is expected to be much smaller than
$10^{14}$GeV by several orders of magnitude. As we have seen, 
the unification scale is generally assumed to be around $10^{16}$ GeV,
and $B$-violating gauge bosons should have masses comparable to this
scale. Baryon-number violating Higgs bosons may have a mass one or two
orders of magnitude less.  For example, in $SU(5)$ the $B$
violating Higgs bosons in the five-dimensional representation that
may have a mass as small as $10^{14}$ GeV. In fact, these Higgs bosons
are more likely than gauge bosons to produce a baryon asymmetry since
it is easier to arrange the requisite $CP$ violation in the Higgs decay.  But even the light $B$-violating Higgs bosons are
expected to have masses larger than the inflaton mass, and it would be
kinematically impossible to create them directly in $\phi$ decay, $\phi\rightarrow X\overline{X}$. This is because one expects
\be
M_\phi\ll M_X.
\ee

\subsubsection{Thermal production of heavy particles}

 One might think that the $X$ bosons could be created by thermal
scattering during the stage of thermalization of the decay products of
the inflaton field. Indeed, the reheat temperature is best regarded as the temperature below which
the Universe becomes radiation dominated.  In this regard it has a
limited meaning.  For instance, it {\it should not} be interpretated
as the maximum temperature obtained by the universe during reheating.
The maximum temperature is, in fact, much larger than $T_{RH}$.  One
implication of this is that it is incorrect, to assume that the
maximum abundance of a massive particle species $X$  produced after
inflation is suppressed by a factor of $\exp(-M_X/T_{RH})$ \cite{ckr} and therefore it is incorrect to conclude   that GUT baryogenesis is imcompatible with models of inflation where the reheating temperature is much smaller than the GUT scale and, in general, than the mass of the $X$ particles, $T_{RH}\ll M_X$.
Particles of mass much greater than the eventual reheating
temperature $T_{RH}$ may be created by the thermalized decay products
of the inflaton.  Indeed, a stable particle
species $X$ of mass $M_X$ would be produced in the reheating process
in sufficient abundance that its contribution to closure density today
would be approximately $M_X^2 \langle \sigma |v|\rangle(g_*/10^3)
(M_X/10^4T_{RH})^7$, where $g_*$ is the number of effective degrees of
freedom of the radiation energy density and $\langle \sigma|v|
\rangle$ is the thermal average of the $X$ annihilation cross section
times the M{\o}ller flux factor. Thus, particles of mass as large as
$10^4$ times the reheating temperature may be produced in interesting
abundance \cite{ckr}. The  
 number density  $n_X$ of particles $X$ after  freeze out and reheating may be easily computed  \cite{ckr} 
\begin{equation}
\label{ph}
\frac{n_X}{n_\gamma}\simeq 3\times 10^{-4}\left(\frac{100}{g_*}\right)^{3/2}
\left(\frac{T_{RH}}{M_X}\right)^7\left(\frac{\mpl}{M_X}\right)
\end{equation}
and is not exponentially suppressed. 
It is easy to check that for such small values of $T_{RH}$, the ratio (\ref{ph}) is always much larger than the equilibrium value
\be
\left(\frac{n_X}{n_\gamma}\right)_{{\rm EQ}}=
\left(\frac{\pi^{1/2}}{\xi(3)}\right)
\left(\frac{M_X}{2T_{RH}}\right)^{3/2}\:e^{\frac{-M_X}{T_{RH}}}.
\ee 
This result is crucial for the out-of-equilibrium decay scenarios of baryogenesis. For instance, as we shall see,    
in theories where $B-L$ is a spontaneously broken local symmetry, as suggested by $SO(10)$ unification, the cosmological baryon asymmetry can be generated by the out-of-equilibrium decay of the lightest  heavy Majorana right-handed neutrino $N_1^c$, whose typical mass is about $10^{10}$ GeV \cite{fy}. For reheat temperatures of the order of $10^9$ GeV, the number density of the right-handed neutrino is about $3\times 10^{-2}\:n_\gamma$ and one can estimate the final bayon number to be of the order of $B\sim (n_{N_1^c}/n_\gamma)(\epsilon/g_*)\simeq 10^{-4}\epsilon$, where $\epsilon$ is the coefficient containing one-loop suppression factor and $CP$ violating phases. The observed value of the baryon asymmetry, $B\sim 10^{-10}$, is then obtained without any fine tuning of parameters.

\vspace{0.5cm}
\centerline{{\it Exercise 2}}
\vspace{0.5cm}
Compute the maximum temperature during the process of reheating. {\it Hint}: Consider the
early-time solution for radiation ({\it i.e.} when $H \gg
\Gamma_\Phi$ and before a significant fraction of the comoving
coherent energy density is converted to radation).

\subsubsection{The gravitino problem}

 There is one more problem associated with GUT baryogenesis in the old theory of reheating, namely the problem of relic gravitinos \cite{gravitino}. If one has to invoke supersymmetry to preserve the flatness of the inflaton potential, it is mandatory to consider the cosmological implications of the gravitino -- a spin-(3/2) particle which appears in the  extension of global supersymmetry to local supersymmetry -- or supergravity \cite{gelmann}. The gravitino is
 the fermionic superpartner of the graviton and has interaction strength with the observable sector -- that is the SM particles and their superpartners -- inversely proportional to the Planck mass. One usually associates the scale of supersymmetry breaking with the electroweak scale in order to handle the hierarchy problem \cite{nilles} and the mass of the gravitino is of order of the weak scale, $m_{3/2}={\cal O}(1)$ TeV. The decay rate of the gravitino is given by
\be
\Gamma_{3/2}\sim \frac{m_{3/2}^3}{\mpl^2}\sim (10^5\:{\rm sec})^{-1}\:\left(
\frac{m_{3/2}}{{\rm TeV}}\right)^3.
\ee
The slow decay rate of the gravitinos
is the essential source of the cosmological problems because the  decay products of the gravitino  will destroy the $^4$He and D nuclei by photodissociation, 
and thus successful nucleosynthesis predictions. The most stringent bound comes from the resulting 
overproduction of D $+$ $^3$He, which would require that the gravitino abundance is smaller than $\sim 10^{-10}$ relative to the entropy density at the time of reheating after inflation \cite{kaw}
\be
\label{lll}
\frac{n_{3/2}}{s}\lsim (10^{-10}-10^{-11}).
\ee
The Boltzmann equation governing  the number density of gravitinos $n_{3/2}$  during the thermalization stage after inflation is
\be
\label{l}
\frac{d n_{3/2}}{dt}+3H n_{3/2}\simeq\langle \Sigma_{{\rm tot}} v\rangle n^2_{{\rm light}},
\ee
where $\Sigma_{{\rm tot}}\propto 1/\mpl^2$ is the total cross section determining the rate of production of gravitinos and $n_{{\rm light}}\sim T^3$ represents the number density of light particles in the thermal bath. The number density of gravitinos at thermalization is readily obtained solving Eq.  (\ref{l}) and reads
\be
\label{ll}
\frac{n_{3/2}}{s}\simeq 10^{-2}\:\frac{T_{RH}}{\mpl}.
\ee
Comparing Eqs. (\ref{lll}) and (\ref{ll}), one may obtain an upper bound on the reheating temperature after inflation 
\be
T_{RH}\lsim (10^{10}-10^{11})\: {\rm GeV}.
\ee
Therefore, if $T_{RH}\sim M_{{\rm GUT}}$, gravitinos would be abundant during nucleosynthesis and destroy the good agreement of the theory with observations. However, if the initial state after inflation was free from gravitinos, the reheating temperature seems to be  too low to create superheavy $X$ bosons that eventually decay and produce the baryon asymmetry -- even taking into account the previous considerations about the fact that the maximum temperature during reheating is not $T_{RH}$ \cite{book,ckr}. 

\subsection{Inflation and reheating: the new wisdom}

The outlook for GUT baryogenesis has brightened recently with the
realization that reheating may differ significantly from the simple
picture described above \cite{explosive,KT1,KT2,KT3,KT4}.  In the
first stage of reheating, called {\it preheating} \cite{explosive},
nonlinear quantum effects may lead to an extremely effective
dissipational dynamics and explosive particle production even when
single particle decay is kinematically forbidden. Particles can be
produced in the regime of a broad parametric resonance, and it is
possible that a significant fraction of the energy stored in the form
of coherent inflaton oscillations at the end of inflation is released
after only a dozen or so oscillation periods of the inflaton. What is
most relevant for these lectures is that preheating may play
an extremely important role for baryogenesis \cite{klr,alr,krt} and, in particular, for GUT generation of the baryon
asymmetry. Indeed, it was shown in \cite{klr,krt} that the baryon
asymmetry can be produced efficiently just after the preheating era,
thus solving many of the problems that GUT baryogenesis had to face in
the old picture of reheating.

The presence of a preheating stage at the beginning of the reheating process is based on the fact that, for some  parameter ranges,
there is a new decay channel that is non-perturbative: 
due to the coherent oscillations of the inflaton field
stimulated emissions of bosonic particles
into energy bands with large occupancy numbers are induced \cite{explosive}. The modes in these
bands can be understood as Bose condensates, and they behave like
classical waves. The back-reaction of these modes on the homogeneous
inflaton field and the rescattering among themselves produce a state
that is far from thermal equilibrium and may induce very interesting
phenomena, such as non-thermal phase transitions \cite{nt1,nt2,nt3}
with production of a stochastic background of
gravitational waves \cite{KT4} and of heavy particles in a state
far from equilibrium, which may 
constitute today the dark matter in our Universe \cite{dm1,dm2}. 

 The idea of preheating is relatively simple, the
oscillations of the inflaton field induce mixing of positive and
negative frequencies in the quantum state of the field it couples to because of the {\it time-dependent} mass of the quantum field. Let us focus -- for sake of simplicity -- to the case of 
chaotic inflation, with a massive inflaton $\phi$ with quadratic potential $V(\phi)=\frac{1}{2}M_\phi^2\phi^2$, $M_\phi\sim 10^{13}$ GeV,  and coupled
to a massless scalar field $\chi$ via the quartic coupling $g^2\phi^2\chi^2$. 

The evolution equation for the
Fourier modes of the $\chi$ field with momentum $k$ is
\be
\ddot X_k + \omega_k^2 X_k=0,
\ee
with
\begin{eqnarray}
X_k&=&a^{3/2}(t)\chi_k,\nonumber\\
\omega_k^2 &=& k^2/a^2(t) + g^2\phi^2(t).
\end{eqnarray}
This Klein-Gordon equation may be cast
in the form of a Mathieu equation
\be
X_k'' + [A(k) - 2q\cos2z]X_k =0,
\ee
where $z=M_\phi t$ and
\begin{eqnarray}
\label{pa}
A(k)&=&\frac{k^2}{a^2 M_\phi^2}+2q,\nonumber\\
q&=&g^2\frac{\Phi^2}{4 M_\phi^2},
\end{eqnarray} 
where $\Phi$ is the
amplitude and $M_\phi$ is the frequency of inflaton oscillations,
$\phi(t)=\Phi(t)\sin(M_\phi t)$. Notice that, at least initially when $\Phi=c \mpl\lsim \mpl$ 
\be
g^2\frac{\Phi^2}{4 M_\phi^2}\sim g^2 \: c^2 \frac{\mpl^2}{M_\phi^2}\sim g^2\:c^2\times 10^{12}\gg 1
\ee
and the resonance is broad. 
For certain values of the parameters $(A,q)$
there are exact solutions $X_k$ and the corresponding number density $n_k$ that grow exponentially with time because they belong to an instability band of the Mathieu
equation (for a recent comprehensive review on preheating
  after chaotic inflation \cite{Kofman} and references
  therein)
\be
X_k\propto e^{\mu_k M_\phi t}\Rightarrow n_k\propto e^{2\mu_k M_\phi t},
\ee
where the parameter $\mu_k$ depends upon the instability band and, in the broad resonance case, $q\gg 1$, it is $\sim 0.2$.

These instabilities can be interpreted as coherent
``particle'' production with large occupancy numbers. One way of
understanding this phenomenon is to consider the energy of these modes
as that of a harmonic oscillator, $E_k = |\dot X_k|^2/2 + \omega_k^2
|X_k|^2/2 =  \omega_k (n_k + 1/2)$.  The occupancy number of
level $k$ can grow exponentially fast, $n_k\sim\exp(2\mu_k M_\phi t)\gg1$,
and these modes soon behave like classical waves.  The parameter $q$ during preheating determines
the strength of the resonance. It is possible that the model
parameters are such that parametric resonance does {\it not} occur,
and then the usual perturbative approach would follow, with decay rate
$\Gamma_\phi$. In fact, as the Universe expands, the growth of the scale
factor and the decrease of the amplitude of inflaton oscillations
shifts the values of $(A,q)$ along the stability/instability chart of
the Mathieu equation, going from broad resonance, for $q\gg 1$, to
narrow resonance, $q\ll 1$, and finally to the perturbative decay of
the inflaton.  

It is important to notice that, after the short period of preheating, the Universe is likely to enter a long period of matter domination where the biggest contribution to the energy density  of the Universe is provided by 
the residual small amplitude oscillations of the classical inflaton field and/or by the inflaton quanta produced during the back-reaction processes. This period will end when the age of the Universe becomes of the order of the perturbative lifetime of the inflaton field, $t\sim \Gamma_\phi^{-1}$. At this point, the Universe will be reheated up to a temperature $T_{RH}$ given in (\ref{eq:TRH}) obtained applying the old theory of reheating described in the previous section.

\subsection{GUT baryogenesis and preheating}

A crucial observation for baryogenesis is that even particles with
mass larger than that of the inflaton may be produced during
preheating. To see how this might work, let us assume that the
interaction term between the superheavy bosons and the inflaton field
is of the type $g^2\phi^2|X|^2$.  During preheating, quantum
fluctuations of the $X$ field with momentum $\vec{k}$ approximately
obey the Mathieu equation where now
\be
A(k) = \frac{k^2 + M_X^2}{ M_\phi^2} + 2q.
\ee
  Particle
production occurs above the line $A = 2 q$.  The width of the
instability strip scales as $q^{1/2}$ for large $q$, independent of
the $X$ mass.  The condition for broad resonance \cite{explosive,klr}
\be
A-2q \lsim
q^{1/2}
\ee
becomes 
\be
\frac{k^2 + M^2_X}{M_\phi^2} \lsim g
\frac{\Phi}{M_\phi},
\ee
 which yields for the typical energy of $X$ bosons
produced in preheating  
\be
E_X^2 = k^2 + M^2_X \lsim g\Phi M_\phi,
\ee
By the time the resonance develops to the
full strength, $\Phi^2 \sim 10^{-5} \mpl^2$.  The resulting
estimate for the typical energy of particles at the end of the broad
resonance regime for $M_\phi\sim 10^{13}$ GeV is 
\be
E_X \sim 10^{-1}
g^{1/2}\sqrt { M_\phi \mpl} \sim g^{1/2} 10^{15}\:{\rm  GeV}. 
\ee
 Supermassive
$X$ bosons can be produced by the broad parametric resonance for $E_X
> M_X$, which leads to the estimate that $X$ production will be
possible if $M_X < g^{1/2} 10^{15}$ GeV.

For $g^2 \sim 1$ one would have copious production of $X$ particles
(in  this regime the problem
is non-linear from the beginning and therefore $g^2=1$ has to
 be understood as a rough
estimate of the limiting case) as heavy as $10^{15}$GeV, i.e., 100
times greater than the inflaton mass. The only problem here is that for large coupling $g$, radiative
corrections to the effective potential of the inflaton field may
modify its shape at $\phi \sim M_{\rm Pl}$.  However, this problem
does not appear if the flatness of the inflaton potential is protected
by supersymmetry.

This is a significant departure from the old constraints of reheating.
Production of $X$ bosons in the old reheating picture was
kinematically forbidden if $M_\phi< M_X$, while in the new scenario it is
possible because of coherent effects. It is also important to note
that the particles are produced out-of-equilibrium, thus satisfying
one of the basic requirements to produce the baryon asymmetry
\cite{sak}.

Scattering of $X$ fluctuations off the zero mode of the inflaton field
limits the maximum magnitude of $X$ fluctuations to be $\langle
X^2\rangle_{\rm max} \approx M_\phi^2/g^2$ \cite{KT3}.  For example,
$\langle X^2\rangle_{\rm max} \sim 10^{-10} \mpl^2$ in the case
$M_X = 10\:M_\phi$. This restricts the corresponding number density of
created $X$-particles.

A potentially important dynamical effect is that the parametric
resonance is efficient only if the self-interaction couplings of the
superheavy particles are not too large.  Indeed, a self-interaction
term of the type $\lambda |X|^4$ provides a non-thermal mass to the
$X$ boson of the order of $(\lambda \langle X^2\rangle)^{1/2}$, but
this contribution is smaller than the bare mass $M_X$, if $\lambda\lsim
g^2 M_X^2/M_\phi^2$.  Self-interactions may also terminate the resonance
effect because scattering induced by the coupling $\lambda$ may remove
particles from the resonance shells and redistribute their momenta
\cite{explosive}. But this only happens if, again, $\lambda \gg g^2$
\cite{KT2}.

The parametric resonance is also rendered less efficient when the $X$
particles have a (large) decay width $\Gamma_X$, which is essential for the out-of-equilibrium decay to take place.
Roughly speaking, one
expects that the explosive production of particles takes place only if
the typical time, $\tau_e$, during which the number of $X$ bosons
grows by a factor of $e$, is smaller than the decay lifetime
$\tau_X=\Gamma_X^{-1}$.  During the broad resonance regime, typically
$\tau_e \lsim 10\:M_\phi^{-1}$. If we write the decay width by $\Gamma_X=\alpha_X\:M_X$, this requires
$\alpha_X \lsim  0.1 M_\phi/M_X$.  Notice that
smaller values of $\Gamma_X$ are favored not only because particle
production is made easier, but also because the superheavy particles
may remain out-of-equilibrium for longer times, thus enhancing the
final baryon asymmetry.

Using the methods developed in Refs.\ \cite{KT1,KT2,KT3}, one can  
study numerically the production of massive, unstable $X$ particles
in the process of the inflation decay \cite{krt}.
Let us consider a model in which the oscillating inflaton field $\phi$
interacts with a scalar field $X$ whose decays violate baryon number
$B$. As we have learned, the   simplest possibility for the $X$-particle is the Higgs field in
the five-dimensional representation of $SU(5)$. We assume standard kinetic terms, minimal
coupling with gravity, and a very simple potential for the fields of
the form
\begin{equation}
V(\phi,X)={1\over 2} M_\phi^2\phi^2  + {1\over 2} M_X^2 X^2 +
{1\over 2} g^2 \phi^2 X^2.
\label{pot}
\end{equation}

A fundamental parameter in GUT baryogenesis is $n_X$, the number
density of the supermassive leptoquarks whose decays produce the
baryon asymmetry.  It will depend upon the value of $\Gamma$ and $q$.

Since the supermassive bosons are more massive than the inflaton, one expects 
small kinetic energy in the excitations of the $X$ field.  From the potential 
of Eq.\ (\ref{pot}), the square of the
effective mass of the $X$ field is 
\be
(M^{{\rm  EFF}}_X)^2 = M_X^2+g^2\langle \phi^2\rangle
\ee
 and the
energy density in the
$X$ field will be 
\be
\rho_X\simeq(M_X^2+g^2\langle \phi^2\rangle)\langle X^2\rangle.
\ee
Writing $\langle \phi^2\rangle$ as $\phi_0^2 + \langle \delta\phi^2\rangle$,
one  can define an analog of the $X$-particle number density as
\begin{equation}
n_X=\rho_X/M^{\rm  EFF}_X = \left[4q(\phi_0^2 +
\langle\delta\phi^2\rangle)/\phi_0^2(0) +
m_{\chi}^2\right]^{1/2} M_\phi \langle X^2 \rangle,
\label{n}
\end{equation}
where $m_\chi
= M_X/M_\phi$.

Eq.\ (\ref{n}) enables one to calculate the number density of the
created $X$-particles if the variances of the fields, $\langle
X^2\rangle$, $\langle\delta\phi^2\rangle$, and the inflaton zero mode
$\phi_0 (\tau)$ (here $t$ and $\tau$ are related by $M_\phi dt= a(\tau)d\tau$) are known.  

 \begin{figure}
\centering
\leavevmode\epsfysize=3.2in \epsfbox{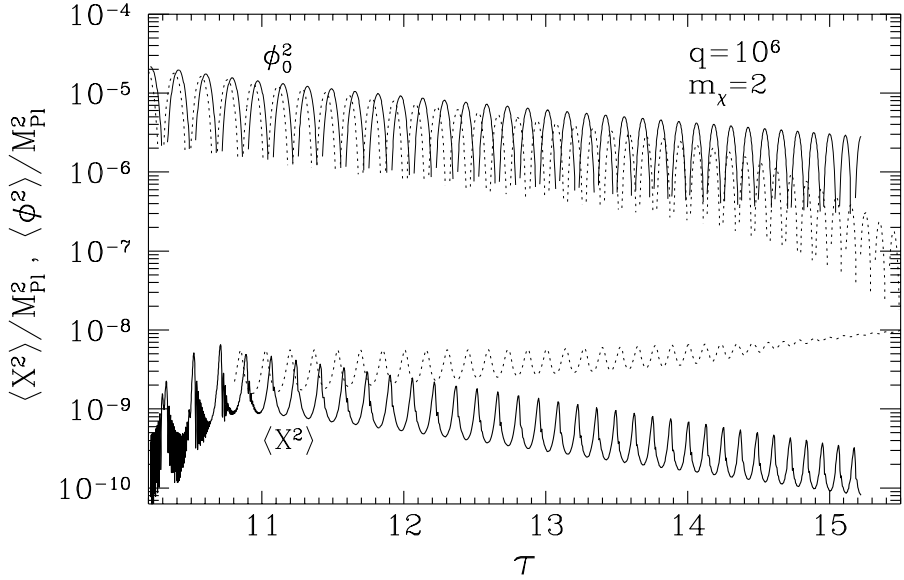}
\caption{The variance of $X$ with model parameters $q=10^6$, $m_\chi=2$, and
$\Gamma \equiv \Gamma_X/M_\phi=6\times10^{-2}$ is shown by the lower solid curve as a
function of time.  The upper solid curve corresponds to the inflaton
zero mode. The dotted curves represent the same quantities for $\Gamma
=0$.  }
\label{fig:Fig4}
\end{figure}
\begin{figure}
\centering
\leavevmode\epsfysize=3.2in \epsfbox{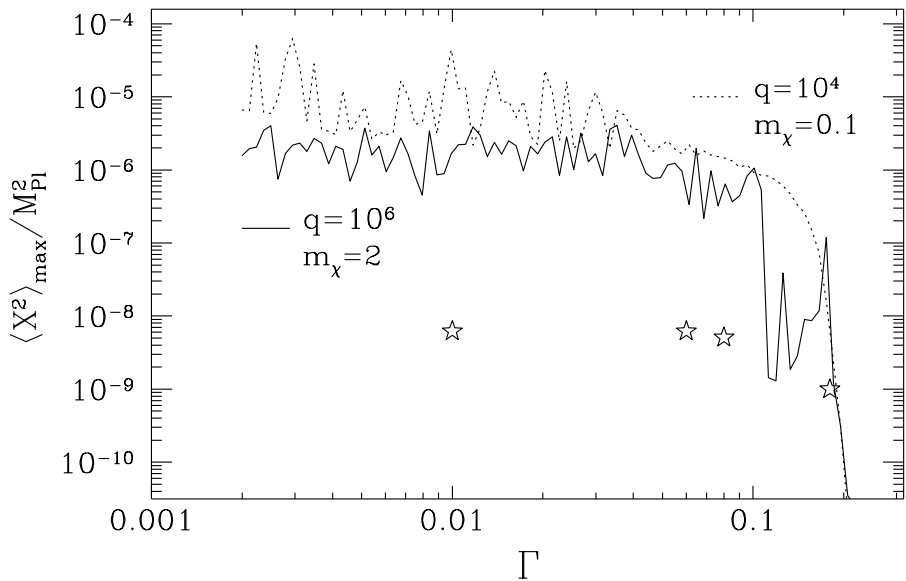}
\caption{ The maximum value of the variance of the $X$-field, $\langle
X_{\rm max}^2 \rangle$, is shown as function of $\Gamma$.  Stars mark
$\langle X_{\rm max}^2\rangle$ obtained in the full non-linear
problem. $\langle X_{\rm max}^2\rangle$ in the Hartree approximation
is shown by the dotted curve for $q=10^4$, $m_\chi=0.1$, and by the
solid curve for $q=10^6$, $m_\chi=2$.  }
\label{fig:Fig5}
\end{figure}

The time evolution of the variance, $\langle X^{2} \rangle$, and of
the inflaton zero mode, $\langle \phi \rangle$, is shown in Fig. 4,  by the solid curves for the case $q=10^6$, $m_\chi=2$,
and $\Gamma=6\times10^{-2}$.  We see that the particle creation
reaches a maximum at $\tau \approx 10.8$ when $\langle X^{2} \rangle
\approx 10^{-9}$ in the ``valleys'' between the peaks. At later times, $\tau >
10.8$, particle creation by the oscillating inflaton field can no
longer compete with $X$-decays due to the non-zero value of
$\Gamma$. For comparison, we show in the same figure the case $\Gamma
=0$ represented by the dotted curves \cite{KT3}. In the $\Gamma =0$
case, particle creation is able to compete with the expansion of the
universe so that $\langle X^{2} \rangle$ remains roughly
constant.

Using Eq.\ (\ref{n}), one finds  for the maximum number density of created
$X$-particles 
\be
n_X=\left[4q\phi_0^2(10.8)/\phi_0^2(0) +
m_{\chi}^2\right]^{1/2} M_\phi \langle X^2 \rangle \approx \left[10^3 +
m_{\chi}^2\right]^{1/2} M_\phi \langle X^2 \rangle \approx 30 M_\phi \langle
X^2\rangle.
\ee
It is easy to understand that if we increase the value of $\Gamma$, the
parametric resonance will not be able to compete with the decay of $X$
at earlier times.  Moreover, for sufficiently large values of
$\Gamma$, the resonance will be shut off in the linear
regime. 

In exploration of parameter space it turns out more convenient to  go  to the Hartree approximation
 which requires much less
computing resources. The maximum value of the variance of $X$ reached
during the time evolution of the fields in the Hartree approximation
is shown in Fig.\ \ref{fig:Fig5} as a function of the parameters of
the model. Here the stars also show the maximum of $\langle
X^{2}(\tau) \rangle$ in the full non-linear problem for a few values
of $\Gamma$.  At small $\Gamma$ the Hartree approximation
overestimates $\langle X^{2} \rangle$ significantly \cite{KT2,
KT3}. Nonetheless, at large values of $\Gamma$ it is a quite reliable
approach.  One may  see that $\langle X^{2} \rangle$ drops sharply when
$\Gamma > 0.2$, and  this critical value of
$\Gamma$ does not depend significantly upon $m_X$ or $q$ \cite{krt}.

The most relevant case with $q=10^8$, where $X$-bosons as massive as
ten times the inflaton mass can be created, is shown in Fig. 6 in the
Hartree approximation. Note, that two lower curves which correspond to
$\Gamma$ equal to 0.08 and 0.12 never reach the limiting value
$\langle X^2\rangle_{\rm max} \sim 10^{-10} \mpl^2$, which is
imposed by rescattering \cite{KT3}, and the Hartree approximation
ought to be reliable in this cases.

As outlined above, one may  consider a three part reheating process,
with initial conditions corresponding to the frozen universe at the
end of inflation.  The first stage is explosive particle production,
where a fraction $\delta$ of the energy density at the end of
preheating is transferred to $X$ bosons, with $(1-\delta$) of the
initial energy remaining in $\phi$ coherent oscillation energy. We
assume that this stage occurs within a few Hubble times of the end of
inflation.  The second stage is the $X$ decay and subsequent
thermalization of the decay products.  We assume that decay of an
$X$--${{\overline{X}}}$ pair produces a net baryon number $\epsilon$,
as well as entropy.  Reheating is brought to a close in the third
phase when the remaining energy density in $\phi$ oscillations is
transferred to radiation.

\begin{figure}
\centering
\leavevmode\epsfysize=3.6in \epsfbox{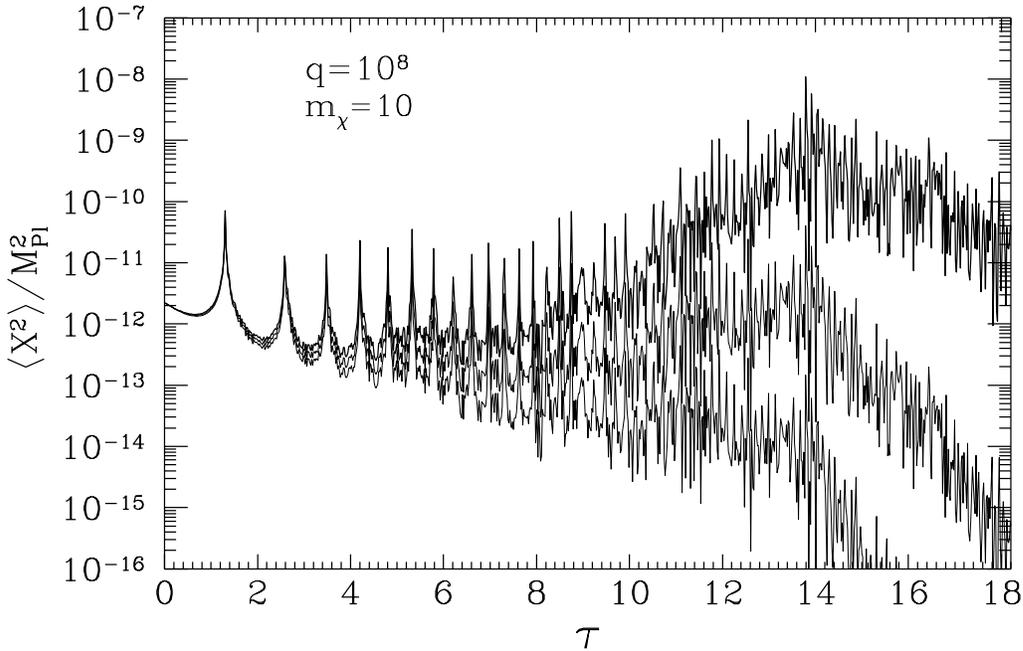}
\caption{ The time dependence of the variance of $X$ in the Hartree 
approximation with
model parameters $q=10^8$,
 $m_\chi=10$ and for three values of $\Gamma$, from top to bottom:
 0.04, 0.08, 0.12.}
\label{fig:Fig6}
\end{figure}

\begin{figure}
\centerline{ \epsfysize=3.6in  \epsfbox{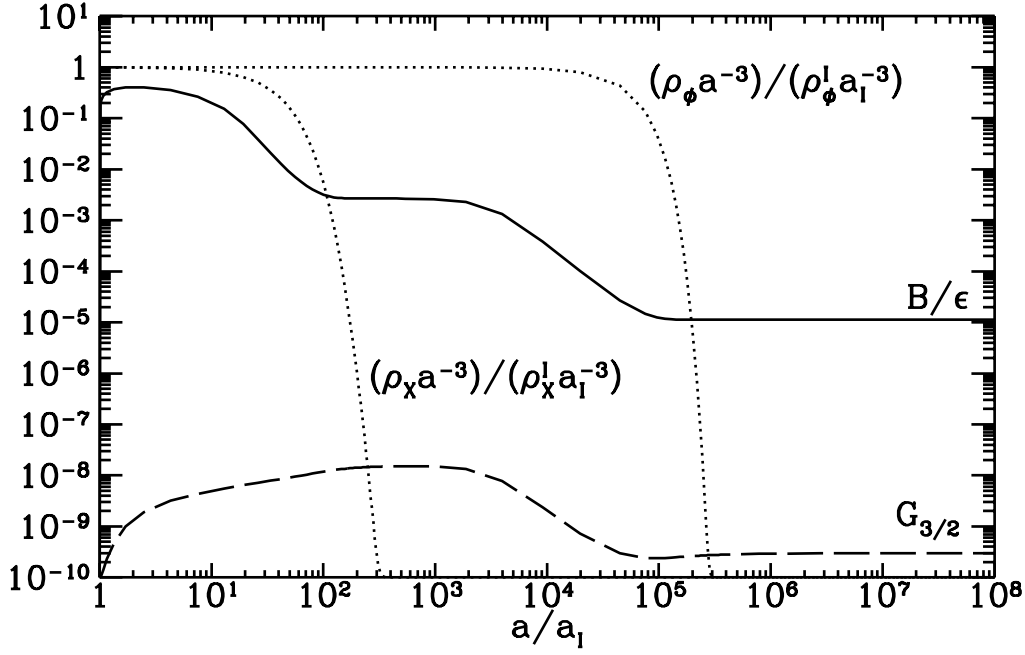} }
\caption{  The evolution of the baryon
 number, the $X$ number density, the energy density in $\phi$
 oscillations, and the gravitino-to-entropy ratio as a function
 of the scale factor $a$.}
\end{figure}

The final baryon asymmetry depends linearly upon the ratio $\delta$
between the energy stored in the $X$ particles at the end of the
preheating stage and the energy stored in the inflaton field at the
beginning of the preheating era \cite{klr}. 
The description simplifies if we assume zero initial kinetic energy of
the $X$s.   One may also a assume that there are fast interactions that
thermalize the massless decay products of the $X$.  Then in a
co-moving volume $a^3$, the total number of $X$ bosons, $N_X=n_Xa^3$,
the total baryon number, $N_B=n_Ba^3$, and the dimensionless radiation
energy, $R=\rho_Ra^4$, evolve according to
\begin{eqnarray}
\label{network}
\dot{N}_X & = & - \Gamma_X \left( N_X - N_X^{{\rm EQ}} \right); \quad
 \dot{R}    =   - a M_X \dot{N}_X;
\nonumber \\
\dot{N}_B & = & - \epsilon \dot{N}_X
       - \Gamma_X N_B \left( N_X^{{\rm EQ}} / N_0 \right) .
\end{eqnarray}
$N_X^{{\rm EQ}}$ is the total number of $X$s in thermal equilibrium at
temperature $T \propto R^{1/4}$, and $N_0$ is the equilibrium number
of a massless degree of freedom in a comoving volume.

Fig.\ 7 shows the results of an integration of Eqs. (\ref{network}) in
a toy model with $M_\phi = 10^{13}$GeV, $M_X = 10^{14}$GeV, $\Gamma_X
= 5 \times 10^{-6} M_X$, $\Gamma_\phi = 5 \times 10^{-10} M_\phi$ ,
and two degrees of freedom ($b$ and ${{\overline{b}}}$).  Initial
conditions were chosen at $a=a_I$ to be $\rho_X = \rho_{\phi }
\sim 10^{-4} M_\phi^2 \mpl^2$, and $R = N_B = 0$.  The $\rho_X =
\rho_{\phi }$ assumption corresponds to $\delta=1/2$. 
 The baryon
number $B = n_B/s$ rapidly rises.  However $B$ decreases as entropy is
created and $X$ inverse reactions damp the baryon asymmetry.  After
most of the energy is extracted from the initial $X$ background, the
baryon number is further damped as entropy is created during the decay
of energy in the $\phi$ background.  
One can also numerically integrate the equation governing the number
density of gravitinos $n_{3/2}$. The result for
$G_{3/2}=n_{3/2}/s$ is shown in Fig. 7. Notice that, even though
gravitinos are copiously produced at early stages by scatterings of
the decay products of the $X$, $G_{3/2}$ decreases as entropy is
created during the subsequent decay of energy in the $\phi$
background.  

Since the
number of $X$ bosons produced is proportional to $\delta$, the final
asymmetry is proportional to $\delta$ and $B/\epsilon \sim 10^{-9}$ can be
obtained for $\delta$ as small as $10^{-6}$.
One  can
estimate  this ratio as
\begin{equation}
\delta\simeq 3\times 10^6\, \sqrt{\frac{q}{10^6}} \,
m_\chi\, \frac{\langle X^2\rangle}{\mpl^2}.
\end{equation}
Therefore, for $q=10^8$ and $m_\chi=10$, $\delta$ is of the order of
$3\times 10^{8} \langle X^2\rangle/\mpl^2$. Since
the final baryon asymmetry scales approximately as $\Gamma^{-1}$ and
is given by $B\simeq 5\times
10^{-4}\:\delta\:\epsilon\:(\Gamma/5\times 10^{-5})^{-1}$ \cite{klr},
where $\epsilon$ is an overall parameter accounting for
$CP$ violation, one can  see that the observed baryon asymmetry
$B\simeq 4\times 10^{-11}$ may be explained by the phenomenon of GUT
baryogenesis after preheating if
\begin{equation}
\frac{\langle X^2\rangle}{\mpl^2}\simeq 5\times 10^{-13}
\left(\frac{10^{-2}}{\epsilon}\right)\left(\frac{\Gamma}{5\times 10^{-5}}
\right).
\end{equation}
From Fig.\ 6 we can read that this only may happen if 
\beq
\Gamma_X\lsim 10^{-3} \:M_X.
\ee  
This result may be considered very
comfortable since we can conclude that whenever the resonance
develops, i.e., when $\Gamma_X \lsim 10^{-1} M_\phi=10^{-2} M_X$,
GUT baryogenesis after preheating is so efficient that the right
amount of baryon asymmetry is produced for almost the entire range of
values of the decay rate $\Gamma_X$. In other words, provided that
superheavy $X$-bosons are produced during the preheating stage, they
will be {\it ineffective} in producing the baryon asymmetry {\it only}
if their decay rate falls in the range $10^{-3} M_X \lsim\Gamma_X \lsim  10^{-2} M_X$. GUT baryogenesis after preheating solves many of the
serious drawbacks of GUT baryogenesis in the old theory of reheating
where the production of superheavy states after inflation was
kinematically impossible. Moreover, the out-of-equilibrium condition
is naturally attained in our scenario since the distribution function
of the $X$-quanta generated at the resonance is far from a thermal
distribution.  This situation is considerably different from the one
present in the GUT thermal scenario where superheavy particles usually
decouple from the thermal bath when still relativistic and then decay
producing the baryon asymmetry.
It is quite intriguing that out of all possible ways the parametric
resonance may develop, Nature might have chosen only those ways
without instantaneous thermalization and also with a successful
baryogenesis scenario.

\section{The baryon number violation in the Standard Model}

In this section we will be concerned with the violation of the baryon and lepton number in the SM. It is well-known that by considering the most general  Lagrangian invariant under the SM gauge group $SU(3)_C\otimes SU(2)_L\otimes U(1)_Y$ and assuming that the Higgs fields are color singlets, the Lagrangian is automatically invariant under global abelian symmetries which may be identified with the baryonic and leptonic symmetries. They are -- therefore -- accidental symmetries. As a result, it is not possible to violate $B$ and $L$ at the tree-level and at any order of perturbation theory: the proton is stable in the SM and any perturbative process which violates $B$ and/or $L$ in Grand Unified Theories is necessarily suppressed by powers of $M_{{\rm GUT}}/M_W$. 
Nevertheless, in many cases the perturbative expansion does not describe all the dynamics of the theory and -- indeed -- in 1976 't Hooft \cite{ho}
realized that nonperturbative effects (instantons) may give rise to processes which violate the combination $B+L$, but not the orthogonal combination $B-L$. The probability of these processes to occur is exponentially suppressed, $\sim
{\rm exp}(-4\pi/\alpha_W)\sim 10^{-150}$ where $\alpha_W=g_2^2/4\pi$ is the weak gauge coupling, and probably irrelevant today. In more extreme situations -- like the primordial Universe at very high temperatures \cite{ds,km,krs} -- baryon and lepton number violation processes may be fast enough to play a significant role in baryogenesis. This will be the subject of the present section.

\subsection{The $B+L$ anomaly}

The violation of the baryonic number within the SM is due to the fact that the current corresponding to the global abelian group $U_{B+L}$ -- even though it is conserved at the classical level -- is not conserved
at the quantum level, that is the $U_{B+L}$ is {\it anomalous}.
Let us consider in the euclidean space the generating function ${\rm exp}[-Z]=
\int\:{\cal D}\psi\:{\cal D}\overline{\psi}\:{\rm exp}[-S]$ \cite{fu}. The most general phase  transformation onto the Dirac field $\psi$ with mass $m$
\be
\label{rot}
\psi(x)\rightarrow e^{i(a+b\:\gamma_5)\theta(x)}\:\psi(x)
\ee
induces an additional term in the action given by
\be
\label{s0}
\delta S_0=-\int\: d^4x\:\left[\overline{\psi} \:m\:\left(e^{2ib\gamma_5\theta(x)}-1\right)\:\psi +\overline{\psi}\gamma^\mu(a+b\gamma_5)\psi\partial_\mu\theta(x)\right].
\ee
The rotation (\ref{rot}) gives rise to a nontrivial jacobian due to  the noninvariance of the measure ${\cal D}\psi\:{\cal D}\overline{\psi}$. This may be expressed as an additional contribution to the action
\be
\delta S_1=i\:\int\:d^4x\:\theta(x)\left[\frac{(a-b)}{8\pi^2}\:{\rm
Tr}\:F^{(L)\mu\nu}\widetilde{F}^{(L)}_{\mu\nu}-\frac{(a+b)}{8\pi^2}\:{\rm
Tr}\:F^{(R)\mu\nu}\widetilde{F}^{(R)}_{\mu\nu}\right],
\ee
where $F^{(L)\mu\nu}$ ($F^{(R)\mu\nu}$) is the field strength which couples to the left-handed (right-handed) current of the field $\psi$, while $\widetilde{F}_{\mu\nu}=\frac{1}{2}\epsilon_{\mu\nu\sigma\rho}F^{\sigma\rho}$. Notice that we  have absorbed the gauge coupling into the definitions of 
$F^{(L,R)\mu\nu}$ and the traces are over the group indices. 

If the rotation (\ref{rot}) corresponds to the baryon number rotation, then 
\be
a=\frac{1}{3}\:\:\:{\rm and}\:\:\: b=0.
\ee
Integrating by parts (\ref{s0}) and requiring that the generating function is invariant under  the baryonic number transformation, we obtain that the baryonic current $J^\mu_B=\sum_q\frac{1}{3}\overline{q}\gamma^\mu q$ is anomalous
\be
\label{anomaly}
\partial_\mu J^\mu_B=i\frac{N_F}{32\pi^2}\left(-g_2^2 F^{a\mu\nu}\widetilde{F}^{a}_{\mu\nu}+ g_1^2 f^{\mu\nu}\widetilde{f}_{\mu\nu}\right),
\ee
where $N_F$ is the number of fermionic families, $F^{a}_{\mu\nu}$ is the field strength of $SU(2)_L$ and $f_{\mu\nu}$ that of $U(1)_Y$ with coupling constants $g_2$ and $g_1$, respectively, and we have made use of the fact that ${\rm Tr}(T^a T^b)=\frac{1}{2} \delta_{ab}$ for the $SU(2)_L$ generators and of the values $Y=1/6$, 2/3, $-1/3$ for $Q_L$, $u_R$ and $d_R$, respectively. 

Analogously, if we consider the rotation associated to the lepton number
\be
\label{b-l}
a=1 \:\:\:{\rm and}\:\:\: b=0,
\ee
we obtain
\be
\partial_\mu J^\mu_B=\partial_\mu J^\mu_L,
\ee
where $J^\mu_L=\sum_\ell(\overline{\ell}\gamma^\mu \ell+\overline{\nu}_\ell\gamma^\mu \nu_\ell)$ is the leptonic current. 
The relation (\ref{b-l}) shows explicitly that the current associated to $B-L$ is conserved. In fact -- since each quark and leton family gives the same contribution  to the anomaly  and each leptonic flavor is conserved in the SM at the classical level -- the conserved charges are actually three
\be
\ell_i\equiv\frac{1}{3} B-L_i,
\ee
where $L_i$ $(i=e$, $\mu$, $\tau$) are the leptonic flavors and $\sum_i L_i=L$.

\subsection{Topology of $SU(2)_L$ and baryon number violation}

In the previous subsection we have described how the chiral anomaly induces the nonconservation of the baryonic current. We now wish to understand what is the physical significance of the terms in the righ-hand side of the Eq. (\ref{anomaly}). First, we note that these terms may be reexpressed as 
\be
\partial_\mu J^\mu_B=i\frac{N_F}{32\pi^2}\left(
-g_2^2 \partial_\mu K^\mu +g_1^2 \partial_\mu k^\mu\right),
\ee
where
\begin{eqnarray}
K^\mu &=& 2\epsilon^{\mu\nu\rho\sigma}\left(\partial_\nu A_\rho^a A_\sigma^a-\frac{1}{3} g_2 \epsilon_{abc}A_\nu^a A_\rho^b A_\sigma^c\right),\nonumber\\
k^\mu &=& 2\epsilon^{\mu\nu\rho\sigma}\left( \partial_\nu B_\rho B_\sigma\right),
\end{eqnarray}
and the variation $\Delta B$ of the total baryon number
\be
B=i\: \int\:d^3 x\: J^0_B
\ee 
in the time interval $\Delta t$ is related to the quantity $N_{{\rm CS}}$ and $n_{{\rm CS}}$, called the Chern-Simons numbers, 
in the same time interval
\be
\label{delta}
\Delta B=N_F(\Delta N_{{\rm CS}}-\Delta n_{{\rm CS}}),
\ee
where
\begin{eqnarray}
\label{kkk}
N_{{\rm CS}}&=& -\frac{g_2^2}{16\pi^2}\:\int\:d^3x\:2\epsilon^{ijk} \:{\rm Tr} \left[
\partial_i A_j A_k+i\frac{2}{3} g_2 A_i A_j A_k\right],\nonumber\\
n_{{\rm CS}}&=& -\frac{g_1^2}{16\pi^2}\:\int\:d^3x\:\epsilon^{ijk} \:\partial_iB_j B_k,
\end{eqnarray}
where we have defined $A_i\equiv A_i^a\sigma^a/2$.

Now, each $U(1)_Y$ gauge transformation
\be
B_i\rightarrow B_i+\frac{i}{g_1}(\partial_i U_Y)U_Y^{-1},
\ee
with $U_Y(x)=e^{i\alpha_Y(x)}$ leaves $n_{{\rm CS}}$ unchanged. On the contrary, there exist $SU(2)_L$ gauge transformations
\be
A_i\rightarrow U A_i U^{-1} +\frac{i}{g_2}(\partial_i U)U^{-1},
\ee
which induce a nonvanishing variation of $N_{{\rm CS}}$
\be
\delta N_{{\rm CS}}=\frac{1}{24\pi^2}\:\int\: d^3 x \:{\rm Tr}\left[
(\partial_i U)U^{-1}(\partial_j U)U^{-1}(\partial_k U)U^{-1}\right] \epsilon^{ijk}.
\ee
This is due to the topological properties of $SU(2)$. The most generic $2\times 2$ unitary matrix with determinant equal to unity may be expressed as $a{\bf 1}+ibi_i\sigma^i$, with the condition $a^2+|b|^2=1$. Therefore the topology of $SU(2)$ is the same as $S^3$, the surface of the hypersphere in four dimensions (three-sphere) and the gauge transformations are maps from the euclidean space onto $SU(2)\sim S^3$.
To clarify this point further, we recall that classically, the ground state must correspond to {\it time-independent} field configuration with vanishing energy density. We have therefore $F_{\mu\nu}^a\equiv 0$, which means that the field ${\bf A}$ is a pure gauge, ${\bf A}_{{\rm vac}}=(i/g)(\nabla U)U^{-1}$ and we are working in the gauge $A_0=0$. Furthermore, we may restrict ourselves to the transformations $U$ that have the same limit in all spatial directions. We may take this limit to be the identity  in the group, $U\rightarrow {\bf 1}$ as $|\vec{x}|\rightarrow \infty$. Under these circumstances, all the configurations ${\bf A}_{{\rm vac}}$ may be regarded as describing a ground state. We have seen that $SU(2)$ is isomorphic to the three-demensional sphere $S^3$. On the other hand, the whole three-dimensional space with all points at infinity identified is also topologically equivalent to $S^3$. Therefore the gauge transformation $U(x)$ associated with each vacuum is a mapping from $S^3$ onto $S^3$. According to the homotopy theory, such mappings fall into equivalence classes. Two mappings $\vec{x}\rightarrow U_1(\vec{x})$ and $\vec{x}\rightarrow U_2(\vec{x})$ belong to the same class if there exists a continuous transformation from $U_1(\vec{x})$ to $U_2(\vec{x})$. In the case at hand, the classes are labeled by positive or negative integer called the winding number. 

We may consider some standard maps
\begin{eqnarray}
U^{(0)}(x)&=& 1,\nonumber\\
U^{(1)}(x)&=&\frac{x_0+i\vec{x}\cdot\vec{\sigma}}{r}, \:\:\:r=(x_0+|\vec{x}|^2)^{1/2},\nonumber\\
&\vdots&\nonumber\\
U^{(n)}(x)&=& \left[U^{(1)}\right]^n,\nonumber\\
&\vdots &
\end{eqnarray}
It is easy to check that $\delta N_{{\rm CS}}$ vanishes for $U^{(0)}$  and any continous transformation of  $U^{(0)}$
\be
U(x)= U^{(0)}(x)\left({\bf 1}+i\epsilon^a(x)\sigma^a\right),
\ee
where $\epsilon^a(x)\rightarrow 0$ when $|\vec{x}|\rightarrow\infty$. On the other hand, $\delta N_{{\rm CS}}$  does not vanish if we consider the transformation $U^{(1)}$
\be
\delta N_{{\rm CS}}\left(U^{(1)}\right)=1.
\ee
The same result is obtained considering continuous deformations of $U^{(1)}$. It is also possible to show that -- in general -- the transformations $U^{(n)}$ and their continuous tranformations give rise to 
\be
\delta N_{{\rm CS}}\left(U^{(n)}\right)=n.
\ee
Therefore, the gauge transformations of $SU(2)$ may be divided in two categories, those which do not change the Chern-Simons number, and those which change the Chern-Simons number by $n$, the winding number.

Let us now consider the SM in the limit in which the mixing angle is zero, {\it i.e.} the theory is pure gauge  $SU(2)_L$ theory coupled to the Higgs field $\Phi$. If we choose the gauge $A_0=0$, we may go from the classical vacuum defined as
\be
{\cal G}_{{\rm vac}}^{(0)}=\left\{A_i^{(0)}=0, \Phi^{(0)}=(0,v), N_{{\rm CS}}=0\right\},
\ee
to an infinite number of other vacua
\be
{\cal G}_{{\rm vac}}^{(n)}=\left\{A_i^{(n)}=(i/g_2)(\nabla U^{(n)})(U^{(n)})^{-1}, \Phi^{(n)}=U^{(n)}\Phi^{(0)}, N_{{\rm CS}}=n\right\},
\ee
which are classically degenerate and have different Chern-Simons number. Here we have denoted the vacuum expectation value (VEV) of the Higgs field by $v$.

If we now go back to eq. (\ref{delta}), we are able to understand the connection among  the baryonic chiral anomaly, the topological structure of $SU(2)$ and the baryon number violation. If the system is able to perform a transition from the vacuum ${\cal G}_{{\rm vac}}^{(n)}$ to the closest one
${\cal G}_{{\rm vac}}^{(n\pm 1)}$, the Chern-Simons number is changed by one unity and 
\be
\Delta B=\Delta L=N_F.
\ee
Each transition creates 9 left-handed quarks (3 color states for each generation) and 3 left-handed leptons (one per generation)
\be
3\sum_{i=1}^3 Q^i_L +\sum_{i=1}^3 \ell^i_L\leftrightarrow 0.
\ee

\subsection{The sphaleron}

To quantify  the probability of transition between two different vacua, it is important to understand the properties of the field configurations which interpolate the two vacua and ``help'' the transition. A fundamental result
has been obtained in ref. \cite{km}, where it was found that there exist static configurations (therefore independent from $t$) which correspond to unstable solutions of the equations of motion. These solutions are called {\it sphalerons} (which in greek stands for ``ready to fall'') and correspond to saddle points of the energy functional and posses Chern-Simons number equal to 1/2. The situation is schematically depicted in Fig. 8

\begin{figure}
\centering
\leavevmode\epsfysize=3.2in \epsfbox{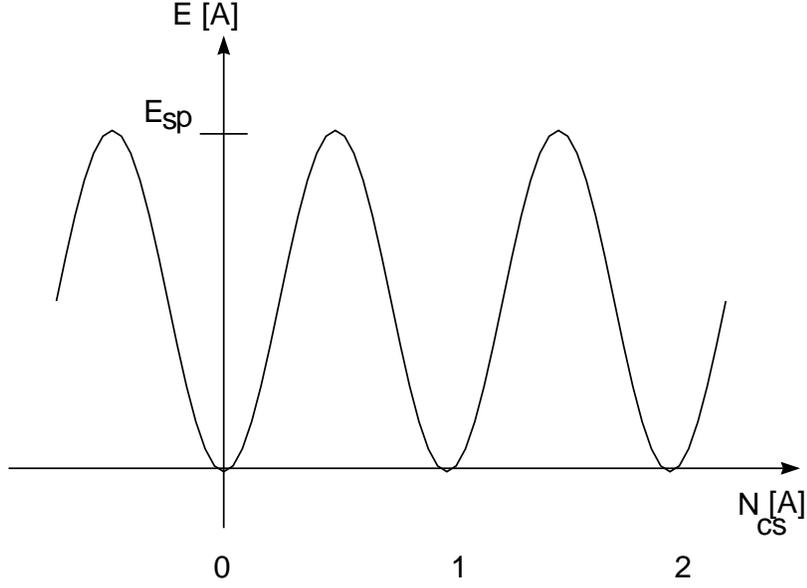}
\caption{Schematical representation of the energy dependence of the gauge configurations as a function of the Chern-Simons number. Sphalerons correspond to the maxima of the curve.}
\label{fig:Fig8}
\end{figure}

The sphaleron may be identified  by considering the minimum energy path among all the paths that, in the configuration space, connect two vacua whose Chern-Simons number differs by one unit. Along this path, the sphaleron is the configuration of maximum energy and is localized is space, even though -- contrary to the case of the soliton -- is unstable.

In the limit of vanishing mixing angle,  $\theta_W\rightarrow 0$, the sphaleron solution has been found explicitly by Klinkhamer and Manton \cite{km} and has the following form
\begin{eqnarray}
\label{kk}
A_i dx^i&=& \frac{i}{g_2} \:f(g_2 v r) \:U^\infty\: d(U^\infty)^{-1},\nonumber\\
\Phi&=& \frac{i v}{\sqrt{2}} \:h(g_2 v r) \:U^\infty\:\left(
\begin{array}{c}
0\\
1
\end{array}\right),
\end{eqnarray}
where $U^\infty(x_1,x_2,x_3)=U^{(1)}(x_0=0,x_1,x_2,x_3)$. The energy functional
\begin{eqnarray}
E&=&\int\:d^3 x\left[\frac{1}{4} F_{ij}^a F_{ij}^a+({\cal D}_i\Phi)^\dagger ({\cal D}_i\Phi)+V(\Phi)\right],\nonumber\\
V(\Phi)&=& \lambda\left(\Phi^\dagger\Phi -\frac{1}{2}v^2\right)^2,
\end{eqnarray}
may be reexpressed as
\begin{eqnarray}
\label{func}
E &=& \frac{4\pi v}{g_2}\:\int_0^\infty\: d\xi\:\left[
4\left(\frac{d f}{d\xi}\right)^2 +\frac{8}{\xi^2}\left[f(1-f)\right]^2+\frac{1}{2}\xi^2 \left(\frac{d h}{d\xi}\right)^2\right.\nonumber\\
&+& \left. \left[h(1-f)\right]^2+\frac{\lambda}{g_2^2}\xi^2(h^2-1)^2\right],\nonumber\\
\xi&=& g_2 vr,
\end{eqnarray}
where $V(\phi)$ is the potential of the Higgs field. 

The functions $f$ and $h$ are the ones which minimize the energy functional 
(\ref{func}) with the boundary conditions
\be
f(\xi)\rightarrow
\left\{
\begin{array}{c}
\sim \xi^2, \:\:\xi\rightarrow 0\\
1, \:\:\xi\rightarrow\infty
\end{array}\right.
\ee
and 
\be
h(\xi)\rightarrow
\left\{
\begin{array}{c}
\sim \xi, \:\:\xi\rightarrow 0\\
1, \:\:\xi\rightarrow\infty.
\end{array}\right.
\ee
The sphaleron is therefore the solution which interpolates between 
${\cal G}_{{\rm vac}}^{(0)}$ (for $\xi\rightarrow 0$) and ${\cal G}_{{\rm vac}}^{(1)}$ (for $\xi\rightarrow \infty$). The energy and the typical dimensions of the sphaleron configuration are basically the result of the competition between the energy of the gauge configuration and the energy of the Higgs field. The latter introduces the weak scale into the problem. From the  quantitative point of view, the potential energy of the Higgs field is less important and for the sphaleron configuration of dimension $\ell$, we have
\begin{eqnarray}
A_i&\sim & \frac{1}{g_2\ell},\nonumber\\
E(A_i) &\sim & \frac{4\pi}{g_2^2\ell},
\end{eqnarray}
while the energy of the Higgs field is
\be
E(\phi)\sim 4\pi v^2\ell.
\ee
Minimizing the sum $E(A_i)+E(\phi)$ we obtain that
the typical dimension of the sphaleron is
\be
\ell_{{\rm sp}}\sim \frac{1}{g_2 v}\sim 10^{-16} \:\:{\rm cm},
\ee
and 
\be
E_{{\rm sp}}\sim \frac{8\pi v}{g_2}\sim 10\:\: {\rm TeV}.
\ee
A more accurate result may be found by means of variational methods \cite{km}
\be
\label{b}
E_{{\rm sp}}=\frac{4\pi v}{g_2}B\left(\frac{\lambda}{g_2}\right),
\ee
where $B$ is a function which depends very weakly on $\lambda/g_2$: 
$B(0)\simeq 1.52$ and $B(\infty)\simeq 2.72$. Including the mixing angle $\theta_W$ changes the energy of the sphaleron at most of 0.2\%.
The previous computation of the sphaleron energy was performed at zero temeprature. The sphaleron at finite temperature -- but still in the broken phase -- was computed in \cite{bk} where it was shown that its energy follows approximately the scaling law
\be
E_{{\rm sp}}(T)=E_{{\rm sp}}\frac{\langle \phi(T)\rangle}{v},
\ee
where $\langle \phi(T)\rangle$ is the VEV of the Higgs field at finite temperature in the broken phase. This energy may be writen as
\be
\label{pp}
E_{{\rm sp}}(T)=\frac{2 m_W(T)}{\alpha_W}B\left(\frac{\lambda}{g_2}\right),
\ee
where $m_W(T)=\frac{1}{2}g_2\langle \phi(T)\rangle$.

The Chern-Simons number of the sphaleron may be explicitly computing by plugging (\ref{kk}) into (\ref{kkk}) \cite{km} and, as mentioned above, one obtains
\be
N_{{\rm CS}}^{{\rm sp}}=\frac{1}{2}.
\ee

\subsection{Baryon number violating transitions}

The probability of baryon number nonconserving processes at zero temperature has been computed by 't Hooft \cite{ho} and, as we have already mentioned, is highly suppressed by a factor  ${\rm exp}(-4\pi/\alpha_W)\sim 10^{-150}$. This factor may be interpreted as the probability of making a transition from one classical vacuum to the closest one by tunneling, by going through the barrier of $\sim 10$ TeV corresponding to the sphaleron. An easy way to evaluate this number is to remember that the field configurations that describe the transitions (sphalerons or instantons in the case of tunneling) are characterized by $A_i\sim 1/(g_2 v)$ and therefore their contribution to the generating function is
\be
\int\:{\cal D}A\:e^{-S[A]}\sim e^{-1/\alpha_W}\ll 1.
\ee
On the other side one might think that baryon number violating transitions may be obtained in physical situations which involve a large number of fields. The contribution to the transition amplitude in processes which involve $N$ fields may be estimated as
\be
\left|A_{B+L}\right|^2\sim\left(\frac{1}{\alpha_W}\right)^N\:e^{-4\pi/\alpha_W}=e^{-4\pi/\alpha_W-N\:{\rm log}\:\alpha_W}.
\ee
Therefore, if the number of fields involved is about $N\sim 1/\alpha_W$, the  transition probability  may become of order unity. The sphaleron may be produced by collective and coherent excitations containing $N\gsim 1/\alpha_W$ quanta with wavelength of the order of $\ell_{{\rm sp}}\sim 1/M_W$. At temperatures $T\gg M_W$, these modes essentially obey statistical mechanics and the transition probability may be computed via classical considerations. Note also that at temperatures $T\ll M_W$ it is no longer possible to deal with classical considerations because the Compton wavelength of the thermal excitations $\sim T^{-1}$ is much larger than the size of the sphaleron.  

\subsubsection{Baryon number violation below the electroweak phase transition}

After (or during) the electroweak phase transition \cite{mariano} by which the SM gauge group 
$SU(3)_C\otimes SU(2)_L\otimes U(1)_Y$ breaks down to $U(1)_{em}$, the calculation of the baryon number violation rate can be done by using the semiclassical approximations \cite{thermal}. The vacuum expectation value of the Higgs field $\langle \phi(T)\rangle$ is nonvanishing and the sphaleron configuration may be explicitly written down. 

We now want to estimate   the transition probability between two different vacua havig Chern-Simons number which differ by one unity.
One may use an useful analogy. Let us consider a pendulum of mass $m$ and be $\theta$ the angle which determines the position of the pendulum with respect to the position at rest, $\theta=0$. It is clear that 
the transformation
\be
\theta\rightarrow \theta+2\pi,
\ee
may be consider a sort of gauge transformation since the position $\theta$ and 
$\theta+2\pi$ are indistinguishable.  The periodic 
 potential reads
\be
V(\theta)=(m g h)(1-\cos\theta),
\ee
wheer $h$ is the pendulum length. The energy of the corresponding ``sphaleron'' -- the saddle point solution to the equation of motion -- is $V(\pi)=2mgh$. According to the classical theory, for energies smaller than $V(\pi)$, only oscillations around $\theta=0$ are possible. However, quantum theory predicts a nonvanishing probability of tunneling through the barrier separating $\theta=0$ from $\theta=2\pi$, {\it i.e.} a complete rotation of the pendulum. 

Since the solution to the Schrodinger equation reads
\be
\psi(\theta)=A\:e^{-\int_0^\theta\:d\theta^\prime\:\sqrt{2m V(\theta^\prime)}}
\ee
the density probability ${\cal P}$ for penetration from $\theta=0$ to $\theta=2\pi$ is
\be
{\cal P}\sim \left|\psi(2\pi)\right|^2=|A|^2 e^{-2\int_0^{2\pi}\:d\theta\:\sqrt{2m V(\theta)}}
\ee
and the quantum tunneling is exponentially suppressed. Imagine now to raise up the temperature of the system, so that the pendulum coupled to the thermal bath becomes excited at higher and higher energies. As the temperature becomes
of the order of $V(\pi)$, it becomes possible for the pendulum to reach the position $\theta=\pi$ and to roll down to $\theta=2\pi$. The transition rate
is therefore
\be
\Gamma(T)\propto e^{-V(\pi)/T},
\ee
and becomes unsuppressed as long as $T\gg V(\pi)$. 

More formally, one has to remember that one of the fundamental objects in statistical thermodynamics is the partition function
\be
Z={\rm Tr}\:e^{-\beta \hat{H}},
\ee
where $ \hat{H}$ is the Hamiltonian operator and $\beta=T^{-1}$. For a scalar field, one may introduce the field eingenstates $|\phi(\vec{x}),t\rangle$ of the Heisenberg picture field operator $\hat{\phi}(\vec{x},t)$
\be
\hat{\phi}(\vec{x},t)|\phi(\vec{x}),t\rangle=\phi(\vec{x})|\phi(\vec{x}),t
\rangle.
\ee
Then the partition function may be written as ``summation'' over the eigenstates
\be
Z=\sum_{\phi(\vec{x})}\langle \phi(\vec{x}),t=0|e^{-\beta \hat{H}}|\phi(\vec{x}),t=0\rangle.
\ee
We can now make the analogy with the zero temperature case where in the language of field theory
\begin{eqnarray}
\langle \phi^{\prime\prime}(\vec{x}),t^{\prime\prime}|\phi^{\prime}(\vec{x}),t^{\prime}\rangle &=& \langle \phi^{\prime\prime}(\vec{x}),t=0|e^{-i\hat{H}(t^{\prime\prime}-t^\prime)}
|\phi^\prime(\vec{x}),t=0\rangle\nonumber\\
&\propto& \int\:{\cal D}\phi\:{\cal D}\pi \:{\rm exp}\left[
i\int_{t^\prime}^{t^{\prime\prime}}\: dt\:\int\:d^3x\left(\pi\dot{\phi}-
{\cal H}(\pi,\phi)\right)
\right],
\end{eqnarray}
where the path integral is over all the conjugate momenta of $\phi$, $\pi$ and over all the functions satisfying the boundary conditions $\phi=\phi^{\prime\prime}(\vec{x})$ at $t^{\prime\prime}$ and $\phi=\phi^{\prime}(\vec{x})$ at $t^{\prime}$. If, heuristically we introduce a variable
\be
\tau\equiv it
\ee
and take the limit of integration
\be
t^\prime=0\:\:\:{\rm and}\:\:\: t^{\prime\prime}=-i\beta,
\ee
we obtain
\begin{eqnarray}
& & \langle\phi^{\prime\prime}(\vec{x}),t=0| e^{-\beta \hat{H}}|\phi^\prime(\vec{x}),t=0\rangle\nonumber\\
&\propto& \int\:{\cal D}\phi\:{\cal D}\pi \:{\rm exp}\left[
\int_{0}^{\beta}\: d\tau\:\int\:d^3x\left(i\pi\frac{\partial\phi}{\partial\tau}-
{\cal H}(\pi,\phi)\right)
\right],
\end{eqnarray}
where now the new boundary conditions are given by
\be
\phi(\beta,\vec{x})=\phi^{\prime\prime}(\vec{x})\:\:\:{\rm and}\:\:\:
\phi(0,\vec{x})=\phi^{\prime}(\vec{x}).
\ee
By integrating out the conjugate momenta and identifying the boundary conditions, we may compute the partition function
\be
Z\propto \int\:{\cal D}\phi\:{\rm exp}\left[
\int_{0}^{\beta}\: d\tau\:\int\:d^3x\:
{\cal L}\:(\phi,\overline{\partial}\phi)
\right],
\ee
where 
\be
\overline{\partial}\phi\equiv\left(i \frac{\partial\phi}{\partial\tau},\nabla\phi\right)
\ee
and
\be
{\cal L}(\phi,\overline{\partial}\phi)=-\frac{1}{2}\left(\frac{\partial\phi}{\partial\tau}\right)^2-\frac{1}{2}\left(\nabla\phi\right)^2-V(\phi).
\ee
The theory at finite temperature may be therefore interpreted   as a theory
in $3+1$ dimensions in the euclidean space with periodic boundary conditions
on the time coordinate and period $\beta=1/T$. 

In the limit of very high temperatures only the zero mode of the expansion
$\phi=\beta^{-1}\sum_n e^{-i\omega_n\tau}\widetilde{\phi}(\omega_n,\vec{x})$, where $\omega_n=2\pi n/\beta$,  is important and the action reduces to  
$-S_3/T$ where $S_3$ is euclidean three-dimensional action
\be
S_3=\int\:d^3 x\left[\frac{1}{2}(\nabla\phi)^2 +V(\phi)\right].
\ee
The transition probability per unit time and unit volume at finite temperature  between two different minima at $\phi_1$ and $\phi_2$  of a given potential $V(\phi)$ for a generic scalar field $\phi$ is  therefore given at finite temperature  by \cite{thermal}
\be
\frac{\Gamma}{V}\sim A(T)\:e^{-S_3/T},
\ee
where  $A(T)$ is a prefactor which, on dimensional argument, is ${\cal O}(T^4)$ and the  three-dimensional action must be computed for the field configuration (bounce solution) which interpolates between the two vacua
\begin{eqnarray}
\frac{d^2\phi}{dr^2}+\frac{2}{r}\frac{d\phi}{dr}&=&V^\prime(\phi),\nonumber\\
{\rm lim}_{r\rightarrow\infty}\:\phi(r)&=&\phi_1,\nonumber\\
\left.\frac{d\phi}{dr}\right|_{r=0}&=&0,\nonumber\\
\phi(0)&=&\phi_2,\nonumber\\
r&=&|\vec{x}|.
\end{eqnarray} 
This configuration is a bubble whose interior is characterized by the value of the scalar field $\phi_2$ and the exterior by $\phi_1$. 

\vspace{0.5cm}
\centerline{{\it Exercise 3}}
\vspace{0.5cm}
Estimate the typical size $R_c$ of the bounce solution in the limit  of thick  bubbles.

After this long disgression, we are ready to estimate the topological transition rate. Since the transition which violates the baryon number
is  is sustained by the sphaleron configuration, one gets $S_3=E_{{\rm sp}}(T)$. The prefactor was computed in \cite{carson} as
\be
\label{ratebr}
\Gamma_{{\rm sp}}\sim 2.8\times 10^{5}\:T^4\:\left(\frac{\alpha_W}{4\pi}\right)^4\:\kappa\left[\frac{E_{{\rm sp}}(T)}{B\:}\right]^7\:e^{-E_{{\rm sp}}(T)/T},
\ee
where $B$ has been defined in (170) and $\kappa$ is the functional determinant associated to the fluctuations about the sphaleron. It has been estimated to be in the range $10^{-4}\lsim \kappa\lsim 10^{-1}$ \cite{dhs}.

\subsubsection{Baryon number violation above the electroweak phase transition}

At temperatures above the electroweak phase transition, the vacuum expectation value of the Higgs field is zero, $\langle\phi(T)\rangle=0$, the Higgs field decouples and the sphaleron configuration ceases to exist. The relevant configuration, at this point, are of the form
\begin{eqnarray}
\Phi&=&0,\nonumber\\
A_i dx^i&=& \frac{i}{g_2}\:f \:U^\infty\: d(U^\infty)^{-1}.
\end{eqnarray}
Let us estimate the rate $\Gamma_{{\rm sp}}$ on dimensional grounds. As we mentioned,  at high
 temperature $T$ the Higgs field decouples from the dynamics and it 
suffices to consider a pure $SU(2)$ gauge theory. Topological 
transitions take  place through the creation of non-perturbative,
 nearly static, magnetic field configurations that generate a
 change in the  Chern-Simons number $\Delta N_{{\rm CS}}$ 
with a corresponding  baryon number  generation
 $\Delta B= N_f\Delta N_{{\rm CS}}$. 

If the field configuration responsible for the transition has a typical scale $\ell$, a change $\Delta N_{{\rm CS}}\simeq 1$
requires 
\be
\Delta N_{{\rm CS}}\sim g_2^2\:\ell^3\:\partial A_i A_i\sim g_2^2
\ell^3\:\frac{A_i}{\ell} A_i\sim 1\Rightarrow A_i\sim \frac{1}{g_2\ell}.
\ee
This means that the typical energy of the configuration is 
\be
E_{{\rm sp}}\sim \ell^3\:(\partial A_i)^2\sim \frac{1}{g_2^2\ell}.
\ee
To evade the Boltzmann
 suppression factor
this   energy should not be larger than the temperature $T$,
which requires
\be 
\ell\gsim \frac{1}{g_2^2 T}.
\ee
Such a  length scale corresponds to the one of the  dynamically generated    magnetic mass 
of order $g_2^2 T$ which behaves as a cut off for  
 the maximum coherence length of the system. The rate of one unsuppressed transition per volume $\ell^3$ and time $t\sim \ell$ is therefore
\be
\label{rateun}
\Gamma_{{\rm sp}}\sim \frac{1}{\ell^3 t}\sim (\alpha_W\:T)^4.
\ee
This simple scaling argument has been recently criticized in refs. \cite{arnold1,arnold2} where it has been argued that damping effects in the plasma suppress the rate by an extra power of $\alpha_W$ to give $\Gamma_{{\rm sp}}\sim \alpha_W^5 T^4$. Indeed, since the transition rate  involves physics at soft energies $g_2^2T$ that are small compared to the typical  hard energies $\sim T$ of  the thermal excitations in the plasma, the simplest way of analyzing the problem  is to consider an effective  theory for the soft modes, where  the hard  modes have been integrated  out and to keep the dominant contributions,  the so-called hard thermal loops \cite{hard}. It is the    resulting typical  frequency $\omega_c$  of a gauge field configuration immersed  in the plasma and  spatial with extent  $(g^2 T)^{-1}$ that determines the change of baryon number per unit time  and unit volume. This frequency $\omega_c$ has been estimated to be $\sim g_2^4T$ when  taking into account  the damping effects  of the hard modes \cite{arnold1,arnold2}. This gives  $\Gamma_{{\rm sp}}\sim \frac{\omega_{{\rm c}}}{\ell^3 }\sim \alpha_W^5 T^4$. The effective dynamics of soft nonabelian gauge fields at finite temperature has been recently addressed also in \cite{bod}, where it was found that $\Gamma_{{\rm sp}}\sim \alpha_W^5 T^4\:{\rm ln}(1/\alpha_W)$. 
Lattice simulations with hard-thermal loops included have been performed \cite{moore} and seem to indicate the $\Gamma_{{\rm sp}}\sim 30\alpha_W^5 T^4$, which is not far from $\alpha_W^4 T^4$. In order to see whether these predictions are reliable, one should write down an effective classical hamiltonian for the soft modes of the gauge configurations after having integrated out also the   soft loops between 
magnetic fields. These soft loops result to be crucial since 
they  not only renormalize the effective $\alpha_W$ coupling to
non perturbative values,  but also
falsify the naive dimensional arguments about the typical time scale of the sphaleron-like fluctuations. From now on, we will parametrize the sphaleron rate as 
\be
\Gamma_{{\rm sp}}=\kappa(\alpha_W\:T)^4.
\ee

\subsection{The wash-out of $B+L$}

Let us suppose -- for sake of simplicity -- that all the charges
which are conserved by the interactions of the particles in the plasma  ($Q$, $L_i$, $B-L$, $\ell_i=B/3-L_i$, $\cdots$) are zero. If we  introduce a chemical potential for the charge $B+L$, $\mu_{B+L}$, 
the free energy  density of the system (femions) is given by
\be
F=T\:\int\:\frac{d^3k}{(2\pi)^3}\left[{\rm log}\:\left(
1+e^{-(E_k-\mu_{B+L})/T}\right)+(\mu_{B+L}\rightarrow -\mu_{B+L})\right].
\ee
The charge density of $B+L$ may be expressed in terms of the chemical potential by
\be
n_{B+L}\sim \mu_{B+L}T^2
\ee
and -- therefore -- we may relate the free energy with $n_{B+L}$
\be
F\sim \mu_{B+L}^2 T^2+{\cal O}(T^4)\sim \frac{n_{B+L}^2}{T^2}+{\cal O}(T^4).
\ee
The free energy increases quadratically with the fermion number density and the transitions which increase $n_{B+L}$ are energetically disfavoured with respect to the ones that decrease the fermion number. If these transitions are active for a long enough period of time, the system relaxes to the state of minimum energy, {\it i.e.} $n_{B+L}=0$: {\it any initial asymmetry in $B+L$ relaxes  to zero}.

To address this issue more quantitatively, one has to consider the ratio between the transitions with $\delta N_{{\rm CS}}=+1$ and the ones with $\delta N_{{\rm CS}}=-1$
\be
\frac{\Gamma_{+}}{\Gamma_{-}}=e^{-\Delta f/T},
\ee
wheer $\Delta f$ is the free energy difference between the two vacua. If we define $\Gamma_{{\rm sp}}$ to be the average between $\Gamma_{+}$ and $\Gamma_{-}$, we may compute the rate at which the baryon number is washed out \cite{bs}
\be
\label{master}
\frac{dn_{B+L}}{dt}=\Gamma_{+}-\Gamma_{-}\simeq -\frac{13}{2}\:N_F\:\frac{\Gamma_{{\rm sp}}}{T^3}\:n_{B+L}.
\ee
Equation (\ref{master})  is crucial to discuss the fate of the baryon asymmetry
generated at the GUT scale and is called {\it Master} equation. 

Let us now consider temperatures much above the electroweak phase transition, $T\gg M_W$. Baryon number violation processes are active at very  high temperatures if the rate \ref{rateun}) is smaller than the expansion of the Universe
\be
\frac{\Gamma_{{\rm sp}}}{T^3}\gsim H\Rightarrow T\lsim\alpha_W^4\:\frac{\mpl}{g_*^{1/2}}\sim 10^{12}\:{\rm GeV}.
\ee
If so, any preexisting asymmetry in $B+L$ is erased exponentially with a typical time scale 
$\tau \sim 2N_FT^3/13 \Gamma_{{\rm sp}}$.

Let us now consider temperatures $T\sim M_W$ when the electroweak phase transition is taking place and the Higgs VEV $\langle \phi(T)\rangle$ is not zero. Baryon number violation processes are out-of-equilibrium if, again,  the
rate (\ref{ratebr}) is smaller than the expansion rate of the Universe. This translates into the bound on $E_{{\rm sp}}(T)$ \cite{bs}
\be
\label{bb}
\frac{E_{{\rm sp}}(T_c)}{T_c}\gsim 45,
\ee
wheer we have indicated by $T_c$ the critical temperature at which the electroweak phase transition is taking place. Using the relation (\ref{pp}) this bound may be translated into a bound on $\langle \phi(T_c)\rangle$
\be
\label{kop}
\frac{\langle \phi(T_c)\rangle}{T_c}\gsim 1.
\ee
Any generation of the baryon asymmetry at the electroweak phase transition requires -- therefore --  a strong enough phase transition, that is  able to produce a VEV for the Higgs field larger than the critical temperature. We will come back  to this point later on.

\subsubsection{A crucial point}

In all the considerations leading to Eq. (\ref{master}) we have been assuming that all the charges which are conserved by the interactions of the particles in the plasma  ($Q$, $B-L$,  $L_i$, $\ell_i=B-L_i/3$, $\cdots$) are vanishing. Suppose now that these charges -- let us denote them generically by $Q_i$ -- are not zero. Define by $(B+L)_{{\rm EQ}}$ the value of the number density associated to 
the  $B+L$ charge when the sphaleron transitions are in {\it equilibrium} in the plasma (ideally, when the sphaleron rate $\Gamma_{{\rm sp}}\rightarrow \infty$).   In such a case, it is possible to show (see Exercise 4) that $(B+L)_{{\rm EQ}}$ is not vanishing in the plasma  \cite{kl}
\be
\label{mean}
(B+L)_{{\rm EQ}}=\sum_i \:c_i\:Q_i,
\ee
where the numerical coefficients $c_i$ depend upon which interactions are in equilibrium in the plasma and the particle content of the theory.

Eq. (\ref{mean}) tells us that  anomalous baryon number violating processes do not wash out completely the combination $B+L$ if at least one of the charges which are conserved by the interactions of the plasma, {\it e.g.} $(B-L)$ is nonvanishing. Eq. (\ref{master}) changes accordingly (see Exercise 4):
\be
\label{master1}
\frac{d n_{B+L}}{dt}\propto -\frac{\Gamma_{{\rm sp}}}{T}\frac{\partial F}{\partial (B+L)}\propto -\frac{\Gamma_{{\rm sp}}}{T}\left[n_{(B+L)}-n_{(B+L)}^{{\rm EQ}}\right].
\ee
The interpretation of this equation is straightforward. It is  a Boltzmann equation in the sense that,  in the limit $\Gamma_{{\rm sp}}\rightarrow \infty$, the solution is $B+L=(B+L)_{{\rm EQ}}$. If the conserved charges are zero, then any $B+L$ is washed-out, see Eq. (\ref{mean}). However, if $(B+L)_{{\rm EQ}}$ is not zero, then 
sphalerons transitions will act on the system  until  the $(B+L)$ charge has been reduced to its equilibrium value. The latter is not necessarily zero if some other conserved charge, like $B-L$,  is not zero. In other words, sphaleron transitions push the system towards the state of minimum free energy, which is characterized by a nonvanishing $B+L$ if other conserved charges are non zero.
This is a crucial point to keep in mind when we will talk about electroweak baryogenesis.

\vspace{0.5cm}
\centerline{{\it Exercise 4}}
\vspace{0.5cm}
{\it a)} Consider the two Higgs doublet model in the broken phase. 
The Higgs doublets are defined as $H_1=(H_1^0,H^-)^T$ and $H_2=(H^+,H_2^0)^T$ and couple to the down-type and up-type quarks, respectively.
By considering all the processes in thermal equilibrium (but the ones mediated by  light
quark Yukawa interactions, Cabibbo suppressed gauge interactions and sphalerons transitions), identify the charges which are conserved by the interactions. {\it Hint}: one of them is $B+L$; {\it b)} assuming that also the sphaleron transitions are in equilibrium, compute the relation between
the corresponding equilibrium value of the $B+L$ charge (call it $(B+L)_{{\rm EQ}}$),  and the other conserved charges.  {\it c)} Compute the free energy of the system and show that it scales like $\left[(B+L)-(B+L)_{{\rm EQ}}\right]^2$.

\subsection{Baryon number violation within the SM and GUT baryogenesis}
 
At this point, we are ready to discuss the implications of the baryon number violation in the early Universe for the baryogenesis scenarios discussed so far. The basic lesson we have learned in the previous subsections is that  any asymmetry $B+L$ is rapidly erased by sphaleron transitions as soon as the temperatures drops down $\sim 10^{12}$GeV. Now, we can always write the baryon number $B$ as
\be
B=\frac{B+L}{2}+\frac{B-L}{2}.
\ee
This equation seems trivial, but is dense of physical significance! Sphaleron transitions only erase the combination $B+L$, but leave untouched the orthogonal combination $B-L$. This means that the only chance for a GUT baryogenesis scenario to work is to produce at high scale an asymmetry in $B-L$.  
In section 4 -- however -- we have learned that there is no possibility of generating such an asymmetry in the framework of $SU(5)$. This is because  the fermionic content of the theory is the one of the SM and there is no violation of $B-L$. Sphaleron transitions are therefore the killers of any GUT baryogenesis model based on the supersymmetric version of $SU(5)$ with $R$ parity conserved (The non-supersymmetric version is already ruled out by experiments on the proton decay lifetime). This is a striking result.  

\subsubsection{Baryogenesis via leptogenesis}

The fact that the combination $B-L$ is left unchanged by sphaleron transitions
opens up the possibility of generating the baryon asymmetry from a lepton asymmetry. 
This was suggested by Fukugita and Yanagida \cite{fy}. The basic idea is that, if an asymmetry in the lepton number is produced, sphaleron transition will
reprocess it and convert (a fraction of) it into baryon number. This is because $B+L$ must be vanishing all the times and therefore the final baryon asymmetry results to be 
$B\simeq -L$.
The primordial lepton asymmetry is generated by the out-of-equilibrium
decay of heavy right-handed Majorana neutrinos $N_L^c$. Once the lepton number is produced, the processes in thermal equilibrium distribute the charges in such a way that in the 
 high temperature phase of the standard model the asymmetries of
baryon number $B$ and of $B-L$ are proportional in thermal equilibrium 
\cite{kl} (see Eq. (\ref{mean})
\be
     B=\left(\frac{8 N_F+4N_H}{22 N_F+13N_H}\right)(B-L),
\ee
Where $N_H$ is the
number of Higgs doublets. As we have already stressed, in the standard model, as well as its
unified extension based on the group $SU(5)$, $B-L$ is conserved. Hence,
no asymmetry in $B-L$ can be generated, and $B$ vanishes.
However, a nonvanishing $B-L$ asymmetry may be naturally obtained adding right-handed Majorana neutrinos
to the standard model. This extension of the standard model can be
embedded into GUTs with gauge groups containing
$SO(10)$. Heavy right-handed Majorana neutrinos can also
explain the smallness of the light neutrino masses via the see-saw
mechanism \cite{gelmann}.

The basic piece of the Lagrangian that we need to understand leptogenesis is the coupling between the right-handed neutrino, the Higgs doublet $\Phi$ and the lepton doublet $\ell_L$
\be
{\cal L}=\overline{\ell_L}\:\Phi\:h_\nu\:N_L^c +\frac{1}{2}\overline{N_L^c}\:M\:N_L^c+
{\rm h.c.}
\ee
The vacuum expectation value of the Higgs field $\langle \Phi\rangle$ generates Dirac masses $m_D$ for neutrinos $m_D=h_\nu \langle \Phi\rangle$, which are assumed to be much smaller than the Majorana masses $M$. When the  Majorana right-handed neutrinos decay into leptons and Higgs scalars, they violate the lepton number ( right-handed neutrino fermionic lines do not have any preferred arrow) 
\begin{eqnarray}
N_L^c &\rightarrow & \overline{\Phi}+\ell,\nonumber\\
N_L^c &\rightarrow &\Phi+\overline{\ell}.
\end{eqnarray}
The interference between the tree-level and the one-loop amplitudes, see Fig. 9(a),  yields
a $CP$ asymmetry equal to 
\be
  \frac{1}{8\pi v^2\left(m_D^{\dag}m_D\right)_{ii}}\sum_j
  {\rm Im}\left[\left(m_D^{\dagger}m_D\right)_{ij}^2\right]\:f
  \left({M_j^2\over M_i^2}\right),
\ee
where 
\be
   f(x)=\sqrt{x}\left[1-(1+x)\ln\left(\frac{1+x}{x}
  \right)\right]
\ee
and the index $i$ is summed over all the three species of right-handed neutrino. The final baryon asymmetry has been computed by several authors \cite{luty,luty1,covi,buch} and it has been shown to be of the order of
\be
     B\simeq (0.6-1)\times 10^{-10}.
\ee
\begin{figure}
\centering
\leavevmode\epsfysize=3.6in \epsfbox{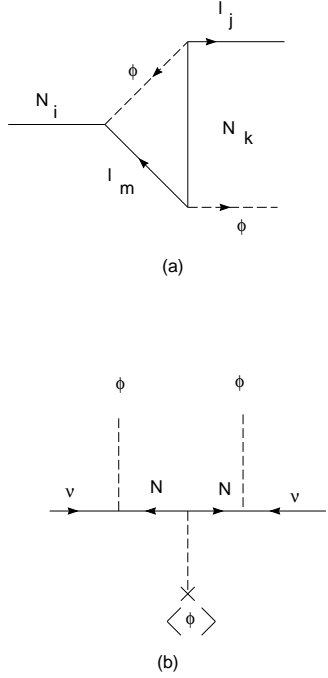}
\caption{(a) One-loop interference giving rise to the lepton asymmetry; (b) Diagram giving rise to the 5-dimensional operator of Eq. (216). }
\end{figure}

However, one has to avoid a large lepton number violation at intermediate temperatures which may potentially dissipate away the baryon number in combination with the sphaleron transitions. Indeed, the diagram of Fig. 9(b), induced by the exchange of a heavy right-handed neutrino,  gives rise to  a $\Delta L=2$ interaction of the form
\be
\label{zzz}
\frac{m_\nu}{\langle \Phi\rangle^2}\ell_L\ell_L\Phi\Phi +{\rm h.c.},
\ee
where $m_\nu$ is the mass of the light left-handed neutrino. The rate of lepton number violation induced by thi interaction is therefore $\Gamma_L\sim (m_\nu^2/\langle \Phi\rangle^4) T^3$. The requirement of harmless letpon number violation, $\Gamma_L\lsim H$ imposes an interesting bound on the neutrino mass
\be
m_{\nu}\lsim 4\:{\rm eV}\left(\frac{T_X}{10^{10}\:{\rm GeV}}\right)^{-1/2},
\ee
where $T_X\equiv {\rm Min}\left\{T_{B-L},10^{12}\:{\rm GeV}\right\}$ and 
$T_{B-L}$ is the temperature at which the $B-L$ number production takes place and $\sim 10^{12}$ GeV is the temperature at which sphaleron transitions enter in equilibrium. One can also reverse the argument and  study leptogenesis assuming a similar pattern of mixings and masses for leptons and quarks, as suggested by $SO(10)$ unification \cite{buch}. This implies that $B-L$ is broken at the unification scale $\sim 10^{16}$ GeV, if $m_{\nu_{\mu}} \sim 3\times  10^{-3}$ eV as preferred by the MSW explanation of the solar neutrino deficit \cite{wolf,ms}.

\section{Electroweak baryogenesis}

So far, we have been assuming that the departure from thermal equilibrium, necessary to generate any baryon asymmetry, is attained by late decays of heavy particles. In this section, we will focus on a different mechanism, namely the departure from equilibrium during first order phase transitions. 

A first order phase transition is defined to occur if some thermodynamic quantities change discontinuously. This happens because there exist two separate thermodynamic states that are in thermal equilibrium at the time of the phase transition. The thermodynamic quantity that undergoes such a discontinuous change is generically called the order parameter $\phi$. Whether a phase transition is of the first order or not depends upon the parameters of the theory and it may happen that, changing those parameters, the order parameter becomes continuous at the time of the transition. In this case, the latter is said to be of the second order
at the point at which the transition becomes continuous and a continuous crossover at the other points for which all physical quantities undergo no changes. In general, we are interested in systems for which the high temperature ground state of the theory is at $\phi=0$ and the low temperature phase is at  $\phi\neq 0$ \cite{book,mariano}. 

For a first order phase transition, the extremum at $\phi=0$ becomes separated from a second local minimum of the potential  by an energy barrier. At the critical temperature $T_c$ both phases are equally favoured energetically and at later times the minimum at $\phi\neq 0$ becomes the global minimum of the theory. 
The phase transition  proceeds by nucleation of bubbles. Initially, the  bubbles are not large enough for their volume energy to overcome the competing surface tension, they shrink and disappear. However, at the nucleation temperature, critical bubbles form, 
{\it i.e.} bubbles which are just large enough to be nucleated and to  grow. As the bubble walls separating the broken from the unbroken phase pass each point in space, the order parameter changes rapidly, leading to a significant departure from thermal equilibrium. 

The critical bubbles of the broken (Higgs) phase have a typical profile
\be
\phi(r)=\frac{\langle\phi(T_c)\rangle}{2}\left[1+{\rm tanh}\left(\frac{r}{L_\omega}\right)\right],
\ee
where $r$ is th spatial coordinate, $L_\omega$ is the bubble wall width and 
$\langle\phi(T_c)\rangle$ is the VEV of the Higgs field inside the bubble. 

\begin{figure}
\centering
\leavevmode\epsfysize=3.6in \epsfbox{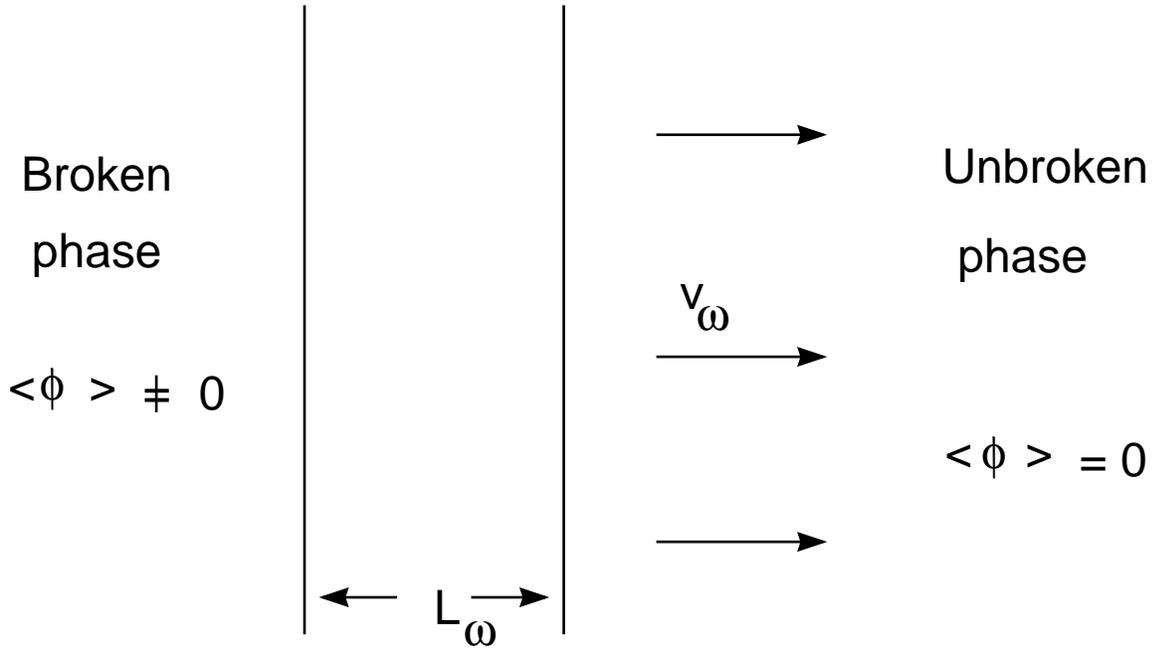}
\caption{Schematic picture of the propagating bubble separating the broken from the unbroken phase during the electroweak phase transition. }
\end{figure}

Bubbles expand with velocity $v_\omega$  until the fill  the Universe; local departure from thermal equilibrium takes place in the vicinity of the expanding bubble walls, see Fig. 10. 

{\it The fundamental idea} of electroweak baryogenesis  is to produce asymmetries in some local charges which are (approximately) conserved by the interactions inside the bubble walls, where local departure from thermal equilibrium is attained. These local charges will then diffuse into the unbroken phase where  baryon number violation is active thanks to the unsuppressed sphaleron transitions.  The latter convert the asymmetries into baryon asymmetry, because the state of minimum free energy is attained for nonvanishing baryon number, see eq. (\ref{master1}). Finally, the baryon number flows into the broken phase where it remains as a remnant of the electroweak phase transition if the sphaleron transitions are suppressed in the broken phase. The recipe for electroweak baryogenesis is therefore the following:  

-- Look for 
 those  charges which are
approximately conserved in the symmetric phase, so that they
can efficiently diffuse in front of the bubble where baryon number
violation is fast, and non-orthogonal to baryon number,
so that the generation of a non-zero baryon charge is energetically
favoured.

-- Compute the $CP$ violating  currents  of the plasma locally induced by the passage of the bubble wall. 

-- Write and solve a set of coupled differential diffusion equations for the local particle densities, including the $CP$ violating  source terms derived from the computation of the current at the previous  step   and the particle number changing reactions. The solution to these equations gives a net baryon number which is produced in the symmetric phase and then transmitted into the interior of the bubbles of the broken phase, where it is not wiped out if the first transition is strong enough.

\subsection{Electoweak baryogenesis in the SM}

Since $C$ and $CP$ are known to be violated by the electroweak interactions, it is possible -- in principle --to satisfy all Sakharov's conditions within the SM if the electroweak phase transition leading to the breaking of $SU(2)_L\otimes U(1)_Y$ is of the first order \cite{kl}. There are  very good reviews on electroweak baryogenesis and the reader is referred to them for more details \cite{review1,review2,review3,review4}.

 The asymmetry   flowing  inside the bubbles of the broken phase  will survive  if sphaleron transitions are frozen out and baryon number violation is inefficient.   As we have learned in the previous section, baryon number violation is out-of-equilibrium inside the bubble wall only of $\frac{\langle \phi(T_c)\rangle}{T_c}\gsim 1$, {\it i.e.} if the electoweak phase transition is strong first order. Let us now understand as this condition translates into a {\it upper} bound on the Higgs mass $m_h$.  

In general, given an order parameter $\phi$ and a set of particles $i$ with masses $m_i(\phi)$ in the $\phi$ background, plasma masses $\pi_i(T)$ and degrees of freedom $n_i$, the effective one-loop improved  potential at finite temperature is given by \cite{mariano}
\be
\label{pott}
\Delta V^{{\rm bos}}(\phi,T)=\sum_i\: n_i\left\{
\frac{m_i^2(\phi)}{24}T^2-\frac{T}{12\pi}\left[m_i^2(\phi)+\Pi_i(T)\right]^{3/2}-\frac{m_i^4(\phi)}{64\pi^2}\:{\rm log}\:\frac{m_i^2(\phi)}{A_B T^2}\right\}
\ee
if the particles are bosons  and
\be
\Delta V^{{\rm fer}}(\phi,T)=\sum_i\: n_i\left\{
\frac{m_i^2(\phi)}{48}T^2+\frac{m_i^4(\phi)}{64\pi^2}\:{\rm log}\:\frac{m_i^2(\phi)}{A_F T^2}\right\}
\ee
if they are fermions. Here $A_B=16\:A_F=16\pi^2\:{\rm exp}(3/2-2\gamma_E)$, $\gamma_E\simeq 0.5722$.

One can therefore  write the total  one-loop  effective potential of the SM Higgs field at finite temperature as as \cite{mariano}
\be
\label{potential}
V(\phi,T)=D(T^2-T_0^2)\phi^2-ET\phi^3+\frac{\lambda(T)}{4}\phi^4,
\ee
where 
\begin{eqnarray}
D&=& \frac{2 M_W^2+M_Z^2+ 2m_t^2}{8 v^2},\nonumber\\
E&=& \frac{2 M_W^3+M_Z^3}{4\pi v^3},\nonumber\\
T_0&=& \frac{m_h^2-8B v^2}{4D},\nonumber\\
B&=& \frac{3}{64\pi^2 v^4}(2 M_W^4+M_Z^4 -4m_t^4),\nonumber\\
\lambda(T)&=&\lambda-\frac{3}{16\pi^2 v^4}\left(2M_W^4\:{\rm Log}\frac{M_W^2}{A_B T^2}+M_Z^4\:{\rm Log}\frac{M_Z^2}{A_B T^2}-
4 m_t^4 \:{\rm Log}\frac{M_t^2}{A_F T^2}\right),
\end{eqnarray}
where $m_t$ is the mass of the top-quark.

It is now easy to see that, when the minimum $\phi=0$ becomes metastable, {\it i.e} at the temperature $T_c$ when $V(0,T_c)=V(\phi(T_c),T_c)$, one has
\be
\frac{\phi(T_c)}{T_c}=\frac{2 E T_c}{\lambda(T_c)}\simeq \frac{4 E v^2}{m_h^2},
\ee
where we have used the fact that $m_h^2=2\lambda v^2$. The condition (\ref{kop}) is therefore satisfied only if
\be
m_h\lsim \sqrt{\frac{4 E}{1.3}}\sim 42\:{\rm GeV}.
\ee
 On the other hand, the current lower bound on $m_h$ comes from combining the results of DELPHI, L3 and OPAL experiments and is $m_h> 89.3$ GeV \cite{bock}.
A simple one-loop computation shows, therefore, that the electroweak phase transition is too weakly first order
to assure the preservation of the generated baryon asymmetry at
the electroweak phase transition in the SM. More complete perturbative  and
non-perturbative analyses \cite{review2} have shown that the electroweak phase transition is first order if the mass of the Higgs $m_h$ is smaller than about 80 GeV and for larger masses becomes a smooth crossover. 
Let us now briefly analyzed the issue of $CP$ violation within the SM. Because of  $CP$ violation in the kaon system, it is of great interest to see whether enough $CP$ violation is present in the SM to generate the baryon asymmetry at the observed level. 

A very rough (and optimistic) estimate of the amount of $CP$ violation necessary to generate $B\simeq 10^{-10}$ can be obtained as follows. Since the baryon number violation rate in the symmetric phase is proportional to $\alpha_W^4\simeq 10^{-6}$, if we indicate by $\delta_{CP}$ the suppression factor due to $CP$ violation, we get
\be
B\simeq \frac{\alpha_W^4 \:T^3}{s}\:\delta_{CP}\simeq 10^{-8}\:\delta_{CP}.
\ee
Even neglecting all the suppression factors coming from the dynamics of the electroweak phase transition, we discover that
\be
\delta_{CP}\gsim 10^{-3}.
\ee
A naive estimate suggests that, since $CP$ violation vanishes in the SM if any two quarks of the same charge have the same mass, the measure of $CP$ violation
 should be the Jarlskog invariant
\begin{eqnarray}
\frac{A_{CP}}{J}&=& \left(M_t^2-M_c^2\right)\left(M_c^2-M_u^2\right)\left(M_u^2-M_t^2\right)
\nonumber\\
& & \left(M_b^2-M_s^2\right) \left(M_s^2-M_d^2\right)\left(M_d^2-M_b^2\right),
\end{eqnarray}
where $J$ is twice the area of the unitarity triangle. The quantity $A_{CP}$ has dimension twelve. In the limit of high temperature,  $T$ much larger than the quark masses $M_q$, the only mass scale in the problem is the temperature itself. Therefore, the dimensionless quantity $\delta_{CP}$ is 
\be
\delta_{CP}\simeq \frac{A_{CP}}{T_c^{12}}\simeq 10^{-20},
\ee
far too small for the SM to explain the observed baryon asymmetry.

This admittedly too naive reasoning has been questioned by Farrar and Shaposhnikov \cite{fr} who have pointed out that for quarks having momentum 
$p\sim T\gg M_q$, the above estimate is certainly correct since light quarks are effectively degenerate in mass and the GIM suppression is operative; on the other side, this is no longer true when quarks have a momentum $p\sim M_q$. Since the mass jump through the bubble wall is just $M_q$, quarks coming from the symmetric phase and with momentum $p<M_q$ are reflected off from the wall, while the ones with momentum $p> M_q$ are partially reflected and partially trasmitted. In the reflection processes quarks and antiquarks acquire different probabilities of penetrating the bubble wall. In such a way, it might be possible to produce a net baryon number flux from outside to  inside  the bubble wall. For instance, considering momenta between $M_d$ and $M_s$, then all the strange quarks might be  reflected off, while down quarks have a nonvanishing probability of being transmitted. However,  this effect is largely suppressed by the fact that fermions, when they propagate in the plasma, acquire a damping rate $\gamma\sim 0.1\:T\gg M_s$ and the quark energy and momenta cannot be defined exactly, but have a spread of the order of $\gamma\gg (M_s-M_d)$. In other words, the lifetime of the quantum packet is much shorter than the typical reflection time from the bubble wall $(\sim 1/M_s)$: $CP$ violation, which is based on coherence and needs   at least a time $\sim 1/M_s$ to be built up, cannot be efficient \cite{gavela,huet}. Therefore, the common wisdom is that electroweak baryogenesis is not possible within the SM.

\subsection{Electoweak baryogenesis in the MSSM}

 The most promising and  
well-motivated framework for electroweak baryogenesis beyond the SM  seems to be supersymmetry (SUSY) \cite{haber,ah}. Let us remind the reader only a few notions about the MSSM that will turn out to be useful in the following.

Let us consider the MSSM superpotential
\be
\label{superpot}
W=\mu\hat{H}_1\hat{H}_2+h^u\hat{H}_2\hat{Q}\hat{u}^c+h^d\hat{H}_1
\hat{Q}\hat{d}^c+h^e\hat{H}_1\hat{L}\hat{e}^c,
\ee
where we have omitted the generation indices. The Higg sector contains
the two Higgs doublets
\be
H_1=\left(
\begin{array}{c}
H_1^0\\
H^-
\end{array}
\right)\:\:\:{\rm and}\:\:\: 
H_2=\left(
\begin{array}{c}
H^+\\
H_1^0
\end{array}
\right).
\ee
The lepton Yukawa matrix $h^e$ can be always taken real and diagonal while $h^u$ and $h^d$ contain the KM phase. 

What is relevant for baryogenesis is to identify possible new sources of $CP$ violation. They emerge from the operators which break softly supersymmetry

{\it i) Trilinear couplings:}

\be
\Gamma^u H_2\widetilde{Q}\widetilde{u}^c+\Gamma^d H_1\widetilde{Q}\widetilde{d}^c+\Gamma^e H_1\widetilde{L}\widetilde{e}^c+{\rm h.c.},
\ee
where we have defined
\be
\Gamma^{(u,d,e)}\equiv m_{3/2}\:A^{(u,d,e)}\cdot h^{(u,d,e)}.
\ee
Generally, in supergravity models the matrices $A^{(u,d,e)}$ are assumed to be proportional to the identiy matrix
\be
A^{(u,d,e)}(M_{{\rm GUT}})=A\cdot {\bf 1},
\ee
where the $A$ parameter can be complex.

{\it ii) bilinear couplings:}

\be
\label{bmu}
\mu B H_1 H_2+{\rm h.c.}
\ee

{\it iii) Majorana gaugino masses:}

\be
\frac{1}{2}\left(M_1\lambda_1\lambda_1+M_2\lambda_2\lambda_2+
M_3\lambda_3\lambda_3\right) +{\rm h.c.}
\ee
At the GUT scale it is usually assumed that
\be
M_1=M_2=M_3=M.
\ee
{\it iv) Scalar soft masses:}

\be
m_{ab}^2\widetilde{z}_a^* \widetilde{z}+{\rm h.c.}
\ee
The new contributions to explicit violation of $CP$ are given in the phases of the complex parameters $A$, $B$, $M_i$ ($i=1,2,3)$ and by the parameter $\mu$ in the superpotential (\ref{superpot}). Two phases may be removed by redifining the phase of the superfield $\hat{H}_2$ in such a  way that the phase of $\mu$ is opposite to that of $B$. The product $\mu B$ in (\ref{bmu}) is therefore real. It is also possible to remove the phase of the gaugino mass $M$ by an $R$ symmetry transformation. The latter leaves all the other supersymmeric couplings invariant and only modifies the trilinear ones, which get multiplied by ${\rm exp}(-\phi_M)$ where $\phi_M$ is the phase of $M$. 

The phases which are left are therefore
\be
\phi_A={\rm arg}(A M)\:\:\:{\rm and}\:\:\:\phi_\mu=-{\rm arg}(B).
\ee
The two new phases $\phi_A$ and $\phi_\mu$ will be crucial for the generation of the baryon asymmetry.

Electroweak  
baryogenesis in the framework of the Minimal Supersymmetric Standard Model  
(MSSM) has  attracted much attention in the past years, with 
particular emphasis on the strength of the phase transition ~\cite{early1,early2,early3,early4} and  
the mechanism of baryon number generation \cite{nelson,noi,higgs,cpt,cpt1,reviewbau,ck}.

Recent  analytical \cite{r1,r2,r3,bod1,losada1,losada2,r4,r5,losada3} and  lattice  
computations  \cite{r6,r7,r8,r9} have  revealed   that the phase transition can be sufficiently strongly  
first order if  the
ratio of the vacuum expectation values of the two neutral Higgses $\tan\beta$  
is smaller than $\sim 4$. Moreover, taking into account all the experimental bounds 
as well as those coming from the requirement of avoiding dangerous
 color breaking minima,  the lightest Higgs boson should be  lighter than about  $105$ GeV,
 while the right-handed stop mass might  be close to the present experimental bound and should
 be smaller than, or of the order of, the top quark mass \cite{r5}. 

Moreover, as we have seen, the MSSM contains additional sources  
of CP-violation  besides the CKM matrix phase. 
These new phases are essential  for the generation of the baryon number since  large  
$CP$ violating  sources may be  locally induced by the passage of the bubble wall separating the broken from the unbroken phase during the electroweak phase transition.   Baryogenesis is fuelled  when transport properties allow the $CP$ violating   
charges to efficiently diffuse in front of the advancing bubble wall where
anomalous electroweak baryon violating processes are not suppressed.
The new phases   
appear    in the soft supersymmetry breaking  
parameters associated to the stop mixing angle and to  the gaugino and  
neutralino mass matrices; large values of the stop mixing angle
are, however, strongly restricted in order to preserve a
sufficiently strong first order electroweak phase transition. 
Therefore, an acceptable baryon asymmetry from the stop sector
may only be generated through a delicate balance between the values
of the different soft supersymmetry breaking parameters contributing
to the stop mixing parameter, and their associated $CP$ violating  
phases \cite{noi}. As a result, the contribution to the final baryon asymmetry from the stop sector turns out to be negligible.   On the other hand, charginos and neutralinos may be responsible for the observed baryon asymmetry if   the phase of the parameter $\mu$ is    large enough \cite{noi,ck}. Yet, 
this is true within the MSSM. If the strength of the
 electroweak phase transition is enhanced by the presence of some
new degrees of freedom beyond the ones contained in the MSSM, {\it
e.g.} some extra standard model gauge singlets,
 light stops (predominantly the 
right-handed ones) and charginos/neutralinos are expected to 
give quantitatively the
same contribution to the final baryon asymmetry.  

\subsubsection{The electroweak phase transition in the MSSM} 

As discussed above, a strongly first order electroweak
phase transition
can  be achieved in the presence of a top squark
lighter than the top quark~\cite{r5}.
In order to naturally
suppress its contribution to the parameter $\Delta\rho$ and hence
preserve a good agreement with the precision measurements at LEP,
it should be mainly right handed. This can be achieved if the left
handed stop soft supersymmetry breaking mass $m_Q$
is much larger than $M_Z$.

The stop mass matrix is given by
\be
{\cal M}_{\widetilde{t}}=\left(
\begin{array}{cc}
M_{LL}^2 & M_{LR}^2\\
M_{LR}^{*2} & M_{RR}^2
\end{array}\right),
\ee
where
\begin{eqnarray}
M_{LL}^2 &\simeq & m_Q^2+h_t^2 \left|H_2^0\right|^2,\nonumber\\
M_{RR}^2 &\simeq & m_U^2+h_t^2 \left|H_2^0\right|^2,\nonumber\\
M_{LR}^2&=& h_t\left(A_t H_2^0-\mu^* H_1^0\right).
\end{eqnarray}
For moderate mixing,
the lightest stop mass is then approximately given by
\be
\label{app}
\mstop^2  \simeq m_U^2  + m_t^2(\phi) \left( 1  -
\frac{\left|\widetilde{A}_t\right|^2}{m_Q^2}
\right)
\ee
where $\widetilde{A}_t = A_t - \mu^{*}/\tan\beta$ is the
particular combination appearing in the off-diagonal terms of
the left-right stop squared mass matrix and $m_U^2$ is
the soft supersymmetry breaking squared mass parameter
of the right handed stop.  Notice that the Higgs sector contains two neutral $CP$ even states, $H_1^0$ and $H_2^0$. However, in the limit in which $m_A \gg T_c$, where $m_A$ is the mass of the 
pseudoscalar particle of the Higg sector, only one neutral Higgs survives
\be
\phi=\cos\beta H_1^0+\sin\beta H_2^0,
\ee
where $\tan\beta=\langle H_2^0\rangle/\langle H_1^0\rangle$, and the low-energy potential reduces to the one-dimensional SM-like potential $V(\phi)$.
 
The preservation of the baryon number asymmetry requires  the
order parameter $\langle \phi(T_c)\rangle /T_c$ to be larger than one.
The latter is bounded from above
\be
\frac{\langle \phi(T_c)\rangle}{T_c} < \left(\frac{\langle \phi(T_c)\rangle}{T_c}\right)_{\rm SM}
+ \frac{2 \; m_t^3  \left(1 -
\widetilde{A}_t^2/m_Q^2\right)^{3/2}}{ \pi \; v \; m_h^2}\ ,
\label{totalE}
\ee
where $m_t = \overline{m}_t(m_t)$ is the on-shell running top
quark
mass in the $\overline{{\rm MS}}$ scheme.
The first term on the right hand side of
expression ~(\ref{totalE}) is the Standard Model contribution
\be
\left(\frac{\langle \phi(T_c)\rangle}{T_c}\right)_{\rm SM}
\simeq \left(\frac{40}{m_h[{\rm GeV}]}\right)^2,
\ee
and the second term is
the contribution that would be obtained if the right handed
stop plasma mass vanished at the critical
temperature (see Eq.~(\ref{plasm})). Remember that in the expression for the one-loop effective potential (\ref{potential}), the parameter $E$
gets contributions from boson fields. So, the difference between the SM and the MSSM is that light stops may give a large contributions to the effective potential in the MSSM.

In order to overcome the Standard Model
constraints, the stop contribution must be therefore large.
The stop contribution strongly depends
on the value of $m_U^2$, which must be small in magnitude, and
negative, in order to induce a sufficiently strong first order phase
transition. Indeed, large stop contributions
are always associated with small values of the right handed stop
plasma mass
\begin{equation}
m^{\rm eff}_{\;\widetilde{t}} = -\widetilde{m}_U^2 + \Pi_R(T),
\label{plasm}
\end{equation}
where $\widetilde{m}_U^2 = - m_U^2$, $\Pi_R(T) \simeq 4 g_3^2
T^2/9+h_t^2/6[2-
\widetilde{A}_t^2/m_Q^2]T^2$  is the finite
temperature
self-energy contribution to the right-handed
squarks. Moreover, the
trilinear mass term, $\widetilde{A}_t$,
must be $\widetilde{A}_t^2 \ll m_Q^2$
in order to avoid the suppression of  the stop contribution
to $\langle \phi(T_c)\rangle/T_c$.

Although large values of $\widetilde{m}_U$, of order of the critical
temperature,
are useful to get a strongly first order phase transition, they may
also induce charge and color breaking minima. Indeed, if the
effective plasma mass at the critical temperature vanished, the
universe would be driven to a charge and color breaking minimum at
$T \geq T_c$ . Hence, the upper bound on $\langle \phi(T_c)\rangle/T_c$,
Eq.~(\ref{totalE}) cannot be reached in realistic scenarios. A
conservative
bound on $\widetilde{m}_U$ may be obtained by demanding that
the electroweak symmetry breaking minimum should be lower
than any
color-breaking minima induced by the presence of
$\widetilde{m}_U$ at
zero temperature, which yields the condition
\begin{equation}
\widetilde{m}_U \leq \left(\frac{m_H^2 v^2 g_3^2}{12}\right)^{1/4}.
\label{colorbound}
\end{equation}
It can be shown that this condition is sufficient to prevent dangerous
color breaking minima at zero and finite temperature for any value of
the mixing parameter $\widetilde{A}_t$. A more general analysis is provided in \cite{r5}. 

In order to obtain values of $\langle \phi(T_c)\rangle/T_c$ larger than one,
the Higgs mass must take  small values, close to the present
experimental bound.  Numerically,
an upper bound, of order 80 GeV, can be derived.
For small mixing, the one-loop Higgs mass has a very simple form
\begin{equation}
m_h^2 = M_Z^2 \cos^2 2\beta + \frac{3}{4\pi^2}
\frac{\overline{m}_t^4}{v^2}
\log\left(\frac{m_{\widetilde{t}}^2 m_{\widetilde{T}}^2}
{\overline{m}_t^4}\right)\left[1
+ {\cal{O}}\left(\frac{\widetilde{A}_t^2}{m_Q^2}\right)
\right],
\end{equation}
where $m_{\widetilde{T}}^2 \simeq m_Q^2 + m_t^2$, is the heaviest
stop
squared mass. Hence, $\tan\beta$ must take values close to one. The
larger the left handed stop mass, the closer to one $\tan\beta$
must be. This implies that
the left handed stop effects decouple at the critical temperature
and hence,
different values of $m_Q$ mainly affect the baryon asymmetry through
the resulting Higgs mass.

Values of the $CP$-odd Higgs mass $m_A \lsim 200$ GeV
are associated with a weaker first order phase transition. Fig.~11
shows the behaviour of the order parameter $\langle \phi(T_c)\rangle/T_c$ in the
$m_A$-$\tan\beta$ plane, for $\widetilde{A}_t = 0$, $m_Q = 500$
GeV and values of $\widetilde{m}_U$ close to its upper bound,
Eq.~(\ref{colorbound}).

\begin{figure}
\centering
\leavevmode\epsfysize=3.5in \epsfbox{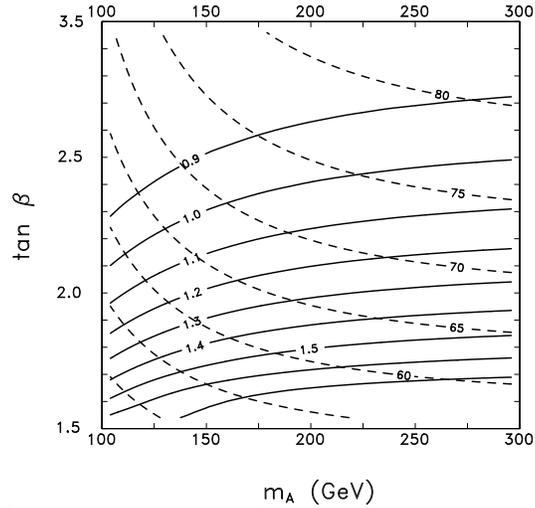}
\caption{Contour plots of constant values of $\langle \phi(T_c)\rangle/T_c$ (solid
lines) and $m_h$ in GeV (dashed lines) in the plane
$(m_A,\tan\beta)$. We have fixed $m_t=175$ GeV and the values
of sypersymmetric parameters: $m_Q=500$ GeV, $m_U=m_U^{\rm crit}$
fixed by the charge and color breaking constraint, and
$A_t=\mu/\tan\beta$. }

\end{figure}

In order to correctly interpret the results
of Fig.~11 one should remember that the Higgs mass bounds are somewhat
weaker for values of $m_A \lsim 150$ GeV. However, even for values
of $m_A$ of order 80 GeV, in the low $\tan\beta$ regime the lower
bound on the Higgs mass is of order 60 GeV. Hence, it follows
from Fig.~11 that, to obtain a sufficiently strong first order
phase transition the CP-odd Higgs mass
$m_A \gsim 150$ GeV. When two-loop QCD corrections \cite{r3,r4} associated with stop loops  are included, one finds that $m_A \gsim 120$  GeV   $m_h \lsim 85$ GeV. This region will be explored at LEP2 very soon.

\vspace{0.5cm}
\centerline{{\it Exercise 5}}
\vspace{0.5cm}
Obtain Eq. (\ref{totalE}).

\subsubsection{How to produce the baryon asymmetry in the MSSM} 

As we have previously learned, the first step in the computation of the baryon number asymmetry is to identify  
 those  charges which are
approximately conserved in the symmetric phase, so that they
can efficiently diffuse in front of the bubble where baryon number
violation is fast, and non-orthogonal to baryon number,
so that the generation of a non-zero baryon charge is energetically
favoured according to the Master equation (\ref{master1}). 

Charges with these characteristics in the MSSM 
are the axial stop charge and the Higgsino charge,
which may be produced from the interactions of squarks and
charginos
and/or neutralinos with the bubble wall,
provided a source of $CP$-violation is present in these sectors. This  is exactly the case, since both the parameters $A_t$ and $\mu$ may carry a physical phase \cite{ah}. The idea is that, if nonvanishing  $CP$ violating sources for the right-handed stop and higgsino numbers are induced in the bubble wall, the scattering among particles as well as diffusion will generate an asymmetry in the left-handed fermion asymmetries in the unbroken phase. The asymmetry -- in turn -- will fuel baryogenesis because sphaleron transitions will push the system towards the state of minimum free energy, which is the one with nonvanishing baryon asymmetry. In the next subsection, we will give some indications of how to compute the $CP$-violating sources. Let us now investigate the dynamics of electroweak baryogenesis a little bit further.

One has to start with a set of coupled differential
equations describing the effects of diffusion, particle number
changing reactions and $CP$-violating source terms. Major simplifications of the diffusion equations take place when neglecting all the couplings except for gauge interactions and the top Yukawa coupling. Neglecting the weak sphalerons (in the first step) allows to forget about leptons in the diffusion equations and will turn out to be a good approximation when computing Higgs and quark densities.

If the system is near thermal equilibrium and particles interact weakly,
the particle number densities $n_i$ may be expressed as (see Eq. (\ref{chem})) 
$n_i=k_i\mu_iT^2/6$ where
$\mu_i$ is the local chemical potential, and $k_i$ are statistical
factors of the order of 2 (1) for light bosons (fermions) in thermal
equilibrium, and Boltzmann suppressed for particles heavier than $T$.

What really determines which are the  interactions in equilibrium is the typical time scale for the passage of the bubble wall through a given point, $\tau_\omega\sim L_\omega/v_\omega$. If interactions are faster than $\tau_\omega$ they are in equilibrium, otherwise, they are not. 

The particle densities we
need to include are 

-- the left-handed top
doublet $q\equiv(t_L+b_L)$,

-- the right-handed top quark $t\equiv t_R$,

--  the Higgs particle
$h\equiv(H_1^0, H_2^0, H_1^-, H_2^+)$, and the superpartners
$\widetilde{q}$, $\widetilde{t}$ and $\widetilde{h}$.

The interactions able to change the particle numbers are

-- the top
Yukawa interaction with rate
$\Gamma_t$, 

-- the top quark mass interaction with rate $\Gamma_m$,

-- the Higgs self-interactions in the broken phase
with rate $\Gamma_h$,

--  the strong sphaleron interactions
with rate $\Gamma_{{\rm ss}}$. The axial vector current of QCD $\sum_i\overline{q}^i\gamma^\mu\gamma_5 q^i=-\sum_i\overline{q}_L^i\gamma^\mu q^i_L+\sum_i\overline{q}_R^i\gamma^\mu q^i_R$ where the sum is over the quarks, 
has a triangle anomaly and therefore one may expect axial charge violation due to topological
transitions analogous to the case of sphaleron transitions
\be
\frac{dQ_5}{dt}=-\frac{12\cdot 6}{T^3}\:\Gamma_{{\rm ss}}\:Q_5,
\ee
where $Q_5$ is the axial charge, the factor 12 comes from the total number of quark chirality states and the factor 6 from the relation between the asymmetry in the quark number density and the chemical potential, $n_i\sim \mu_i T^2/6$, see Eq. (\ref{chem}). The rate of these processes at high temperature is expected to be \cite{larry}
\be
\Gamma_{{\rm ss}}=\frac{8}{3}\left(\frac{\alpha_S}{\alpha_W}\right)\:\Gamma_{{\rm sp}},
\ee
where $\alpha_S$ is the strong fine structure leading to the characteristic time of order of
\be
\tau_{{\rm ss}}\simeq \frac{1}{192\:\kappa\:\alpha_S\: T}\lsim \tau_\omega.
\ee
The effect of QCD sphalerons may be represented by the operator
\be
\Pi_{i=1}^3\:(u_L\:u_R^\dagger\:d_L\:d_R^\dagger)_i,
\ee
where $i$ is the generation index. Assuming that these processes are in equilibrium \cite{mz,gs}, we get the following equation for the chemical potentials
\be
\sum_{i=1}^3(\mu_{u^i_L}-\mu_{u^i_R}+\mu_{d^i_L}-\mu_{d^i_R})=0.
\ee
This equation contains the chemical potential for all the quarks and imposes that the total right-handed baryon number is equal to the total left-handed one. In other words, including strong QCD sphalerons allow the generation of the right-handed bottom quark as well as the generation of the first and second family quarks,

-- the weak anomalous interactions with rate $\Gamma_{\rm sp}$,

--  the gauge interactions.

We shall assume that the supergauge interactions are in
equilibrium, that is
\be
\frac{q}{k_q}=\frac{\widetilde{q}}{k_{\widetilde{q}}}=\frac{t}{k_t}=
\frac{\widetilde{t}}{k_{\widetilde{t}}}=\frac{h}{k_h}=\frac{\widetilde{h}}{k_{\widetilde{h}}}.
\ee

Under these assumptions the system may be described by the
densities
$ Q = q + \widetilde{q}$,
$ T=t+\widetilde{t}$ and $ H=h+\widetilde{h}$.
$CP$-violating interactions with the advancing bubble wall produce
source terms $\gamma_{\widetilde{H}}$ for Higgsinos and
$\gamma_R$
for right-handed stops, which tend to push the system out of
equilibrium.
Ignoring the curvature of the bubble wall, any quantity becomes a
function of the coordinate
${\bf z}=z_3+v_{\omega}z$, the coordinate normal
to the wall surface, where we assume the bubble wall is moving along the
$z_3$-axis.

When including the strong sphalerons, right-handed bottom quarks are generated as well as the quarks of the first two families. However, since strong sphalerons are the only processes which produce the first two generation quarks and all quarks have nearly the same diffusion constants, we may constrain the densities algebrically
 in terms of $B\equiv b_R+\widetilde{b}_R$
\be
Q_{1L}=Q_{2L}=-2U_R=-2 D_R=-2 S_R=-2 C_R =-2B=2(Q+T),
\ee
where the last equality comes from imposing that strong sphalerons are in equilibrium. 

Particle transport is treated by including a diffusion term. Taking all the quarks and squarks with the same diffusion constant $D_q$ and Higgs and Higgsinos with diffusion constant $D_h$, onecan write the folowing set of diffusion equations
\begin{eqnarray}
\dot{Q}&=& D_q\nabla^2 Q-\Gamma_t[Q/k_Q-H/k_H-T/k_T]-\Gamma_m[Q/k_Q-T/k_T]\nonumber\\
&-& 6\Gamma_{{\rm ss}}[2 Q/k_Q-T/k_T+9(Q+T)/k_B]+ \gamma_{\widetilde{t}},
\nonumber\\
\dot{T}&=& D_q\nabla^2 T-\Gamma_t[-Q/k_Q+H/k_H+T/k_T]-\Gamma_m[-Q/k_Q+T/k_T]\nonumber\\
&+& 3\Gamma_{{\rm ss}}[2 Q/k_Q-T/k_T+9(Q+T)/k_B]+ \gamma_{\widetilde{t}},
\nonumber\\
\dot{H}&=& D_h\nabla^2 h-\Gamma_t[-Q/k_Q+H/k_H+T/k_T]-\Gamma_h H/k_H + \gamma_{\widetilde{h}},
\end{eqnarray}
where we have inserted the $CP$ violating sources.

Assuming that the rates $\Gamma_t$ and $\Gamma_{{\rm ss}}$ are
fast so that $Q/k_q-
H/k_H- T/k_T={\cal O}(1/\Gamma_t)$ and
$2Q/k_q- T/k_T+
9( Q+ T)/k_b={\cal O}(1/\Gamma_{{\rm ss}})$,
one can find the equation governing the Higgs density
\begin{equation}
\label{equation}
v_{\omega}H^\prime-\overline{D}
H^{\prime\prime}+\overline{\Gamma}H-
\widetilde{\gamma}=0,
\end{equation}
where the derivatives are now with respect to ${\bf z}$,
$\overline{D}$ is the effective diffusion constant,
$\widetilde{\gamma}$ is an effective source term
in the frame of the
bubble wall and $\overline{\Gamma}$ is the effective decay
constant~\cite{nelson}.
An analytical solution to Eq.~(\ref{equation}) satisfying the
boundary conditions $ H(\pm\infty)=0$ may be found in the
symmetric
phase (defined by ${\bf z}<0$) using a ${\bf z}$-independent
effective diffusion constant  and a step function for the effective decay rate
$\overline{\Gamma}= \widetilde{\Gamma} \theta({\bf z})$. A more realistic
form of $\overline{\Gamma}$ would interpolate smoothly between the
symmetric and the broken phase values. The values of $\overline{D}$ and $\overline{\Gamma}$ in
(\ref{equation}) of course depend on the particular
values of supersymmetric parameters. For the
considered range one typically finds $\overline{D}\sim 0.8\ {\rm GeV}^{-1}$,
$\overline{\Gamma}\sim 1.7$ GeV.

The tunneling processes from the symmetric phase to the true minimum in the first order phase transition of the Higgs field in the MSSM has been recently analyzed in \cite{moreno} including the leading two-loop effects. It was shown that  the Higgs profile along the bubbles at the time when the latter  are formed has a typical thickness $L_\omega\sim (20-30)/T$. In general, however,  the value of $L_\omega$ when the bubbles are moving through the plasma with some velocity $v_\omega$  is different from the value at bubble nucleation. Indeed, the motion of the bubble wall is determined by two main factors, namely the pressure difference between inside and outside the bubble --leading to the expansion-- and the friction force, proportional to $v_\omega$, accounting for the collisions of the plasma particles off the wall. The equilibrium between these two forces imples a steady state with a final velocity $v_\omega$. If bubbles are rather thick, thermodinamical conditions are established inside the wall and for the latter is no longer possible to loose energy by thermal dissipation. Under these conditions the bubble wall is accelerated until slightly out-of-equilibrium conditions and the friction forces are reestablished. As we shall see, the total amount of the baryon asymmetry is proportional to $\Delta\beta$ --the change in the ratio of the Higgs vacuum expectation values
$\beta=\langle H_2^0\rangle /\langle H_1^0\rangle$ from ${\bf z}=0$ to inside the bubble wall. 
This quantity tends to zero for large values of $m_A$,
and takes small values, of order $10^{-2}$ for values of the pseudoscalar mass 
$m_A = 150$--200 GeV \cite{moreno}. 

The solution of Eq. (\ref{equation})
for ${\bf z} < 0$ is
\begin{equation}
\label{h1}
 H({\bf z})={\cal A}\:{\rm e}^{{\bf z}v_{\omega}/\overline{D}},
\end{equation}
and for ${\bf z} >0$ is 
\begin{eqnarray}
\label{h3}
 H({\bf z}) & = & \left( {\cal B}_{+} -
\frac{1}{\overline{D}(\lambda_+  - \lambda_-)}
\int_0^{{\bf z}} du \widetilde \gamma(u) e^{-\lambda_+ u} \right)
e^{\lambda_{+} {\bf z}}
\nonumber\\
&+& \left( {\cal B}_{-} -
\frac{1}{\overline{D}(\lambda_-  - \lambda_+)}
\int_0^{\bf{z}} du \widetilde \gamma(u) e^{-\lambda_- u} \right)
e^{\lambda_{-} {\bf z}}.
\end{eqnarray}
where
\begin{equation}
\lambda_{\pm} = \frac{ v_{\omega} \pm
\sqrt{v_{\omega}^2 + 4 \widetilde{\Gamma}
\overline{D}}}{2 \overline{D}}.
\end{equation}
Imposing the continuity of $H$ and
$H'$ at the boundaries, we find
\begin{equation}
\label{h2}
{\cal A}= {\cal B}_{+}\left(1-\frac{\lambda_-}{\lambda_+}\right)=
{\cal B}_{-}\left(\frac{\lambda_+}{\lambda_-}-1\right)=
\frac{1}{\overline{D} \; \lambda_{+}} \int_0^{\infty} du\;
\widetilde \gamma(u)
e^{-\lambda_+ u}.
\end{equation}
From the form of the above equations one can see that $CP$ violating
densities are non zero for a time $t\sim \overline{D}/ v_{\omega}^2$
and the assumptions leading to the analytical
form of $ H({\bf z})$ are valid
provided that the interaction rates $\Gamma_{t}$ and 
$\Gamma_{{\rm ss}}$  are larger than $v_{\omega}^2/\overline{D}$ \cite{nelson,noi}.

The equation governing
the baryon asymmetry $n_B$ is given by~\cite{nelson}
\begin{equation}
\label{h4}
D_q n_B^{\prime\prime}-v_{\omega} n_B^\prime-
\theta(-{\bf z})N_f\Gamma_{{\rm sp}}n_L=0,
\end{equation}
where 
$n_L$ is the total number density of
left-handed weak doublet fermions and
we have assumed that the baryon asymmetry gets
produced only in the symmetric phase.
Expressing $n_L({\bf z})$ in terms of the Higgs number density
\begin{equation}
n_L=\frac{9k_q k_T-8k_b k_T
-5 k_b k_q}{k_H(k_b+9 k_q+9 k_T)}\:H
\end{equation}
and making use of Eqs.~(\ref{h1})-(\ref{h4}), we find that
\begin{equation}
\label{final}
\frac{n_B}{s}=-g(k_i)\frac{{\cal A}\overline{D}\Gamma_{{\rm sp}}}
{v_{\omega}^2 s},
\end{equation}
where  
$g(k_i)$
is a numerical coefficient depending upon the light degrees of
freedom present in the thermal bath.

Eq. (\ref{final}) summarizes  all the ingredients we need to produce  a baryon asymmetry in electroweak baryogenesis: {\it 1)} (the integral of ) a $CP$ violating source  ${\cal A}$, {\it 2)} baryon number violation provided by the sphaleron transitions with rate $\Gamma_{{\rm sp}}$ and {\it 3)}  out-of-equilibrium conditions 
provided by the expanding bubble wall. 

\subsubsection{Out-of-equilibrium field theory with a broad brush}

The next step in the computation of the baryon asymmetry is the evaluation of the $CP$ violating sources for the right-handed stop number and the higgsino number.

Non-equilibrium Quantum Field Theory  provides us with   the necessary    tools  to  write down a set of quantum Boltzmann equations (QBE's) describing the local particle densities and automatically incorporating the $CP$ violating sources. The most appropriate extension of the field theory
to deal with these issues is to generalize the time contour of
integration to a closed time-path (CTP).  The CTP formalism is a powerful Green's function
formulation for describing non-equilibrium phenomena in field theory, it leads to 
 a complete
non-equilibrium quantum kinetic theory approach and   to   a   rigorous   computation of  the $CP$ violating  sources for the stop and the Higgsino numbers \cite{cpt,cpt1,reviewbau}. 
What is more relevant, though, is that 
the $CP$ violating sources-- and more generally the particle number changing interactions--  built up from the CTP formalism are characterized by  ``memory'' effects which are typical of  the  quantum transport theory \cite{dan,henning}.
$CP$ violating sources are built up when right-handed stops and Higgsinos scatter off the advancing Higgs bubble wall and CP is violated at the vertices of interactions.
In  the classical kinetic theory the ``scattering term'' does not include any integral over the past history of the system. This   is equivalent to assuming that any collision in the plasma  does not depend upon the previous ones.   On the contrary, the quantum approach reveals that the $CP$ violating source is manifestly non-Markovian.

We will   now briefly present    some of the  basic features of the  non-equilibrium quantum field theory based on the Schwinger-Keldysh formulation \cite{sk1,sk2}. The interested reader is referred to the excellent review by Chou   {\it et} al. \cite{chou} for a more comprehensive discussion.

Since  we
need the temporal evolution of the particle asymmetries with definite initial conditions and not
simply the transition amplitude of particle reactions, 
the ordinary equilibrium quantum field theory at finite temperature   is not the appropriate tool. 
The most appropriate extension of the field theory
to deal with nonequilibrium phenomena amounts to generalize the time contour of
integration to a closed-time path. More precisely, the time integration
contour is deformed to run from $-\infty$ to $+\infty$ and back to
$-\infty$, see Fig. 12. 

\begin{figure}
\centering
\leavevmode\epsfysize=3.5in \epsfbox{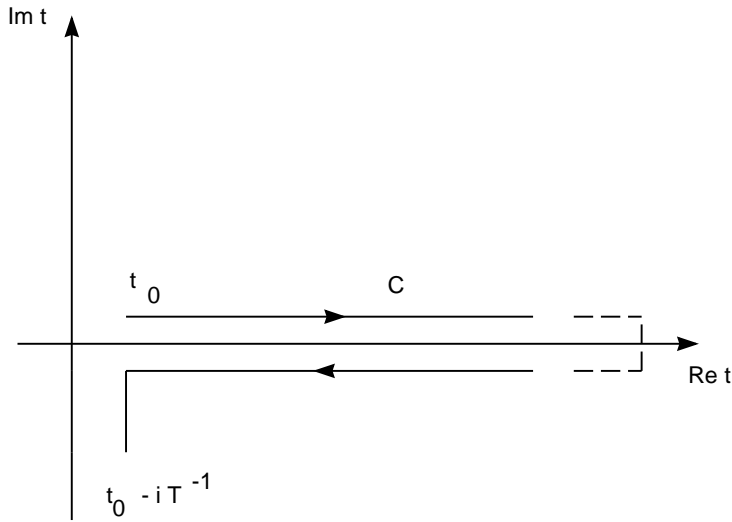}
\caption{The contour C in the $({\rm Re}\:t, {\rm Im}\: t)$ plane proper to the CTP formalism }

\end{figure}

The CTP formalism (often  dubbed as in-in formalism) is a powerful Green's function
formulation for describing non-equilibrium phenomena in field theory.  It
allows to describe phase-transition phenomena and to obtain a
self-consistent set of quantum Boltzmann equations.
The formalism yields various quantum averages of
operators evaluated in the in-state without specifying the out-state. On the contrary, the ordinary quantum field theory (often dubbed as in-out formalism) yields quantum averages of the operators evaluated  with an in-state at one end and an out-state at the other. 

Because of the time contour deformation, the partition function in the in-in formalism for a {\it complex} scalar field is defined to be
\begin{eqnarray}
Z\left[ J, J^{\dagger}\right] &=& {\rm Tr}\:\left[ T\left( {\rm exp}\left[i\:\int_{\rm C}\:\left(J\phi+J^{\dagger}\phi^{\dagger} \right)\right]\right)\rho\right]\nonumber\\
&=& {\rm Tr}\:\left[ T_{+}\left( {\rm exp}\left[ i\:\int\:\left(J_{+}\phi_{+}+J^{\dagger}_{+}\phi^{\dagger}_{+} \right)\right]\right)\right.
\nonumber\\
&\times&\left.  T_{-}\left( {\rm exp}\left[ -i\:\int\:\left(J_{-}\phi_{-}+J^{\dagger}_{-}\phi^{\dagger}_{-} \right)\right]\right) \rho\right],
\end{eqnarray}
where the suffic C in the integral denotes that the time integration contour runs from minus infinity to plus infinity and then back to minus infinity again. The symbol $\rho$ represents the initial density matrix and the fields are in the Heisenberg picture  and  defined on this closed time contour. 
As with the Euclidean time formulation, scalar (fermionic) fields $\phi$ are
still periodic (anti-periodic) in time, but with
$\phi(t,\vec{x})=\phi(t-i\beta,\vec{x})$, $\beta=1/T$.
The temperature appears   due to boundary
condition, but time is now  explicitly present in the integration
contour.

We must now identify field
variables with arguments on the positive or negative directional
branches of the time path. This doubling of field variables leads to
six  different real-time propagators on the contour \cite{chou}.  These six
propagators are not independent, but using all of them simplifies the notation. 
For a generic bosonic charged  scalar field $\phi$ they are defined as 
\begin{eqnarray}
\label{def1}
G_{\phi}^{>}\left(x, y\right)&=&-i\langle
\phi(x)\phi^\dagger (y)\rangle,\nonumber\\
G_{\phi}^{<}\left(x,y\right)&=&-i\langle
\phi^\dagger (y)\phi(x)\rangle,\nonumber\\
G^t _{\phi}(x,y)&=& \theta(x,y) G_{\phi}^{>}(x,y)+\theta(y,x) G_{\phi}^{<}(x,y),\nonumber\\
G^{\bar{t}}_{\phi} (x,y)&=& \theta(y,x) G_{\phi}^{>}(x,y)+\theta(x,y) G_{\phi}^{<}(x,y), \nonumber\\
G_{\phi}^r(x,y)&=&G_{\phi}^t-G_{\phi}^{<}=G_{\phi}^{>}-G^{\bar{t}}_{\phi}, \:\:\:\: G_{\phi}^a(x,y)=G^t_{\phi}-G^{>}_{\phi}=G_{\phi}^{<}-G^{\bar{t}}_{\phi},
\end{eqnarray}
where the last two Green functions are the retarded and advanced Green functions respectively and $\theta(x,y)=\theta(t_x-t_y)$ is the step function.  For a generic fermion field $\psi$ the six different propagators are analogously defined as
\begin{eqnarray}
\label{def2}
G^{>}_{\psi}\left(x, y\right)&=&-i\langle
\psi(x)\bar{\psi} (y)\rangle,\nonumber\\
G^{<}_{\psi}\left(x,y\right)&=&+i\langle
\bar{\psi}(y)\psi(x)\rangle,\nonumber\\
G^{t}_{\psi} (x,y)&=& \theta(x,y) G^{>}_{\psi}(x,y)+\theta(y,x) G^{<}_{\psi}(x,y),\nonumber\\
G^{\bar{t}}_{\psi} (x,y)&=& \theta(y,x) G^{>}_{\psi}(x,y)+\theta(x,y) G^{<}_{\psi}(x,y),\nonumber\\
G^r_{\psi}(x,y)&=&G^{t}_{\psi}-G^{<}_{\psi}=G^{>}_{\psi}-G^{\bar{t}}_{\psi}, \:\:\:\: G^a_{\psi}(x,y)=G^{t}_{\psi}-G^{>}_{\psi}=G^{<}_{\psi}-G^{\bar{t}}_{\psi}.
\end{eqnarray}
For equilibrium phenomena, the brackets $\langle \cdots\rangle$ imply a thermodynamic average over all the possible states of the system. While for homogeneous systems in equilibrium, the Green functions
depend only upon the difference of their arguments $(x,y)=(x-y)$ and there is no dependence upon $(x+y)$,  for systems out of equilibrium, the definitions (\ref{def1}) and (\ref{def2}) have a different meaning. The concept of thermodynamic averaging  is now ill-defined. Instead, the bracket means the need to average over all the available states of the system for the non-equilibrium distributions. Furthermore, the arguments of the Green functions $(x,y)$ are {\it not} usually given as the difference $(x-y)$. For example, non-equilibrium could be caused by transients which make the Green functions
depend upon $(t_x,t_y)$ rather than $(t_x-t_y)$. 

For interacting systems whether in equilibrium or not, one must define and calculate self-energy functions. Again, there are six of them: $\Sigma^{t}$, $\Sigma^{\bar{t}}$, $\Sigma^{<}$, $\Sigma^{>}$, 
$\Sigma^r$ and $\Sigma^a$. The same relationships exist among them as for the Green functions in  (\ref{def1}) and (\ref{def2}), such as
\begin{equation}
\Sigma^r=\Sigma^{t}-\Sigma^{<}=\Sigma^{>}-\Sigma^{\bar{t}}, \:\:\:\:\Sigma^a=\Sigma^{t}-\Sigma^{>}=\Sigma^{<}-\Sigma^{\bar{t}}. 
\end{equation}
The self-energies are incorporated into the Green functions through the use of  Dyson's equations. A useful notation may be introduced which expresses four of the six Green functions as the elements of two-by-two matrices \cite{craig}

\begin{equation}
\widetilde{G}=\left(
\begin{array}{cc}
G^{t} & \pm G^{<}\\
G^{>} & - G^{\bar{t}}
\end{array}\right), \:\:\:\:
\widetilde{\Sigma}=\left(
\begin{array}{cc}
\Sigma^{t} & \pm \Sigma^{<}\\
\Sigma^{>} & - \Sigma^{\bar{t}}
\end{array}\right),
\end{equation}
where the upper signs refer to bosonic case and the lower signs to fermionic case. For systems either in equilibrium or non-equilibrium, Dyson's equation is most easily expressed by using the matrix notation
\begin{equation}
\label{d1}
\widetilde{G}(x,y)=\widetilde{G}^0(x,y)+\int\: d^4x_3\:\int d^4x_4\: \widetilde{G}^0(x,x_3)
\widetilde{\Sigma}(x_3,x_4)\widetilde{G}(x_4,y),
\end{equation}
where the superscript ``0'' on the Green functions means to use those for {\it noninteracting} system.   This equation appears quite formidable; however, some simple expressions may  be obtained for the respective Green functions. It is useful to notice that Dyson's equation can be written in an alternate form, instead of  (\ref{d1}), with $\widetilde{G}^0$ on the right in the interaction terms,
\begin{equation}
\label{d2}
\widetilde{G}(x,y)=\widetilde{G}^0(x,y)+\int\: d^4x_3\:\int d^4x_4\: \widetilde{G}(x,x_3)
\widetilde{\Sigma}(x_3,x_4)\widetilde{G}^0(x_4,y).
\end{equation}
Equations. (\ref{d1}) and (\ref{d2}) are the starting points to derive the quantum Boltzmann equations
describing the temporal evolution of the $CP$ violating particle density asymmetries.

\subsubsection{The quantum Boltzmann equations} 

Our goal now is    to find the QBE   for the  generic bosonic  $CP$ violating  current
\begin{equation}
\langle J_\phi^\mu(x) \rangle \equiv i \langle \phi^{\dagger}(x)\stackrel{\leftrightarrow}{\partial}_x^\mu  \phi(x)\rangle\equiv \left[ n_\phi(x), \vec{J}_\phi(x)\right].
\end{equation}
The zero-component of this current $n_\phi$ represents the number density of particles minus the number density of antiparticles and is therefore the quantity which enters the diffusion equations of supersymmetric electroweak baryogenesis. 

Since  the $CP$ violating  current can be expressed in terms of the Green function
$G^{<}_{\phi}(x,y)$ as
\begin{equation}
\label{c1}
\langle J_\phi^\mu(x) \rangle= - \left. \left(\partial_x^\mu - \partial_y^\mu\right) G^{<}_{\phi}(x,y)\right|_{x=y},
\end{equation}
 the problem is reduced to find the QBE for the interacting  Green function $G^{<}_{\phi}(x,y)$ when the system is not in equilibrium. This equation  can be found from (\ref{d1}) by operating  by
$\left(\stackrel{\rightarrow}{\Box}_x+m^2\right)$ on both sides of the equation. Here $m$ represents the bare  mass term of the field $\phi$. On the right-hand  side, this operator acts only on $\widetilde{G}_\phi^0$
\begin{equation}
\label{f1}
\left(\stackrel{\rightarrow}{\Box}_x+m^2\right)\widetilde{G}_\phi(x,y)=\delta^{(4)}(x,y)\widetilde{I}_4+
\int\:d^4 x_3 \widetilde{\Sigma}_\phi(x,x_3)\widetilde{G}_\phi(x_3,y),
\end{equation}
where $I$ is the identity matrix. It is useful to also have an equation of motion for the other variable $y$. This is obtained from (\ref{d2}) by operating  by
$\left(\stackrel{\leftarrow}{\Box}_y+m^2\right)$ on both sides of the equation. We obtain
\begin{equation}
\label{f2}
\widetilde{G}_\phi(x,y)\left(\stackrel{\leftarrow}{\Box}_y+m^2\right)=\delta^{(4)}(x,y)\widetilde{I}_4+
\int\:d^4 x_3 \widetilde{G}_\phi(x,x_3)\widetilde{\Sigma}_\phi(x_3,y).
\end{equation}
The two equations (\ref{f1}) and (\ref{f2}) are the starting point for the derivation of the QBE for the particle asymmetries. Let us extract from  (\ref{f1}) and (\ref{f2}) the equations of motions for the Green function $G^{<}_{\phi}(x,y)$
\begin{eqnarray}
\left(\stackrel{\rightarrow}{\Box}_x+m^2\right)G^{<}_{\phi}(x,y)&=&
\int\:d^4 x_3\left[ \Sigma^{t}_{\phi}(x,x_3)G^{<}_{\phi}(x_3,y)-\Sigma^{<}_{\phi}(x,x_3)G^{\bar{t}}_{\phi}(x_3,y)\right],\\
G^{<}_{\phi}(x,y)\left(\stackrel{\leftarrow}{\Box}_y+m^2\right)&=&
\int\:d^4 x_3\left[ G^{t}_{\phi}(x,x_3)\Sigma^{<}_{\phi}(x_3,y)-G^{<}_{\phi}(x,x_3)\Sigma^{\bar{t}}_{\phi}(x_3,y)\right].
\end{eqnarray}
If we now substract the  two equations and make the identification  $x=y$, the left-hand side is given by 
\begin{equation}
\left. \partial_\mu^x\left[\left(\partial_x^\mu-\partial_y^\mu\right) G^{<}_{\phi}(x,y)\right]\right|_{x=y}=
-\frac{\partial J_\phi^\mu(X)}{\partial X^\mu}=-\left(\frac{\partial n_\phi}{\partial T}+\stackrel{\rightarrow}{\nabla}
\cdot\vec{j}_\phi\right),
\end{equation}
and the QBE for the particle density asymmetry is therefore obtained to be
\begin{equation}
\label{con}
\frac{\partial n_\phi(X)}{\partial T}+\stackrel{\rightarrow}{\nabla}\cdot 
\vec{j}_\phi(X)=-\left. \int\:d^4 x_3\left[\Sigma^{t}_{\phi} G^{<}_{\phi}-\Sigma^{<}_{\phi} G^{\bar{t}}_{\phi}-G^{t}_{\phi} \Sigma^{<}_{\phi}-G^{<}_{\phi} \Sigma^{\bar{t}}_{\phi}\right]\right|_{x=y},
\end{equation}
where we have defined the  centre-of-mass coordinate system
\begin{equation}
\label{dd}
X=(T,\vec{X})=\frac{1}{2}(x+y),\:\:\:\: (t,\vec{r})=x-y.
\end{equation}
Notice  that $T$ now means the centre-of-mass time and not temperature. The identification $x=y$ in Eq. (\ref{con}) is therefore equivalent to require $t=\vec{r}=0$. 

In order to examine the ``scattering term'' on the right-hand side of Eq. (\ref{con}), the first step is to restore all the variable arguments. Setting  $x=y$ in the original notation of $\Sigma_\phi(x,x_3) G_\phi(x_3,y)$ gives  $(X,x_3)(x_3,X)$ for the pair of arguments
\begin{eqnarray}
\label{s}
\frac{\partial n_\phi(X)}{\partial T}+\stackrel{\rightarrow}{\nabla}\cdot 
\vec{j}_\phi(X)=&-&\int\:d^4 x_3\left[\Sigma^{t}_{\phi}(X,x_3) G^{<}_{\phi}(x_3,X)-\Sigma^{<}_{\phi}(X,x_3) G^{\bar{t}}_{\phi}(x_3,X)\right. \nonumber\\
&+&\left. G^{t}_{\phi} (X,x_3)\Sigma^{<}_{\phi}(x_3,X)-G^{<}_{\phi}(X,x_3) \Sigma^{\bar{t}}_{\phi}(x_3,X)\right].
\end{eqnarray}
The next step is to employ the definitions in (\ref{def1}) to express the time-ordered functions $G^{t}_{\phi}$, $G^{\bar{t}}_{\phi}$, $\Sigma ^t_\phi$, and $\Sigma^{\bar{t}}_{\phi}$ in terms of $G^{<}_{\phi}$, $G^{>}_{\phi}$, 
 $\Sigma^{<}_{\phi}$ and  $G^{>}_{\phi}$.  Then the time integrals are separated into whether
$t_3>T$ or $t_3<T$ and the right-hand side of Eq. (\ref{s}) reads
\begin{eqnarray}
&=&-\int\: d^4 x_3\:\left\{\theta(T-t_3)\left[\Sigma^{>}_{\phi} G^{<}_{\phi}+G^{<}_{\phi}\Sigma^{>}_{\phi}-
\Sigma^{<}_{\phi} G^{>}_{\phi}-G^{>}_{\phi}\Sigma^{<}_{\phi}\right]\right.\nonumber\\
&+&\left. \theta(t_3-T)\left[\Sigma^{<}_{\phi} G^{<}_{\phi}+G^{<}_{\phi}\Sigma^{<}_{\phi}-
\Sigma^{<}_{\phi} G^{<}_{\phi}-G^{<}_{\phi}\Sigma^{<}_{\phi}\right]\right\}.
\end{eqnarray}
The term with $t_3>T$ all cancel, leaving $T>t_3$.  Rearranging these terms gives \cite{cpt,cpt1,reviewbau}
\begin{eqnarray}
\label{aaa}
& &\frac{\partial n_\phi(X)}{\partial T}+\stackrel{\rightarrow}{\nabla}\cdot 
\vec{j}_\phi(X)=-\int\: d^3 \vec{x}_3\:\int_{-\infty}^{T}\: dt_3\left[\Sigma^{>}_{\phi}(X,x_3) G^{<}_{\phi}(x_3,X)\right.\nonumber\\
&-&\left. G^{>}_{\phi}(X,x_3) \Sigma^{<}_{\phi}(x_3,X)
+ G^{<}_{\phi}(X,x_3)\Sigma^{>}_{\phi}(x_3,X)-\Sigma^{<}_{\phi}(X,x_3) G^{>}_{\phi}(x_3,X)\right].
\end{eqnarray}
This equation is the QBE for the particle density asymmetry and it can be explicitly checked that, in the particular case in which interactions conserve the number of particles and the latter are neither created nor destroyed, the number asymmetry $n_\phi$ is conserved and  obeys the equation of continuity $\partial n_\phi/\partial T+\stackrel{\rightarrow}{\nabla}\cdot 
\vec{j}_\phi=0$. 
During the production of the baryon asymmetry, however, particle asymmetries are not conserved. This occurs because the 
interactions themselves do not conserve the  particle number asymmetries and  there is some source of CP violation in the system.  
The right-hand side of Eq. (\ref{aaa}), through the general form of the self-energy $\Sigma_\phi$,  contains all the information necessary to describe the temporal evolution of the particle density asymmetries:  particle number
changing reactions and $CP$ violating  source terms,  which will  pop out from the corresponding self-energy $\Sigma_{CP}$. If the interactions of the system do not violate $CP$,   there will be no $CP$ violating  sources and the final baryon asymmetry produced during supersymmetric baryogenesis will be vanishing. 

The kinetic  Eq. (\ref{aaa}) has an obvious interpretation in terms of gain and loss processes.   
What is unusual, however,   is the presence of the integral over the time: the equation is manifestly non-Markovian. Only  the assumption that the relaxation time scale of the particle asymmetry is much longer than the time scale of the non-local kernels leads to a Markovian description. A further approximation, {\it i.e.} taking the upper limit of the 
time integral to $T\rightarrow \infty$,  leads to the familiar Boltzmann  equation. 
The physical interpretation of the integral over the past history of the system is straightforward: it leads to the typical ``memory'' effects which are observed in quantum transport theory \cite{dan,henning}. In  the classical kinetic theory the ``scattering term'' does not include any integral over the past history of the system which is equivalent to assume that any collision in the plasma  does not depend upon the previous ones. On the contrary,   
quantum distributions posses strong memory effects and the thermalization rate obtained from the quantum transport theory may be substantially longer than the one obtained from the classical kinetic theory. As shown in \cite{cpt,cpt1,reviewbau} , 
memory effects play a fundamental role in the determination of the $CP$ violating  sources which fuel baryogenesis when transport properties allow the $CP$ violating  charges to diffuse in front of the bubble wall separating the broken from the unbroken phase at the electroweak phase transition. 

Notice that so far we have not made any approximation and the computation is therefore  valid for all
shapes and sizes of the bubble wall expanding in the thermal bath during a
 first-order electroweak phase transition.

Let us now focus on the  generic fermionic $CP$ violating  current. It  reads
\begin{equation}
\langle J_\psi^\mu(x) \rangle \equiv  \langle \bar{\psi}(x)\gamma^\mu \psi(x)\rangle\equiv \left[ n_\psi(x), \vec{J}_\psi(x)\right],
\end{equation}
where $\psi$ indicates  a Dirac fermion  and $\gamma^\mu$ represent the usual Dirac matrices. Again, the zero-component of this current $n_\psi$ represents the number density of particles minus the number density of antiparticles and is therefore the relevant quantity  for the diffusion equations of supersymmetric electroweak baryogenesis.

We want to find a couple of  equations of motion for the interacting fermionic Green function $\widetilde{G}_\psi(x,y)$ when the system is not in equilibrium. Such  equations  may be found  by applying  the operators $\left(i\stackrel{\rightarrow}{\not  \partial}_x -M\right)$
and $\left(i\stackrel{\leftarrow}{\not  \partial}_y +M\right)$  on both sides of  Eqs. (\ref{d1}) and (\ref{d2}), respectively. Here $M$ represents the bare mass term of the fermion $\psi$. We find
\begin{eqnarray}
\label{c}
\left(i\stackrel{\rightarrow}{\not  \partial}_x -M\right)\widetilde{G}_\psi(x,y)&=&\delta^{(4)}(x,y)\widetilde{I}_4+
\int\:d^4 x_3 \widetilde{\Sigma}_\psi(x,x_3)\widetilde{G}_\psi(x_3,y),\\
\widetilde{G}_\psi(x,y)\left(i\stackrel{\leftarrow}{\not  \partial}_y +M\right)&=& -\delta^{(4)}(x,y)\widetilde{I}_4
-\int\:d^4 x_3 \widetilde{G}_\psi(x,x_3)\widetilde{\Sigma}_\psi(x_3,y).
\end{eqnarray}
We can  now  take the trace over the spinorial indeces of  both sides of the equations, sum up  the two equations above  and finally extract the equation of motion for the Green function $G^{>}_{\psi}$
\begin{eqnarray}
\label{v}
{\rm Tr} \left\{\left[i\stackrel{\rightarrow}{\not  \partial}_x + i\stackrel{\leftarrow}{\not  \partial}_y\right]
G^{>}_{\psi}(x,y)\right\}&=& \int\:d^4 x_3\:{\rm Tr}\left[\Sigma^{>}_\psi(x,x_3)G^t_\psi(x_3,y)-\Sigma^{\bar{t}}_\psi(x,x_3)G^{>}_\psi(x_3,y)\right.\nonumber\\
&-&
\left. G^{>}_\psi(x,x_3)\Sigma^t_\psi(x_3,y)+G^{\bar{t}}_\psi(x,x_3)\Sigma^{>}_\psi(x_3,y)\right].
\end{eqnarray}
Making use of the centre-of-mass coordinate system,  we can work out the left-hand side of Eq.  (\ref{v})
\begin{eqnarray}
& &\left.{\rm Tr}\left[i\stackrel{\rightarrow}{\not  \partial}_x G^{>}_\psi(T,\vec{X},t,\vec{r})+
G^{>}_\psi(T,\vec{X},t,\vec{r})i\stackrel{\leftarrow}{\not  \partial}_y\right]\right|_{t=\vec{r}=0}\nonumber\\
&=&\left. i\left(  \partial^x_\mu + \partial^y_\mu\right) i \langle
\bar{\psi}\gamma^\mu \psi\rangle\right|_{t=\vec{r}=0}\nonumber\\
&=&-\frac{\partial}{\partial X^\mu} \langle
\bar{\psi}(X)\gamma^\mu \psi(X)\rangle\nonumber\\
&=&-\frac{\partial}{\partial X^\mu} J^\mu_\psi.
\end{eqnarray}
The next step is to employ the definitions in (\ref{def2}) to express the time-ordered functions $G^{t}_{\psi}$, $G^{\bar{t}}_{\psi}$, $\Sigma ^t_\psi$, and $\Sigma^{\bar{t}}_{\psi}$ in terms of $G^{<}_{\psi}$, $G^{>}_{\psi}$, 
 $\Sigma^{<}_{\psi}$ and  $G^{>}_{\psi}$. The computation goes along the same lines as the analysis made in the previous section and we get \cite{cpt,cpt1,reviewbau} 
\begin{eqnarray}
\label{b}
& & \frac{\partial n_\psi(X)}{\partial T}+\stackrel{\rightarrow}{\nabla}\cdot 
\vec{j}_\psi(X)=\int\: d^3 \vec{x}_3\:\int_{-\infty}^{T}\: dt_3\:{\rm Tr}\left[\Sigma^{>}_{\psi}(X,x_3) G^{<}_{\psi}(x_3,X)\right.\nonumber\\  &-& \left. G^{>}_{\psi}(X,x_3) \Sigma^{<}_{\psi}(x_3,X)     
+ G^{<}_{\psi}(X,x_3)\Sigma^{>}_{\psi}(x_3,X)-\Sigma^{<}_{\psi}(X,x_3) G^{>}_{\psi}(x_3,X)\right].
\end{eqnarray}
This is the ``diffusion'' equation describing the temporal evolution of a generic fermionic number asymmetry $n_\psi$. As for the bosonic case, all the information regarding particle number violating interactions and $CP$ violating  sources  are  stored in the self-energy $\Sigma_\psi$. 

\subsubsection{The $CP$ violating source for higgsinos and the final baryon asymmetry}

As we mentioned, a strongly first order electroweak
phase transition
can  be achieved in the presence of a top squark
lighter than the top quark.
In order to naturally
suppress its contribution to the parameter $\Delta\rho$ and hence
preserve a good agreement with the precision measurements at LEP,
it should be mainly right-handed. This can be achieved if the left-handed stop soft supersymmetry breaking mass $m_Q$ 
is much larger than $M_Z$. Under this assumption, however, the   right-handed stop contribution to the baryon asymmetry results to be negligible. We will concentrate, therefore, only on the $CP$ violating source for the Higgsino. 

The Higgs fermion  current associated with  neutral
and charged Higgsinos can be written
as
\be
\label{corhiggs}
J^{\mu}_{\widetilde{H}}=\overline{\widetilde{H}}\gamma^\mu \widetilde{H}
\ee
where $\widetilde{H}$ is the Dirac spinor
\be
\label{Dirac}
\widetilde{H}=\left(
\begin{array}{c}
\widetilde{H}_2 \\
\overline{\widetilde{H}}_1
\end{array}
\right)
\ee
and $\widetilde{H_2}=\widetilde{H}_2^0$ ($\widetilde{H}_2^+$),
$\widetilde{H_1}=\widetilde{H}_1^0$ ($\widetilde{H}_1^-$) for
neutral (charged) Higgsinos. The processes in the plasma which  change the Higgsino number are the ones induced by the top Yukawa coupling and by  interactions with the Higgs profile. 
The interactions among the charginos and the charged Higgsinos which are responsible for the $CP$ violating source in the diffusion equation for the Higgs fermion number read
\begin{equation}
{\cal L}=-g_2\left\{\overline{\widetilde{H}}\left[v_1(x)P_L+{\rm e}^{i\theta_\mu} v_2(x) P_R\right]\widetilde{W}\right\}+{\rm h.c.},
\end{equation}
where $\theta_\mu$ is the phase of the $\mu$-parameter and we have indicated 
$\langle H_i^0(x)\rangle$ by $v_i(x)$, $i=1,2$.

Analogously, the interactions among the Bino, the $\widetilde{W}_3$-ino and the neutral Higgsinos are
\begin{equation}
 {\cal L}=-\frac{1}{2}\left\{\overline{\widetilde{H}^0}\left[v_1(x)P_L+{\rm e}^{i\theta_\mu} v_2(x) P_R\right]\left(g_2\widetilde{W}_3-g_1\widetilde{B}\right)\right\}+{\rm h.c.}
\end{equation}

To compute the source for the Higgs fermion number $\gamma_{\widetilde{H}}$ 
 we   perform a  ``Higgs insertion expansion'' around the symmetric phase. At the lowest level of perturbation, the interactions of the charged Higgsino induce 
a contribution to the self-energy of the form (and analogously for the other component 
$\delta\Sigma^{{\rm CP},>}_{\widetilde{H}}$)
\begin{equation}
\label{qf}
\delta\Sigma^{CP,<}_{\widetilde{H}}(x,y)=g^L_{CP}(x,y)P_L G^{0,<}_{\widetilde{W}}(x,y) P_L+
g^R_{CP}(x,y)P_R G^{0,<}_{\widetilde{W}}(x,y) P_R,
\end{equation}
where 
\begin{eqnarray}
\label{qs}
g^L_{CP}(x,y)&=&g_2^2 v_1(x) v_2(y){\rm e}^{-i\theta_\mu},\nonumber\\
g^R_{CP}(x,y)&=&g_2^2 v_1(y) v_2(x){\rm e}^{i\theta_\mu}.
\end{eqnarray}
We have approximated the exact Green function of winos $G_{\widetilde{W}}$ by the equilibrium Green function in the unbroken phase $G^{0}_{\widetilde{W}}$. This is because any departure from thermal equilibrium distribution
functions
is caused at a given point by the passage of the wall and, therefore,
is  ${\cal O}(v_{\omega})$.  Since we will show that the source  is
already linear in $v_{\omega}$,
working with thermal {\it equilibrium} Green 
functions in the unbroken phase
amounts to ignoring terms of higher order in
$v_{\omega}$. This is 
  accurate as long as  the bubble wall is moving slowly in
the plasma. 
Similar formulae hold for the neutral Higgsinos. 

The dispersion
relations of charginos and neutralinos are changed by high
temperature
corrections~\cite{weldon}. Even though fermionic dispersion
relations
are highly nontrivial,  especially when dealing with Majorana fermions \cite{majorana}, relatively simple expressions
for the equilibrium fermionic spectral functions may be given in the limit
in which the damping rate is smaller than the typical self-energy
of the fermionic excitation ~\cite{henning}. 
If we now insert the expressions (\ref{qf}) and  (\ref{qs}) into the QBE 
(\ref{b}), we get 
the $CP$ violating source \cite{cpt,cpt1,reviewbau} 
\begin{eqnarray}
\gamma_{\widetilde{H}}&=&- \int\: d^3\vec{x}_3\:\int_{-\infty}^{T}\: dt_3\:{\rm Tr}\left[\delta\Sigma^{CP,>}_{\widetilde{H}}(X,x_3) G^{0,<}_{\widetilde{H}}(x_3,X)-
G^{0,>}_{\widetilde{H}}(X,x_3) \delta\Sigma^{CP,<}_{\widetilde{H}}(x_3,X)\right.\nonumber\\
&+&\left. G^{0,<}_{\widetilde{H}}(X,x_3)\delta\Sigma^{CP,>}_{\widetilde{H}}(x_3,X)-\delta\Sigma^{CP,<}_{\widetilde{H}}(X,x_3) G^{0,>}_{\widetilde{H}}(x_3,X)\right],
\end{eqnarray}
which contains in the integrand the following function
\begin{equation}
g^L_{CP}(X,x_3)+g^R_{CP}(X,x_3)-
g^L_{CP}(x_3,X)-g^R_{CP}(x_3,X)=2i\sin\theta_\mu\left[v_2(X)v_1(x_3)-v_1(X)v_2(x_3)\right],
\end{equation}
which vanishes if  ${\rm Im}(\mu)=0$ and if the $\tan\beta(x)$ is a constant  along the Higgs profile. 

In order to deal with analytic expressions, we can work out
the thick wall limit and simplify the expressions obtained above
by performing a derivative expansion
\begin{equation}
\label{expansion}
v_i(x_3)= \sum_{n=0}^{\infty}\frac{1}{n!}\; \frac{\partial^n}
{\partial (X^\mu)^n} v_i(X)\left(x^\mu_3-X^\mu\right)^n .
\end{equation}
The term
with no
derivatives vanishes in the expansion (\ref{expansion}),
$v_2(X)v_1(X)-v_1(X)v_2(X)= 0$, which means that the static
term in the derivative expansion  does not contribute
to the source.
For a smooth Higgs profile, the derivatives with
respect to the time coordinate and $n>1$ are associated with higher
powers of $v_{\omega}/L_{\omega}$,  where $v_{\omega}$ and $L_{\omega}$ are the velocity and the width of the bubble wall, respectively.  Since the typical time scale of the processes giving rise to the source is given by the thermalization time of the higgsinos $1/\Gamma_{\widetilde{H}}$, 
the approximation is good for values of
$L_{\omega}\Gamma_{\widetilde{H}}/v_{\omega} \gg 1$.
In other words, this expansion is valid only when the mean free path of the higgsinos in the plasma 
 is smaller than the scale of variation of the Higgs
background determined by the wall thickness, $L_{\omega}$,
and the wall velocity $v_{\omega}$. The  term corresponding to  $n=1$  in the expansion (\ref{expansion})  gives a contribution to the source proportional to the function 
\begin{equation}
v_1(X)\partial_X^\mu v_2(X)- v_2(X)\partial_X^\mu v_1(X)
\equiv v^2(X) \partial_X^\mu\beta(X),
\end{equation}
which  should vanish smoothly for values of $X$ outside the
bubble wall. Here  we have denoted $v^2\equiv v_1^2+ v_2^2$. 
Since the variation of the Higgs fields is due to  the
expansion of the bubble wall through the thermal bath,
the source $\gamma_{\widetilde{H}}$
will be linear in $v_{\omega}$. The corresponding contribution tot he $CP$ violating source reads 
\begin{equation}
\label{source2}
\gamma_{\widetilde{H}}(X) =
{\rm Im}(\mu)\: \left[ v^2(X)\dot{\beta}(X) \right]
\left[ 3 M_2 \; g_2^2 \; {\cal I}^{\widetilde{W}}_{\widetilde{H}}
 +       M_1 \; g_1^2 \; {\cal I}^{\widetilde{B}}_{\widetilde{H}}
\right],
\end{equation}
where
\begin{eqnarray}
{\cal I}^{\widetilde{W}}_{\widetilde{H}} & = & \int_0^\infty dk
\frac{k^2}
{2 \pi^2 
\omega_{\widetilde{H}} \omega_{\widetilde{W}}} \nonumber\\
&\left[ \phantom{\frac{1}{2^2}} \right.&
 \left(1 - 2 {\rm Re}(f^0_{\widetilde{W}}) \right)
I(\omega_{\widetilde{H}},\Gamma_{\widetilde{H}},
\omega_{\widetilde{W}},\Gamma_{\widetilde{W}})+
\left(1 - 2 {\rm Re}(f^0_{\widetilde{H}}) \right)
I(\omega_{\widetilde{W}},
\Gamma_{\widetilde{W}},\omega_{\widetilde{H}},
\Gamma_{\widetilde{H}}) \nonumber\\
&+&
2 \left( {\rm Im}(f^0_{\widetilde{H}}) +
{\rm Im}(f^0_{\widetilde{W}}) \right)
G(\omega_{\widetilde{H}},
\Gamma_{\widetilde{H}},
\omega_{\widetilde{W}},\Gamma_{\widetilde{W}})
\left.\phantom{\frac{1}{2^2}} \right]
\nonumber\\
\end{eqnarray}
and $\omega^2_{\widetilde{H}(\widetilde{W})}=k^2+ |\mu|^2
(M_2^2)$ while $f^0_{\widetilde{H}(\widetilde{W})} =
1/\left[\exp\left(\omega_{\widetilde{H}(\widetilde{W})}/T
+ i \Gamma_{\widetilde{H}(\widetilde{W})}/T \right)
+ 1 \right]$.
The functions $I$ and $G$ are given by 
\begin{eqnarray}
\label{v1}
I(a,b,c,d) &=& \frac{1}{2}\frac{1}{\left[(a+c)^2 + (b+d)^2 \right]}
\sin\left[ 2{\rm arctan}\frac{a+c}{b+d}\right]\nonumber\\
&+&\frac{1}{2}\frac{1}{\left[(a-c)^2 + (b+d)^2 \right]}
\sin\left[ 2{\rm arctan}\frac{a-c}{b+d}\right],\nonumber\\
G(a,b,c,d)=&-&\frac{1}{2}\frac{1}{\left[(a+c)^2 + (b+d)^2 \right]}
\cos\left[ 2{\rm arctan}\frac{a+c}{b+d}\right]\nonumber\\
&-&\frac{1}{2}\frac{1}{\left[(a-c)^2 + (b+d)^2 \right]}
\cos\left[ 2{\rm arctan}\frac{a-c}{b+d}\right].
\end{eqnarray}
Notice that the function $G(\omega_{\widetilde{H}},\Gamma_{\widetilde{H}},
\omega_{\widetilde{W}},\Gamma_{\widetilde{W}})$ has a peak for $\omega_{\widetilde{H}}\sim\omega_{\widetilde{W}}$.  
This resonant behaviour is associated to the fact that 
the Higgs background  is
carrying a very low momentum (of order of the inverse of the bubble wall
width $L_\omega$) and to the 
possibility of absorption or emission of Higgs quanta by the
propagating supersymmetric particles. The resonance  can only take place when  the higgsino and the wino  do not differ too much in mass.
By using the Uncertainty Principle, it is easy to understand that the
width of this resonance is expected
to be proportional to the thermalization rate  of the particles giving rise to
the baryon asymmetry.

The  damping rate of charged and neutral
Higgsinos is expected to be of the order of $5\times 10^{-2} T$. 
The Bino contribution may be obtained from the above
expressions by replacing $M_2$ by $M_1$. The $CP$ violating source for the Higgs fermion number is enhanced  if   $M_{2}, M_{1}\sim \mu$ and  low momentum particles are transmitted over the distance $L_\omega$. 
This means that    the classical approximation is not  entirely adequate to
describe the quantum interference nature of $CP$ violation and only a quantum approach is suitable for the computation of the building up of the $CP$ violating sources. 
Notice that the source is built up integrating over all the history of the system. This leads  to  ``memory effect'' that are responsible for some enhancement of the final baryon asymmetry.   These memory effects  lead to ``relaxation'' times  for the $CP$ violating sources which   are   typically longer than the ones dictated by the thermalization rates of the particles in the  thermal bath.  In fact, this observation
is valid for all the processes described by the ``scattering'' term in the right-handed side of the quantum diffusion equations. 
The slowdown of the relaxation processes  may help to keep the system out of equilibrium for longer times and therefore enhance the final baryon asymmetry. There are  two more  reasons why one should expect  quantum relaxation times to be  longer than the ones predicted by the classical approach. First, the decay of the Green's functions  as functions of the difference of the time arguments: an exponential decay is found in thermal equilibrium when one ignore the frequency dependence of self-energies in the spectral functions,   {\it e.g.}  $\left| G^{>}({\bf k},t,t^\prime)\right|\sim \left| G^{>}({\bf k})\right| \times {\rm exp}\left[-\Gamma({\bf k},\omega)|t-t^\prime|\right]$. The decay of the Green's functions restrict the range of the time integration for the scattering term, reduces the integrals and, therefore, the change  of the local particle number densities as a function of time.  The second effect is the rather different oscillatory behaviour of the functions $G^{>}$ and $G^{<}$ for a given momentum, as functions of the time argument difference.

As we have previously mentioned,  the final baryon asymmetry (\ref{final}) depends sensitively on the parameter ${\cal A}$.
The parameter ${\cal A}$ computed from the higgsino source is
\begin{eqnarray}
{\cal A}&\propto& \frac{2 f(k_i)\Gamma_{\widetilde{H}}}{\overline{D} \; \lambda_{+}} I,\nonumber\\
I &\equiv &\int_0^{\infty} du\; v^2(u)\frac{d\beta(u)} {d u} e^{-\lambda_+ u}\simeq \int_0^{\infty} du\; v^2(u)\frac{d\beta(u)}{d u},
\end{eqnarray}
where $f(k_i)$ is a coefficient depending upon the number of degrees of freedom present in the thermal bath. The integral $I$ has been computed  including two-loop effects in ref. \cite{mariano} and results to be $I\simeq 10^{-2}$ for  $m_A = 150$--200 GeV. 
The final baryon asymmetry turns out to be \cite{cpt1}
\be
\frac{n_B}{s}\simeq \left(\frac{|\sin(\phi_\mu)|}{10^{-3}}\right)
\:4\times 10^{-11},
\ee
for $v_\omega\simeq 1$. 
It is intriguing that these small  values of the phases
are perfectly  consistent with the constraints from the
electric dipole moment of the neutron and   squarks of the
first and second generation as light as $\sim 100$ GeV may be tolerated.

\section{Conclusions}

In these lectures we have learned that cosmology  provides really strong arguments in favour of the nonconservation of the baryon number. The SM of weak interactions, which is so successfull in explaining the  experimental data obtained at  accelerator machines operating at energy scales of about 100 GeV, seems unable to explain the observed baryon asymmetry of the Universe. This is a very strong indication that there is some new, yet undiscovered, physics beyond the SM. We do not know  whether this is just the low energy supersymmetric extension of the SM. If so, we can draw tight constraints on the Higgs spectrum of the MSSM and 
the next generation of accelerator machines, such as  LHC, will tell us if this is a tenable option.  
It might be that this cosmological  puzzle has been taken care of 
by some new physics at energy scales much higher than the weak scale,  the GUT scale, as suggested by  gauge coupling unification. Even though 
this option 
is not testable at  particle colliders, the most striking evidence of baryon number violation  might come from the detection of proton decay. It is very exciting that  in the next few years we will be able to confirm (or disprove) some  of the theories of baryogenesis.

\newpage

\underline{Acknowledgements}: The author would like to express his appreciation to  the organizers of the School for providing the students and the lecturers with such an excellent
and stimulating environment. He also thanks the students for their questions and  enthusiasm. 
He is  grateful  M. Carena, A. Linde, R. Kolb,   M. Quiros, I. Tkachev, I. Vilja and  C.E.M. Wagner for many fruitful interactions and in particular to  R. Kolb whose never-ending scepticism about he idea of electroweak baryogenesis spurred, is spurring and will always spur his efforts. 

\def\NPB#1#2#3{Nucl. Phys. {\bf B#1}, #3 (19#2)}
\def\PLB#1#2#3{Phys. Lett. {\bf B#1}, #3 (19#2) }
\def\PLBold#1#2#3{Phys. Lett. {\bf#1B} (19#2) #3}
\def\PRD#1#2#3{Phys. Rev. {\bf D#1}, #3 (19#2) }
\def\PRL#1#2#3{Phys. Rev. Lett. {\bf#1} (19#2) #3}
\def\PRT#1#2#3{Phys. Rep. {\bf#1} (19#2) #3}
\def\ARAA#1#2#3{Ann. Rev. Astron. Astrophys. {\bf#1} (19#2) #3}
\def\ARNP#1#2#3{Ann. Rev. Nucl. Part. Sci. {\bf#1} (19#2) #3}
\def\mpl#1#2#3{Mod. Phys. Lett. {\bf #1} (19#2) #3}
\def\ZPC#1#2#3{Zeit. f\"ur Physik {\bf C#1} (19#2) #3}
\def\APJ#1#2#3{Ap. J. {\bf #1} (19#2) #3}
\def\AP#1#2#3{{Ann. Phys. } {\bf #1} (19#2) #3}
\def\RMP#1#2#3{{Rev. Mod. Phys. } {\bf #1} (19#2) #3}
\def\CMP#1#2#3{{Comm. Math. Phys. } {\bf #1} (19#2) #3}


\newpage

\vspace{2cm}
\centerline{{\bf Solution to the Exercises}}

\vspace{2cm}

{\it 1)} Suppose that the baryon number $B$ is conserved by the interactions. This means that the baryon number commutes with the hamiltonian of the system $H$, $[B,H]=0$. Therefore, supposing that $B(t_0)$=0, we have $B(t)\propto
\int_{t_0}^t\:[B,H] d t^\prime=0$ at all times and no baryon number production may take place.

\vspace{1cm}

{\it 2)} Let us consider a model universe with two components: inflaton field
energy, $\rho_\phi$ and  radiation energy density, $\rho_R$.  We
will assume that the decay rate of the inflaton field energy density
is $\Gamma_\phi$.  We will also assume that the light degrees
of freedom are in local thermodynamic equilibrium.  

With the above assumptions, the Boltzmann equations describing the
redshift and interchange in the energy density among the different
components is
\begin{eqnarray}
& &\dot{\rho}_\phi + 3H\rho_\phi +\Gamma_\phi\rho_\phi = 0 
	\nonumber \\
& & \dot{\rho}_R + 4H\rho_R - \Gamma_\phi\rho_\phi=0,
\end{eqnarray}
where dot denotes time derivative.

It is useful to introduce the dimensionless constant, $\alpha_\Phi$
defined in terms of $\Gamma_\phi$ as
\begin{equation}
\Gamma_\phi = \alpha_\phi M_\phi.
\end{equation}
For a reheat temperature much smaller than $M_\phi$, $\Gamma_\phi$
must be small.

 It is also convenient to work with dimensionless
quantities that can absorb the effect of expansion of the universe.
This may be accomplished with the definitions
\begin{equation}
\Phi \equiv \rho_\phi M_\phi^{-1} a^3 \ ; \quad
R    \equiv \rho_R a^4.
\end{equation}
It is also convenient to use the scale factor, rather than time, for
the independent variable, so we define a variable $x = a M_\phi$.
With this choice the system of equations can be written as (prime
denotes $d/dx$)
\begin{eqnarray}
\label{eq:SYS}
\Phi' & = & - c_1 \ \frac{x}{\sqrt{\Phi x + R}}   \ \Phi \nonumber \\
R'    & = &   c_1 \ \frac{x^2}{\sqrt{\Phi x + R}} \ \Phi.
\end{eqnarray}
The constant $c_1$ is  given by 
\begin{equation}
c_1 = \sqrt{\frac{3}{8\pi}} \frac{M_{{\rm Pl}}}{M_\phi}\alpha_\phi.
\end{equation}

It is straightforward to solve the system of equations in Eq.\
(\ref{eq:SYS}) with initial conditions at $x=x_I$ of $R(x_I)=0$
and $\Phi(x_I)=\Phi_I$.  It is convenient to express
$\rho_\phi(x=x_I)$ in terms of the expansion rate at $x_I$, which
leads to
\begin{equation}
\Phi_I = \frac{3}{8\pi} \frac{M_{{\rm P}}^2}{M_\phi^2}
		\frac{H_I^2}{M_\phi^2} x_I^3.
\end{equation}
The numerical value of $x_I$ is irrelevant.

Before solving the system of equations, it is useful to consider the
early-time solution for $R$.  Here, by early time, we mean $H \gg
\Gamma_\Phi$, {\it i.e.}, before a significant fraction of the comoving
coherent energy density is converted to radation.  At early times
$\Phi \simeq \Phi_I$, and $R\simeq X \simeq 0$, so the equation for
$R'$ becomes $R' = c_1 x^{3/2} \Phi_I^{1/2}$.  Thus, the early time
solution for $R$ is simple to obtain:
\begin{equation}
\label{eq:SMALLTIME}
R \simeq \frac{2}{5} c_1 
     \left( x^{5/2} -  x_I^{5/2} \right) \Phi_I^{1/2}
			 \qquad (H \gg \Gamma_\phi) \ .
\end{equation}
Now we may express $T$ in terms of $R$ to yield the early-time
solution for $T$:
\begin{equation}
\frac{T}{M_\phi} \simeq \left(\frac{12}{\pi^2g_*}\right)^{1/4}
c_1^{1/4}\left(\frac{\Phi_I}{x_I^3}\right)^{1/8} 
	\left[ \left(\frac{x}{x_I}\right)^{-3/2} -               
                \left(\frac{x}{x_I}\right)^{-4} \right]^{1/4} 
		\qquad (H \gg \Gamma_\Phi) \ . 
\end{equation}
Thus, $T$ has a maximum value of 
\begin{eqnarray}
\frac{T_{MAX}}{M_\phi}& = & 0.77 	\left(\frac{12}{\pi^2g_*}\right)^{1/4} c_1^{1/4} 		\left(\frac{\Phi_I}{x_I^3}\right)^{1/8} \nonumber \\
   & = & 0.77 \alpha_\phi^{1/4}\left(\frac{9}{2\pi^3g_*}\right)^{1/4}
	\left( \frac{M_{{\rm P}}^2H_I}{M_\phi^3}\right)^{1/4} \ ,
\end{eqnarray}
which is obtained at $x/x_I = (8/3)^{2/5} = 1.48$.  It is also
possible to express $\alpha_\phi$ in terms of $T_{RH}$ and obtain
\begin{equation}
\frac{T_{MAX}}{T_{RH}}=	0.77 \left(\frac{9}{2\pi^3g_*}\right)^{1/4}
		\left(\frac{H_I M_{{\rm P}}}{T_{RH}^2}\right)^{1/4} \ .
\end{equation}

For an illustration, in the simplest model of chaotic inflation $H_I^2
\sim M_{{\rm P}} M_\phi$ with $M_\phi \simeq 10^{13}$GeV, which leads to
$T_{MAX}/T_{RH} \sim 2\times 10^3 (200/g_*)^{1/4}$ for $T_{RH} =
10^9$GeV.

We can see from Eq.\ (\ref{eq:SMALLTIME}) that for $x/x_I>1$, in the
early-time regime $T$ scales as $a^{-3/8}$.  So entropy is created in
the early-time regime.  So if one is producing a massive particle
during reheating it is necessary to take into account the fact that
the maximum temperature is greater than $T_{RH}$, and during the
early-time evolution $T\propto a^{-3/8}$.

\vspace{1cm}

{\it 3)}  The euclidean action is given by
\be
S_3=4\pi \int_0^R \:r^2\:dr\left[\frac{1}{2}\left(\frac{d\phi}{dr}\right)^2 +V(\phi)\right],
\ee
where $R$ is the radius of the bubble. If we now indicate by $\delta R$ the thickness of the bubble wall and $\Delta V=V(\phi_2)-V(\phi_1)<0$, we get 
\be 
S_3\sim 2\pi R^2\delta R\left(\frac{\delta\phi}{\delta R}\right)^2\delta R
+\frac{4\pi R^3\Delta V}{3},
\ee
where $\delta\phi=\phi_2-\phi_1$. Suppose now that the bubbles are thick, that is $\delta R\sim R$. In such a case
\be
\label{thick}
S_3\sim 2\pi R(\delta\phi)^2+\frac{4\pi R^3\Delta V}{3}.
\ee
The critical radius $R_c$ is obtained as the maximum of the action (\ref{thick})
\be
R_c\sim \frac{\delta\phi}{\sqrt{-2\Delta V}}.
\ee

\vspace{1cm}

{\it 4)} We introduce a chemical potential for any
particle which takes part to fast processes, and then reduce the number
of linearly independent chemical potentials by solving the corresponding system
of equations.    Finally, we can express the abundances of any
particle in equilibrium in terms of the remaining linear independent chemical
potentials, corresponding to the conserved charges of the system.

Since strong interactions are in equilibrium inside the bubble wall,  we  can chose the
same chemical potential for quarks of the same flavour but different color, and
set to zero the chemical potential for gluons.
Moreover, since inside the bubble wall $SU(2)_L\times U(1)_Y$ is broken, the
chemical potential for the neutral Higgs scalars vanishes\footnote{This is
true if chirality flip interactions, or processes like $Z\rightarrow Z^* h$,
are
sufficiently fast.}.

The other fast processes, and the corresponding chemical potential equations
are:

{\it i)} top Yukawa:
\beq
\begin{array}{ccrl}
t_L + H_2^0 \leftrightarrow t_R + g, &
\:\:\:\:\:&
(\mu_{t_L} =& \mu_{t_R}),\\
b_L + H^+ \leftrightarrow t_R + g,&
\:\:\:\:\:&
(\mu_{t_R}=&\mu_{b_L} + \mu_{H^+}),
\end{array}\\
\label{chem1}
\eeq
{\it ii)} $SU(2)_L$ flavour diagonal:
\beq
\begin{array}{ccrl}
e_L^i\leftrightarrow \nu_L^i + W^-,&
\:\:\:\:\:&
(\mu_{\nu_L^i} =& \mu_{e_L^i} + \mu_{W^+}),\\
u_L^i\leftrightarrow d_L^i + W^+,&
\:\:\:\:\:&
(\mu_{u_L^i} =& \mu_{d_L^i} + \mu_{W^+}),\\
H_2^0 \leftrightarrow H^+ + W^-,&
\:\:\:\:\:&
(\mu_{H^+} =&\mu_{W^+}),\\
H_1^0 \leftrightarrow H^- + W^+,&
\:\:\:\:\:&
(\mu_{H^-} =&-\mu_{W^+}),\\
\end{array}
(i=1,\:2,\:3).
\label{chem2}
\eeq
Neutral current gauge interactions are also in equilibrium, so we have zero
chemical potential for the photon and the $Z$ boson.

Imposing the above constraints, we can reduce the number of
independent chemical potentials to four, $\mu_{W^+}$, $\mu_{t_L}$,
$\mu_{u_L}\equiv 1/2 \sum_{i=1}^2 \mu_{u_L^i}$, and $\mu_{e_L}\equiv
1/3 \sum_{i=1}^3 \mu_{e_L^i}$. These quantities correspond to the four
linearly independent conserved charges of the system. Choosing the
basis $Q$, $(B-L)$, $(B+L)$, and $BP\equiv B_3 - 1/2
(B_1+B_2)$, where the primes indicate that only particles in
equilibrium contribute to the various charges, and introducing the
respective chemical potentials, we can go to the new basis using the
relations
\beq
\left\{
\begin{array}{ccl}
\mu_{Q} & =& 3 \mu_{t_L} + 2 \mu_{u_L} - 3 \mu_{e_L} + 11 \mu_{W^+},\\
\mu_{(B-L)} &=& 3 \mu_{t_L} + 4 \mu_{u_L} - 6 \mu_{e_L} - 6 \mu_{W^+},\\
\mu_{(B+L)} &=& 3 \mu_{t_L} + 4 \mu_{u_L} + 6 \mu_{e_L},\\
\mu_{BP} &  =& 3 \mu_{t_L} - 2 \mu_{u_L}.
\end{array}
\right.
\label{conserva}
\eeq
If sphaleron transitions were fast, then we could eliminate a further
chemical potential through the constraint
\beq
3 \sum_{i=1}^3 \mu_{u_L^i} +  3\sum_{i=1}^3 \mu_{d_L^i} +  \sum_{i=1}^3
\mu_{e_L^i}
= 0.
\label{sfalc}
\eeq
In this case, the value of $(B+L)$ would be determined by that of the
other three charges according to the relation
\beq
(B+L)_{{\rm EQ}} = \frac{3}{80} Q + \frac{7}{20} BP - \frac{19}{40}
(B-L).
\label{equilibrium}
\eeq
The above result should not come as a surprise, since we already know
 that a non zero value for $B-L$ gives
rise to a non zero $(B+L)$ at equilibrium. Stated in other words,
sphaleron transitions
erase the baryon asymmetry only if any conserved charge of the system
has vanishing thermal average, otherwise the equilibrium point
lies at $(B+L)_{{\rm EQ}} \neq 0$.

At high temperature $(\mu_i \ll T)$ the free energy of the system is
given by
\beq
\begin{array}{rl}
F = \frac{T^2}{12} &\left[3 \mu_{e_L}^2 + 3 \mu_{\nu_L}^2 +
6\mu_{u_L}^2 + 3 \mu_{t_L}^2 + 3\mu_{t_R}^2 + 6\mu_{d_L}^2 +
3\mu_{b_L}^2 \right.\\
&\left.+ 6\mu_{W^+}^2 + 2\mu_{H^+}^2 + 2\mu_{H_1^0}^2 +
2\mu_{H_2^0}^2\right].
\end{array}
\eeq
Using (\ref{chem1}), (\ref{chem2}) and (\ref{conserva}) to express
the chemical potentials in terms of the four conserved charges in
(\ref{conserva}) we obtain the free energy as a function  of the
density of $(B+L)= \mu_{B+L} T^2/6$,
\beq
F\left[(B+L)\right] = 0.46 \frac{\left[(B+L) -
(B+L)_{{\rm EQ}}\right]^2}{T^2} +\:\:\:{\rm  constant} \:\:{\rm terms},
\eeq
where the {\it ``constant terms"} depend on $Q$, $(B-L)$, and $BP$
but not on $(B+L)$,
and $(B+L)_{{\rm EQ}}$ is given by (\ref{equilibrium}). 

From the above expression, we can see that
\be
\frac{d n_{B+L}}{dt}\propto -\frac{\Gamma_{{\rm sp}}}{T}\frac{\partial F}{\partial (B+L)}\propto -\frac{\Gamma_{{\rm sp}}}{T}\left[(B+L)-(B+L)_{{\rm EQ}}\right].
\ee

\vspace{1cm}

{\it 5)}  The contribution to the $E$ parameter of the potential (\ref{potential}) is given generically by, see Eq. (\ref{pott}),
\be
-\frac{T}{12\pi}\sum_i\: n_i\:\left[m_i^2(\phi)+\Pi_i(T)\right]^{3/2},
\ee
for a generic bosonic particle $i$ with plasma mass $\Pi_i(T)$. In the case of the right-handed stop, we get
\be
\delta E=-2\:N_c\:\frac{T}{12\pi}\left[\mstop^2+\Pi_R(T)\right]^{3/2},
\ee
where $N_c=3$ is the number of color and $\mstop^2$ is given in (\ref{app}). The upper bound on the contribution to the $E$ parameter from the right-handed stops is obtained when 
 $m_U^2<0$ and  $m^{\rm eff}_{\;\widetilde{t}}=m_U^2+\Pi_R(T)\simeq 0$. This gives
\be
\delta E\lsim \frac{h_t^3}{2\pi}\left(1 -
\widetilde{A}_t^2/m_Q^2\right)^{3/2}.
\ee
Using now the fact that $m_t=h_t v$ and that $\langle\phi(T_c)\rangle/T_c=2E/\lambda\simeq 4v^2 E/m_h^2$, we get 
Eq. (\ref{totalE}).

\end{document}